\documentclass[11pt,a4paper]{article}
\pdfoutput=1
\usepackage{jcappub}

\usepackage{color}
\usepackage{amsfonts, amsmath}
\usepackage{booktabs}
\usepackage[english]{babel}
\usepackage{colortbl}
\usepackage{epsfig}
\usepackage{epstopdf}
\usepackage{graphicx}
\usepackage{subfig}
\usepackage{lipsum}
\usepackage{hyperref}
\usepackage{ulem}

\usepackage[T1]{fontenc} 
\usepackage[load-configurations=astronomy, range-units=brackets, range-phrase=-, per-mode=reciprocal, mode=math]{siunitx}

\def\Mpc{\, h^{-1} \, {\rm Mpc}}
\newcommand{\vb}[1]{\mathbf{#1}}
\def\kMpc{\, h \, {\rm Mpc}^{-1}}
\newcommand{\eq}[1]{eq.~(\ref{#1})}
\newcommand{\eqnb}[1]{eq.~\ref{#1}}
\newcommand{\citeref}[1]{ref.~\citep{#1}}
\newcommand{\vbunit}[1]{\hat{\vb{#1}}}

\newcommand{\fig}[1]{figure~\ref{#1}}
\newcommand{\tabl}[1]{table~\ref{#1}}
\newcommand{\sect}[1]{section~\ref{#1}}
\newcommand{\appen}[1]{appendix~\ref{#1}}

\newcommand{\HH}{{\mathcal{H}}}
\newcommand{ \bn}{{\bf n }}
\newcommand{\ndv}{v_{\parallel}}
\newcommand{\dotndv}{\dot{v}_{\parallel}}
\newcommand{ \bx}{{\bf x }}
\newcommand{ \bs}{{\bf s }}
\newcommand{ \bk}{{\bf k }}
\newcommand{ \br}{{\bf r }}
\newcommand{ \bv}{{\bf v }}
\newcommand{ \bq}{{\bf q }}
\newcommand{\ndvthree}{v_{\parallel}^{(3)}}

\newcommand{\bea}{\begin{eqnarray}}
\newcommand{\eea}{\end{eqnarray}}
\newcommand{\be}{\begin{equation}}
\newcommand{\ee}{\end{equation}}

\title{Modeling relativistic contributions to the halo power spectrum dipole}

\author[a]{Florian Beutler,}

\author[b,c,d]{Enea Di Dio}

\affiliation[a]{Institute of Cosmology \& Gravitation, University of Portsmouth,
Dennis Sciama Building, Burnaby Road, Portsmouth PO1 3FX, UK}
\affiliation[b]{
Center for Theoretical Astrophysics and Cosmology, Institute for Computational Science, University of Zurich, Winterthurerstrasse 190, CH-8057 Zurich, Switzerland}
\affiliation[c]{Physics Division, Lawrence Berkeley National Laboratory, Cyclotron Rd, Berkeley, CA 94720}
\affiliation[d]{Berkeley Center for Cosmological Physics and Department of Physics, University of California, Berkeley, CA 94720}

\emailAdd{florian.beutler@port.ac.uk}
\emailAdd{enea.didio@uzh.ch}

\abstract{
We study the power spectrum dipole of an N-body simulation which includes relativistic effects through ray-tracing and covers the low redshift Universe up to $z_{\rm max} = 0.465$ (RayGalGroup simulation). We model relativistic corrections as well as wide-angle, evolution, window and lightcone effects. Our model includes all relativistic corrections up to third-order including third-order bias expansion. We consider all terms which depend linearly on $\mathcal{H}/k$ (weak field approximation). We also study the impact of 1-loop corrections to the matter power spectrum for the gravitational redshift and transverse Doppler effect. We found wide-angle and window function effects to significantly contribute to the dipole signal. When accounting for all contributions, our dipole model can accurately capture the gravitational redshift and Doppler terms up to the smallest scales included in our comparison ($k=0.48\kMpc$), while our model for the transverse Doppler term is less accurate. We find the Doppler term to be the dominant signal for this low redshift sample. We use Fisher matrix forecasts to study the potential for the future Dark Energy Spectroscopic Instrument (DESI) to detect relativistic contributions to the power spectrum dipole. A conservative estimate suggests that the DESI-BGS sample should be able to have a detection of at least $4.4\sigma$, while more optimistic estimates find detections of up to $10\sigma$. Detecting these effects in the galaxy distribution allows new tests of gravity on the largest scales, providing an interesting additional science case for galaxy survey experiments.}

\begin{document}
\maketitle

\section{Introduction}
\label{sec:intro}

Galaxy redshift surveys have now matured into one of the most powerful tools to test cosmological models. The Baryon Oscillation Spectroscopic Survey (BOSS~\citep{Dawson2012:1208.0022v3}) used $\sim 1\,000\,000$ Luminous Red Galaxies (LRG) to measure the Baryon Acoustic Oscillation (BAO) scale with \%-level precision~\citep{Alam2016:1607.03155v1} at $z\sim 0.5$ and the extended Baryon Oscillation Spectroscopic Survey (eBOSS~\citep{Dawson2015:1508.04473v2}) has extended such studies to $z>1$~\citep{Ata2017:1705.06373v2} using Emission Line Galaxies (ELGs) and Quasars. Future galaxy surveys like DESI~\citep{DESI2016:1611.00036v2} and Euclid~\citep{Laureijs2011:1110.3193v1} will increase the number of galaxies by more than an order of magnitude, sampling a significant portion of the low redshift Universe. This significant increase in statistical power requires careful modeling of effects that influence the observed galaxy positions (angles and redshifts). Here we will study relativistic corrections to Newtonian halo clustering and quantify the impact of these corrections on the halo power spectrum dipole. The final step of our analysis uses the model for the power spectrum dipole to perform Fisher forecasts for the DESI-BGS sample and quantifies the possibility for a first detection of relativistic effects in halo/galaxy clustering.

The most prominent correction to the distance redshift relation is sourced by the local velocity field, which impacts the measured redshift through a Doppler effect also known as redshift-space distortions (RSD~\citep{Kaiser1987}). Measurements of redshift-space distortions are a sensitive probe of the local matter density including effects of the neutrino mass~\citep{Seljak2006:astro-ph/0604335v4,Reid2009:0907.1659v2,Beutler2013:1312.4611v2,Beutler2014:1403.4599v2,Beutler2016:1607.03150v1,Alam2016:1607.03155v1}. 

Besides such Newtonian corrections, we also have to consider relativistic effects experienced by photons before they reach the telescopes, such as lensing and gravitational redshift. The relativistic effects at first order in perturbation theory have been derived in~\citep{Yoo:2012se,Yoo2010:1009.3021v1,Bonvin2011:1105.5280v3,Challinor2011:1105.5292v2}. Some of these effects are integrated over the distance of the source to the observer, such as the lensing and ISW effects, while others are related to the gravitational potential or velocity of the source and observer. Most of these corrections are suppressed, compared to the Newtonian terms, by factors of $\mathcal{H}/k$, where $\mathcal{H}\equiv aH$ is the conformal Hubble parameter. This limits the impact of these corrections to the largest scales in the power spectrum (small $k$). Moreover, the first non-vanishing corrections to the auto-power spectrum are suppressed by $(\mathcal{H}/k)^2$ making the Newtonian approximation highly accurate for most scales. However, it has been pointed out in~\citep{McDonald2009:0907.5220v1} that the cross-power spectrum between tracers with different mass-halo relations (galaxy bias) contains non-vanishing corrections proportional to $\mathcal{H}/k$ in the imaginary power spectrum. This imaginary part of the power spectrum shows up as odd power spectrum multipoles, like the dipole or octopole, when using the common Legendre multipole expansion. Hence such odd multipoles provide a promising observable for relativistic effects in halo and galaxy clustering.

Here we will use N-body simulations including ray tracing (RayGalGroup simulation~\citep{Breton2018:1803.04294v2}) to test perturbation theory (PT) based models of the large scale power spectrum. Similar comparisons have been performed in configuration space~\citep{Breton2018:1803.04294v2,DiDio:2018zmk}. Given that most relativistic effects are located on very large scales, Fourier-space is the natural choice for such measurements, since large-scale linear modes are independent in the Fourier-basis. Our model includes all relativistic corrections up to third-order including third-order bias expansion. We consider all terms which depend linearly on $\mathcal{H}/k$ (weak field approximation). While we restrict all terms proportional to $v_{\parallel}$ (Doppler term) to linear theory, we include 1-loop corrections to the matter power spectrum for the potential term (gravitational redshift) and transverse Doppler effect~\citep{DiDio:2018zmk,DiDio:2020}.

This paper is organized as follows. In \sect{sec:estimator} we introduce the estimator for the cross-power spectrum multipoles. In \sect{sec:theory} we discuss details about the PT-based power spectrum model we employ in this paper. In \sect{sec:sims} we review details about the RayGalGroup simulation which we will use to test our perturbative model in \sect{sec:analysis}. We discuss our findings in \sect{sec:discussion}. In \sect{sec:forecast} we employ our power spectrum dipole model to perform Fisher matrix forecasts for the DESI experiment and we conclude in \sect{sec:conclusion}.

Whenever transforming redshifts and angles into comoving coordinates we use the cosmological parameters of the RayGalGroup simulation, which is a $\Lambda$CDM model with $h=0.72$, $\Omega_m=0.25733$, $\Omega_b=0.043557099$, $\Omega_r=0.000080763524$, $n_s=0.963$, $\sigma_8=0.80100775$ and $w=-1$. We use the same cosmology when generating our perturbative model, while all Fisher forecasts in \sect{sec:forecast} use the Planck 2018 cosmology~\citep{Collaboration]2018:1807.06209v2}.

\section{Cross-power spectrum estimator}
\label{sec:estimator}

The main analysis of this paper is based on the cross-power spectrum between different mass bins of the RayGalGroup simulation. Here we will outline the cross-power spectrum estimator used for this analysis, which uses the Legendre basis following~\citep{Bianchi2015:1505.05341v2,Scoccimarro2015:1506.02729v2}.

Any halo (or galaxy) sample is defined by a catalog of halo positions as well as a random catalog which characterises the survey window. We bin all halo positions and randoms into 3D grids, which allows us to define the overdensity field 
\begin{equation}
    F(\vb{r}) = \sum_i^{N_{\rm data}}w_i(\vb{r}) - \sum_i^{N_{\rm ran}}w_i(\vb{r}),
\end{equation}
where $w(\vb{r})$ can be any (signal-to-noise) weighting and the grid assignment itself implies a pixel window function. The impact of the pixel window function can be mitigated by using non-trivial mass assignment schemes~\citep{Jing2004:astro-ph/0409240v2} as well as interlacing~\citep{Sefusatti2015:1512.07295v2}. We can now estimate the power spectrum multipoles using the Legendre basis as
\begin{equation}
    P^{XY}_{\ell}(k) = \frac{2\ell + 1}{I}\int\frac{ d\Omega_{\vbunit{k}}}{4\pi}\left[A^X_0(\vb{k})\left(A^{Y}_{\ell}(\vb{k})\right)^*\right],
    \label{eq:pkestimator}
\end{equation}
where the superscript $X$ and $Y$ refer to the different tracers of the density field, the index $\ell$ specifies the order of the Legendre multipole and 
\begin{align}
    A_{\ell}(\vb{k}) &= \int d\vb{r} (\hat{\vb{k}}\cdot\hat{\vb{r}})^{\ell} F(\vb{r})e^{-i\vb{k}\cdot \vb{r}}.
\end{align}
The normalisation of the power spectrum is given by
\begin{align}
    I &= \int d\vb{r} n_X(\vb{r})n_Y(\vb{r})\, ,\\
    &= \alpha_X\sum^{N^X_{\rm ran}}_in_Y(\vb{r}_i)w_i^Y(\vb{r})\, ,\notag\\
    &= \alpha_Y\sum^{N^Y_{\rm ran}}_in_X(\vb{r}_i)w_i^X(\vb{r})\notag
\end{align}
with $\alpha_X = N^X_{\rm data}/N^Y_{\rm ran}$, $\alpha_Y = N^Y_{\rm data}/N^X_{\rm ran}$ and $n_X(\vb{r}_i)$ and $n_Y(\vb{r}_i)$ representing the density of tracer $X$ and $Y$ at position $\vb{r}_i$, respectively. For the analysis in this paper we do not include any weight, $w_i$, since the density of the simulation used here is constant, which means that a standard signal-to-noise weighting like FKP~\citep{Feldman1993:astro-ph/9304022v1} has no effect (ignoring the minor redshift dependence shown in \fig{fig:hist_z}). We also ignore any weighting which would account for redshift evolution of the relativistic signatures (see~\citep{Castorina2019:1904.08859v1} for a possible approach to include such weights).

\section{Theory}
\label{sec:theory}

Here we will discuss the model for the halo cross-power spectrum dipole which we will compare to the RayGalGroup simulation in the next section. Our final model includes all relativistic corrections up to third-order including third-order bias expansion. We will consider all terms which depend linearly on $\mathcal{H}/k$ and neglect higher-order terms (weak field approximation). While we restrict all terms proportional to $v_{\parallel}$ (Doppler term) to linear theory, we include 1-loop corrections to the matter power spectrum for the potential term and transverse Doppler effect. The details about this model can be found in~\citep{DiDio:2018zmk,DiDio:2020}.

\subsection{Linear order}

With a galaxy redshift survey, we directly measure the number of galaxies as a function of redshift and position on the sky $N(z,\vbunit{n})$, which allows us to define the galaxy number overdensity as
\begin{equation}
    \Delta(z, \vbunit{n}) = \frac{N(z,\vbunit{n}) - \bar{N}(z)}{\bar{N}(z)}\, ,
    \label{eq:num_delta}
\end{equation}
where $\bar{N}(z)$ is the mean number of galaxies at redshift $z$, averaged over all directions $\vbunit{n}$.

We can define the galaxy density $\rho(z,\vbunit{n}) = N(z,\vbunit{n})/V(z,\vbunit{n})$, where $V(z,\vbunit{n})$ is the volume at redshift $z$ and direction $\vbunit{n}$, which uses the same redshift and angle pixelisation ($dz$, $d\Omega$) as $N(z,\vbunit{n})$. Now we can define the galaxy overdensity 
\begin{equation}
    \delta(z, \vbunit{n}) = \frac{\rho(z,\vbunit{n}) - \bar{\rho}(\bar z)}{\bar{\rho}(\bar z)}\, ,
\end{equation}
where $\bar z$ denotes the background redshift, i.e.~$1+\bar z = 1/a(t)$. Using Newtonian dynamics, \citeref{Kaiser1987} derived 
\begin{align}
    \Delta(z, \vbunit{n}) &= \delta(z,\vbunit{n}) + \frac{1}{\mathcal{H}}\partial_rv_{\parallel} + \left(2 + \frac{d\ln \phi}{d\ln r}\right)v_{\parallel}\, \notag\\
    &= \delta(z,\vbunit{n}) + \frac{1}{\mathcal{H}}\partial_rv_{\parallel} + \left(\frac{2}{\mathcal{H} r} - b_e\right)v_{\parallel}\, ,
    \label{eq:newton}
\end{align}
where $v_{\parallel} = \vb{v}\cdot\vbunit{n}$ \footnote{Note that we define $\vbunit{n}$ to point from the source to the observer. If $\vbunit{n}$ would point from the observer to the source we would have a sign difference in all velocity terms.}, $r$ is the comoving distance to the galaxy and $\phi$ is the selection function, introduced in \citeref{Kaiser1987} as $\rho_{\rm DM} = n / \phi$. The evolution bias $b_e$ is defined in~\eq{eq:bevolve}.

In addition to the terms in \eq{eq:newton}, relativistic corrections start to matter when approaching horizon scales. In the last decade several works have studied the impact of relativistic effects on halo clustering~\citep{Yoo:2012se,Yoo2010:1009.3021v1,Bonvin2011:1105.5280v3,Challinor2011:1105.5292v2,DiDio:2016ykq,Durrer:2016jzq}. The halo number counts to first order are
\begin{align}
    \begin{split}
        \Delta(z,\vb{n}) =\; &\underbrace{b_1\delta}_{(1)} + \underbrace{\frac{1}{\mathcal{H}}\partial_rv_{\parallel}}_{\text{(2) RSD}} + \underbrace{(5s_m - 2)\int^r_0\frac{r-r'}{2rr'}\Delta_{\Omega}(\Phi + \Psi)dr'}_{\text{(3) Lensing}}\\
        &- \underbrace{{\color{red}\frac{1}{\mathcal{H}}\dot{v}_{\parallel}}}_{\text{(4)}} +\underbrace{\left(\frac{\dot{\mathcal{H}}}{\mathcal{H}^2} - {\color{red}1} + 5s_m\left( 1- \frac{1}{r \HH} \right) +  \frac{2}{r\mathcal{H}} -b_e\right)v_{\parallel}}_{\text{(5) Doppler term}} + \underbrace{{\color{red}\frac{1}{\mathcal{H}}\partial_r\Psi}}_{\text{(6)}}\\
        &+ \mathcal{O}(\mathcal{H}^2/k^2)\, ,
    \end{split}
    \label{eq:firstorder}
\end{align}
where $\Delta_{\Omega}$ is the angular Laplacian and $\Psi$ and $\Phi$ are the Bardeen potentials. 
The evolution bias and magnification bias are given by
\begin{align}
    b_e(a, >\!\!\!\bar{L}) &= \frac{\partial\ln\left[n(a,>\!\!\!\bar{L})\right]}{\partial \ln a} = - (1+z)\frac{\partial \ln n(z,>\!\!\!\bar{L})}{\partial z}\, ,
    \label{eq:bevolve}\\
    s_m(a, >\!\!\!\bar{L}) &= -\left.\frac{2\partial\ln\left[n(a,>\!\!\!\bar{L})\right]}{5\partial \ln L}\right|_{\bar{L}}\, ,
\end{align}
where $>\!\!\!\bar{L}$ denotes the threshold luminosity of a given survey and $n(a)$ represents the comoving density~\footnote{Note that when using a physical (rather than a comoving) density, the evolution bias of dark matter is $b_e=3$.}. The evolution bias describes the fact that a wrong estimate of the redshift (due to peculiar velocities) leads to a wrong estimate of the number count at that redshift. Dark matter has an evolution bias of $b_e=0$, since the co-moving number density of dark matter is constant. Any halo or galaxy sample used as tracer of the matter density field can have evolution in the halo number density leading to a non-zero $b_e$. We see such effects in the high mass bin of the RayGalGroup simulation (see \fig{fig:hist_z}) which we will discuss in \sect{sec:analysis}.

Eq.~(\ref{eq:firstorder}) contains all terms contributing to the galaxy number count at linear order within the weak field approximation where
\begin{enumerate}
    \item[\textbf{(1)}] is the true galaxy density fluctuation.
    \item[\textbf{(2)}] is the Kaiser RSD term.
    \item[\textbf{(3)}] is the lensing term, consisting of two contributions. The first term (proportional to the magnification bias $s_m$) accounts for the fact that some galaxies are only part of the sample because their luminosity has been magnified/demagnified by gravitational lenses along the line-of-sight~\citep{Broadhurst1994:astro-ph/9406052v1,Moessner1997:astro-ph/9708271v1}. The second term (not proportional to $s_m$) is a geometrical effect accounting for the change in the observed solid angle $d \Omega$.
    \item[\textbf{(4)}] describes the time evolution of the velocity field, meaning that assuming the wrong distance also leads to an assumption of the wrong velocity since the velocity field is evolving with time.
    \item[\textbf{(5)}] describes the Doppler term including Newtonian and relativistic contributions (discussed in detail in~\citep{Lepori2017:1709.03523v3,Breton2018:1803.04294v2,Hall2016:1609.09252v3,Yoo2011:1109.0998v2,Irsic:2015nla,Bonvin2013:1309.1321v2,Bonvin2016:1610.05946v2}).
    \item[\textbf{(5a)}] describes the time evolution of the Hubble parameter, meaning that assuming the wrong distance also leads to an assumption of the wrong background expansion since the Hubble parameter is evolving with time.
    \item[\textbf{(5b)}] describes the ``lightcone effect''~\citep{Kaiser2013:1303.3663v2,Bonvin2013:1309.1321v2} meaning the peculiar velocity changes the effective size of the redshift bin.
    \item[\textbf{(5c)}] originates from relativistic fluctuations in the convergence.
    \item[\textbf{(5d + 5e)}] are the Newtonian Doppler contributions already present in~\eq{eq:newton}.
    \item[\textbf{(6)}] describes the change in the effective redshift bin due to gravitational redshift.
\end{enumerate}
We have not included terms which are directly proportional to the Bardeen potentials $\Phi$ and $\Psi$, since such terms are suppressed by ($\mathcal{H}/k)^2$ and only matter on very large scales. We note however that the $1/k^2$ scaling of these terms is very similar to the scale-dependent bias introduced by local primordial non-Gaussianity~\citep{Dalal2007:0710.4560v3}. However, the different redshift evolution can help to disentangle this degeneracy in the matter power spectrum~\citep{Yoo:2012se}. 

The first two red-colored terms in~\eq{eq:firstorder} describe the additional velocity of galaxies sourced by the acceleration (gravitational redshift $\dot{v}_{\parallel}+\mathcal{H}v_{\parallel}$). At linear order, these terms happen to be identical to the gradient of the gravitational potential ($\partial_r\Psi$) which is the source of the acceleration. This is a consequence of the equivalence principle~\citep{Bonvin2014:1409.2224v1}, which leads to the Euler equation
\begin{equation}
    \partial_r\Psi - \dot{v}_{\parallel} - \mathcal{H}v_{\parallel} = 0.
    \label{eq:euler}
\end{equation}
Therefore on linear scales galaxy clustering measurements are not sensitive to gravitational redshift.
Here we include these terms since the cancellation is only present when the relativistic Doppler, as well as potential terms, are included. The RayGalGroup simulation includes different relativistic effects in turn, using different redshift definitions ($z_0$ to $z_5$; see eqs.~\ref{eq:z0} - \ref{eq:z5}). Later we will use these different redshift definitions to study each contribution in turn. For that reason the model for e.g.~the potential term in $z_1$ should include the gravitational redshift at linear order, since the cancellation with $\dot{v}_{\parallel}+\mathcal{H}v_{\parallel}$ only happens after the velocities in $z_2$ are included.

\subsection{Beyond linear theory}

The relativistic galaxy number counts beyond linear theory have been derived 
in~\cite{Yoo:2014sfa,Bertacca:2014dra,DiDio:2014lka} (to second order) 
and in~\cite{DiDio:2018zmk} (to third order). In an accompanying paper, we show how to directly derive the relativistic number counts to any order in perturbation theory~\citep{DiDio:2020}.
Here we summarize the relevant results to second and third-order, which we use in the next section to compute the dipole at 1-loop.

At any order $i$ in perturbation theory we can split the galaxy number counts within the weak field approximation into the standard Newtonian contribution $\Delta_N^{(i)}$ and the relativistic contribution $\Delta_R^{(i)}$ where $\Delta_R^{(i)} \sim \left( \HH/k \right) \Delta_N^{(i)}$.
Therefore the Newtonian and relativistic galaxy number counts in the weak field approximation at second order are 
\begin{align}
    \Delta_N^{(2)}(z,\vb{n}) =\;  
    &\delta_g^{(2)} + \HH^{-1} \partial_r \ndv^{(2)} 
    + \HH^{-1} \partial_r \left( \ndv  \delta_g \right)  + \HH^{-2}  \partial_r \left( \ndv \partial_r \ndv \right) \, ,\\
    \begin{split}
        \Delta_R^{(2)}\left( z, \bn  \right) =\;  &\left( -1 +\frac{\dot \HH}{\HH^2} + \frac{2}{\HH r} - b_e\right)\left(  \ndv^{(2)} +\ndv \delta  \right) 
        - \HH^{-1} \dotndv^{(2)} - 2 \HH^{-2} \partial_r \ndv \dotndv \\
        &+ \left( -2 +3 \frac{\dot \HH}{\HH^2} + \frac{4}{\HH r} - 2 b_e \right) \HH^{-1} \ndv \partial_r \ndv
        - 2 \HH^{-2} \ndv \partial_r \dotndv \\
        &- \HH^{-1} \dotndv \delta - \HH^{-1} \ndv \dot \delta +   \HH^{-1}  v^a \partial_a \ndv  + \HH^{-2}\Psi\partial_r^2 \ndv  + \HH^{-1} \Psi \partial_r \delta \\
        &+ \HH^{-1} \partial_r \Psi^{(2)}  + 2 \HH^{-2} \partial_r \ndv \partial_r \Psi+ \HH^{-1} \delta \partial_r \Psi + \HH^{-2} \ndv \partial_r^2 \Psi
    \end{split}
\end{align}
and at third order
\begin{align}
    \begin{split}
        \Delta_N^{(3)}(\vb{n},z) =\; &{ \delta_g^{(3)} +\frac{{\partial_r \ndvthree}}{\HH}}{ +\left[ \HH^{-1} \partial_r\left(  \ndv\delta_g  \right)\right]^{(3)} + \left[ \HH^{-2}  \partial_r \left( \ndv \partial_r \ndv \right) \right]^{(3)}}\\
        &+ { \frac{1}{6} \HH^{-3}\partial_r^3 \ndv^3 +\frac{1}{2} \HH^{-2} \partial_r^2 \left( \delta_g \ndv^2 \right)} \, ,\label{eq:3or_newt}
    \end{split}\\
    \begin{split}
        \Delta_R^{(3)}\left( \bn , z \right) =\; &\left( -1 +\frac{\dot \HH}{\HH^2} + \frac{2}{\HH r} - b_e\right)\left(  \ndv^{(3)} + \left[ \ndv \delta \right]^{(3)} \right) - \HH^{-1} \dotndv^{(3)} \\
        &- 2 \HH^{-2} \left[  \partial_r \ndv \dotndv \right]^{(3)} + \left( -2 +3 \frac{\dot \HH}{\HH^2} + \frac{4}{\HH r} - 2 b_e \right) \left[  \HH^{-1} \ndv \partial_r \ndv \right]^{(3)} \\
        &-2 \HH^{-2} \left[ \ndv \partial_r \dotndv \right]^{(3)} - \HH^{-1} \left[\dotndv \delta \right]^{(3)} - \HH^{-1} \left[\ndv \dot \delta\right]^{(3)} +   \HH^{-1}  \left[v^a \partial_a \ndv\right]^{(3)} \\
        &+ \HH^{-2} \left[ \Psi\partial_r^2 \ndv \right]^{(3)} + \HH^{-1} \left[ \Psi \partial_r \delta\right]^{(3)} + \HH^{-1} \partial_r \Psi^{(3)}  + 2 \HH^{-2} \left[ \partial_r \ndv \partial_r \Psi \right]^{(3)}\\
        &+ \HH^{-1} \left[ \delta \partial_r \Psi\right]^{(3)} + \HH^{-2} \left[ \ndv \partial_r^2 \Psi\right]^{(3)} + \frac{1}{2 \HH^3} \partial^3_r \left( \ndv^2 \Psi \right) - \frac{1}{2 \HH^3} \partial_t \partial_r^2 \ndv^3 \\
        &+ \frac{1}{\HH^2} \partial_r^2 \left( \Psi \ndv \delta \right) - \frac{1}{2\HH^2} \partial_r^2 \left( \ndv v^2 \right) - \frac{1}{\HH^2} \partial_t \partial_r \left( \delta \ndv^2 \right) - \frac{1}{2 \HH} \partial_r \left( \delta v^2 \right) \\
        &+ \frac{1}{\HH^2} \partial_r \left( \partial_r \ndv \ndv^2 \right) \left( 3 \frac{\dot\HH}{\HH^2} + \frac{3}{\HH r} - \frac{3}{2} b_e \right)\\
        &+ \frac{1}{\HH} \partial_r \left( \delta \ndv^2 \right) \left( - \frac{1}{2} + \frac{3}{2} \frac{\dot\HH}{\HH^2} + \frac{2}{\HH r} - b_e \right) \, .
    \end{split}
\end{align}
Note that here we did not make use of the Euler equation (see \eqnb{eq:euler}) which accounts for the difference between the equation above and \citeref{DiDio:2018zmk}. For the sake of simplicity we have set the magnification bias to zero ($s_m=0$). We stress that this is in agreement with the RayGalGroup simulation where magnification bias is not included. 

We also want to include second and third-order terms in the bias expansion. Following~\cite{Desjacques:2016bnm} (and references therein) we have
\begin{align}
    \delta_g^{(2)} =\; &b_1 \delta^{(2)} + \frac{1}{2} b_2 \left( \delta^2  { -  \langle \delta^2 \rangle} \right) + b_{K^2} \left( \left(K_{ij}\right)^2  { -\langle \left(K_{ij}\right)^2 \rangle }  \right)\, ,
\nonumber\\
    =\; &b_1 \delta^{(2)} + \frac{1}{2} b_2 \delta^2   + b_{K^2}  \left(K_{ij}\right)^2  - { \sigma^2 \left( \frac{b_2}{2} + \frac{2}{3} b_{K^2} \right)}\, ,\\
    \begin{split}
        \delta_g^{(3)} =\; &b_1 \delta^{(3)} + b_2 \delta \delta^{(2)} + \frac{1}{6} b_3 \delta^3 + 2  b_{K^2} K_{ij} K^{(2)}_{ij} +b_{K^3} \left( K_{ij} \right)^3\\
        &+ b_{\delta K^2} \delta  \left(K_{ij}\right)^2 + b_{\rm td} O^{(3)}_{\rm td}\, ,
    \end{split}
\end{align}
where~\footnote{We follow the short notation
$$
K^2 \equiv \left( K_{ij} \right)^2 \equiv {\rm tr} \left( K K \right) = K_{ij} K_{ji} \, .
$$}
\bea
K_{ij} &=& \left[ \frac{k_i k_j}{k^2} -\frac{1}{3} \delta_{ij} \right] \delta \left( \bk \right) \, ,\\
O^{(3)}_{\rm td}  &=& \frac{8}{21} \left( \frac{\left[ \bk_1 \cdot \left( \bk_2 +\bk_3 \right) \right]^2}{k_1^2 \left| \bk_2 + \bk_3 \right|^2} -\frac{1}{3}\right) \left( 1-\frac{3}{2} s_2 \left( \bk_2 , \bk_3 \right)  \right) \delta \left( \bk_1 \right) \delta \left( \bk_2 \right) 
\delta \left( \bk_3 \right)
\eea
and
\be
\sigma^2 \equiv \int \frac{d^3q}{\left( 2 \pi \right)^3} P\left( q \right) .
\ee
Under the assumption that the comoving number of sources is conserved, we derive the time evolution of $b_1$ and $b_2$~\cite{Desjacques:2016bnm}
\bea
\dot b_1 &=& \left( 1 - b_1 \right) f \HH \, ,
\\
\dot b_2 &=&\left( -2 b_2 - \frac{8}{21} + \frac{8}{21} b_1 \right) f \HH 
\eea
and relate higher order biases to $b_1$ and $b_2$
\bea \label{eq:start:biases}
b_{K^2} &=& - \frac{2}{7} \left( b_1 -1 \right), \\
b_{\delta K^2}&=& \frac{1}{21} (7 b_1-6 b_2-7), \\
b_{K^3} &=& \frac{22 (b_1-1)}{63}, \\
b_{\rm td} &=& \frac{23}{42} \left( b_1 - 1 \right) \, .
\label{eq:end:biases}
\eea
We remark therefore that at any order $n$ in perturbation theory we have $n$ independent bias parameters.

\subsection{The theoretical cross-power spectrum dipole}

The odd multipoles of the power spectrum (or correlation function) are sourced by relativistic effects through their different parity along the line of sight with respect to the standard Newtonian terms (see e.g.~\citep{McDonald2009:0907.5220v1,Bonvin:2013ogt,Bonvin2014:1409.2224v1,Bonvin:2015kuc,Gaztanaga:2015jrs,Irsic:2015nla,DiDio:2018zmk}). In particular between the odd mulitpoles, most of the signal is carried by the dipole, which therefore represents the most promising candidate for the detection of relativistic effects with upcoming galaxy redshift surveys.

Before we calculate the theoretical power spectrum dipole in \sect{sec:leading_dipole} and following, we first want to bridge the gap between our theoretical approximations and the power spectrum estimator in~\eq{eq:pkestimator}. Here we will consider effects due to bin averaging (\sect{sec:est_to_theory}) as well as evolution and wide-angle effects (\sect{sec:evo_wa}).

\subsubsection{Expectation value of the dipole estimator }
\label{sec:est_to_theory}

We start by clarifying the Fourier convention adopted in this paper
\bea
f ( \bk ) &=& \int d^3 x f( \bx ) e^{-i \bk \cdot \bx}, \\
f( \bx) &=& \int \frac{d^3 k}{( 2 \pi )^3} f ( \bk ) e^{i \bk \cdot \bx} \, .
\eea
The correlation function is related to the power spectrum as
\begin{equation}
    \xi^{XY}(\vb{s}) = \langle \Delta^X( \vb{r}_X) \Delta^Y(\vb{r}_Y) \rangle = \int \frac{d^3 k }{( 2 \pi)^3} P^{XY}( \bk) e^{i \bk \cdot \bs},
\end{equation}
where $\bs= \br_X - \br_Y$. Therefore the correlation function $\xi^{XY}(\vb{ s})$ is the Fourier transform of the matter power spectrum $P^{XY}(\vb{k})$.
The Fourier wave vector $\bk$ is the conjugate variable of the pair separation $\bs$ from the source $Y$ to the source $X$.

From \eq{eq:pkestimator} we can write the dipole estimator as 
\bea
P_1^{XY}  \left( k \right) &=& \frac{3}{V} \int \frac{d \Omega_{\vbunit{k}} }{4 \pi } \int d^3 r_X d^3 r_Y \Delta \left( \br_X \right) \Delta \left( \br_Y \right) \hat \bk  \cdot {\hat \br_Y} e^{i \bk \cdot \left( \br_Y - \br_X \right)} \, 
\nonumber \\
&=& 
- \frac{3 i }{V}  \int d^3 r_X d^3 r_Y \Delta \left( \br_X \right) \Delta \left( \br_Y \right) j_1 \left( k s \right) \hat \br_Y \cdot \hat \bs \, ,
\eea
where $V$ is the volume of the survey. Therefore the expectation value of the estimator reads as
\bea
\langle P_1^{XY}  \left( k \right) \rangle &=&  - \frac{3 i }{V}  \int d^3 r_X d^3 r_Y \xi^{XY} \left( r_Y, s, \hat \br_Y \cdot \hat \bs\right) j_1 \left( k s \right) \hat \br_Y \cdot \hat \bs \, 
\nonumber \\
&=&  
-  \frac{3 i }{V}  \int d^3 r_Y d^3 s \sum_i \xi^{XY} \left( r_Y, s\right) \mathcal{L}_\ell \left( \hat \br_Y \cdot \hat \bs \right) j_1 \left( k s \right) \hat \br_Y \cdot \hat \bs \, 
 \nonumber \\
&=&
 -  \frac{3 i }{V} 8 \pi^2  \int  d r_Y r_Y^2 d s \,s^2 d \mu_Y \sum_\ell \xi^{XY}_\ell \left( r_Y, s\right) \mathcal{L}_\ell \left( \mu_Y \right) j_1 \left( k s \right) \mu_Y\, 
 \nonumber \\
 &=&
 -   \frac{2 i }{V} 8 \pi^2  \int  d r_Y r_Y^2 d s\, s^2 \xi^{XY}_1 \left( r_Y, s\right)  j_1 \left( k s \right) \, ,
 \label{eq:volume}
\eea
where
\be
\xi^{XY}_1 \left( r_Y, s\right)= \frac{3}{2} \int d \mu_Y \xi^{XY} \left( r_Y, s,\mu_Y\right) \mathcal{L}_1 \left( \mu_Y \right)   
\ee
and $\mu_Y =\hat \br_Y \cdot \hat \bs = - \bn \cdot \hat\bs$ (where in the flat-sky approximation we will consider $\bn =- \hat \br_Y\simeq - \hat \br_X$).
Now we need to relate the dipole of the correlation function with respect to the angle $\mu_Y$ with the dipole of the matter power spectrum with respect to the angle $\mu = - \bn \cdot \bk$.
Any multipole of the correlation function is related to the same multipole of the power spectrum as
\be
\xi_\ell \left( s \right) =
i ^\ell \int \frac{dk}{2 \pi^2} k^2 P_{\ell} \left( k \right) j_{\ell} \left( k s \right ) \, .
\label{eq:hankel}
\ee
Hence, the expectation value of the dipole estimator becomes
\bea
\label{eq:exp_val_estimator}
\langle P_1^{XY}  \left( k \right) \rangle &=&\frac{8}{V}  \int dr_Y\, r_Y^2 ds\, s^2 dq\, q^2 P_1 \left(r_Y, q \right) j_1 \left( q s \right) j_1 \left( k s \right) 
\nonumber \\
&=&
\frac{4 \pi}{V} \int dr_Y\, r_Y^2 P_1^{XY} \left( r_Y, k \right)\, ,\label{eq:zaverage}
\eea
where $r_Y$ determines the redshift at which the matter power spectrum is evaluated.

We note that in the following sections we do not explicitly include the redshift dependence of most quantities for reasons of brevity. All evaluations of the growth rate $f(z)$, the matter density $\Omega_m(z)$ and the power spectrum $P(k,z)$ need to be averaged within the redshift bin as given in \eq{eq:zaverage}, accounting for their redshift evolution.

\subsubsection{Evolution and wide-angle effects}
\label{sec:evo_wa}

So far we have computed the theoretical power spectrum, and its dipole, in the flat-sky approximation. However it is well-known that the redshift evolution of galaxy bias and growth rate within the redshift bins as well as wide-angle effects can be comparable to the relativistic projection effects (see for instance~\cite{Gaztanaga:2015jrs,Bonvin:2015kuc}). In particular, the dipole induced by wide-angle effects with respect to the end-point line-of-sight definition can be measured in current surveys~\cite{Gaztanaga:2015jrs,Beutler2018:1810.05051v3}. 

In order to derive the evolution and wide-angle effects, we start considering the full-sky 2-point correlation function induced by density perturbations and redshift-space distortions. We follow the approach presented and developed in~\cite{Campagne:2017wec,Tansella:2018sld} and we compute
\begin{align}
&
\hspace{-0.5cm}
\langle \Delta_N^X \left( \br_X \right) \Delta_N^Y \left( \br_Y \right) \rangle = 
\nonumber \\
&= \int \frac{d^3k_1}{\left( 2 \pi \right)^3}\frac{d^3k_2}{\left( 2 \pi \right)^3} \left( b_1^X + f_X \mu_1^2 \right) \left( b_1^Y + f_Y \mu_2^2 \right) \langle \delta \left( \bk_1 \right) \delta \left( \bk_2 \right) \rangle
e^{i \bk_1 \cdot \br_X }e^{i \bk_2 \cdot \br_Y }
\nonumber \\
&= 
\int \frac{d^3k}{\left( 2 \pi \right)^3}  \left( b_1^X + f_X \mu^2 \right) \left( b_1^Y + f_Y \mu^2 \right) P \left( k \right) e^{i \bk \cdot \br_X }e^{-i \bk \cdot \br_Y }
\nonumber \\
&= 
\int \frac{d^3k}{\left( 2 \pi \right)^3}  \left( b_1^X - f_X \frac{\partial^2}{\partial\left( k r_X \right)^2 }\right) \left( b_1^Y - f_Y \frac{\partial^2}{\partial\left( k r_Y \right)^2 } \right) P \left( k \right) e^{i \bk \cdot \br_X }e^{-i \bk \cdot \br_Y }
\nonumber \\
&= 
 \int \frac{d k}{2 \pi^2} k^2 \left( b_1^X - f_X \frac{\partial^2}{\partial\left( k r_X \right)^2 }\right) \left( b_1^Y - f_Y \frac{\partial^2}{\partial\left( k r_Y \right)^2 } \right) P \left( k \right) 
 \nonumber \\
 & \;\;\;\;\times
 \sum_\ell \left( 2 \ell+1 \right) \mathcal{L}_\ell \left( \hat \br_X \cdot \hat \br_Y \right) j_\ell \left( k r_X \right)  j_\ell \left( k r_Y \right) 
 \nonumber \\
 &= \int \frac{d k}{2 \pi^2} k^2  P \left( k \right)
 \left( b_1^X -  \frac{f_X\partial^2}{\partial\left( k r_X \right)^2 }\right) \left( b_1^Y - \frac{f_Y\partial^2}{\partial\left( k r_Y \right)^2 } \right) j_0 \left( k \sqrt{r_X^2 + r_Y^2 - 2 \br_X \cdot \br_Y} \right) \, ,
\end{align}
where the suffices $X$ and $Y$ denote that the quantities are evaluated at the positions $r_X$ and $r_Y$, respectively. Now, by following the notation of \citeref{Tansella:2018sld} we have
\begin{equation}
    \begin{split}
        \xi \left( r_X, r_Y , \cos \theta \right) =\; &D\left(r_X \right) D\left(r_Y \right) I^n_\ell \left( s \right)\\
        &\times\sum_{\ell, n}\Big[ \left. X^n_\ell \right|_{\rm den \times den}+\left. X^n_\ell \right|_{\rm den \times rsd} + \left. X^n_\ell \right|_{\rm rsd \times den} + \left. X^n_\ell \right|_{\rm rsd \times rsd}\Big] \, ,
    \end{split}\label{eq:full_sky_2point}
\end{equation}
where $\cos \theta =
\hat \br_X \cdot \hat\br_Y$, $D$ the growth function and 
\be
I^n_\ell \left( s \right) = \int \frac{k^2 dk }{2 \pi^2} \frac{j_\ell \left( k s \right) }{\left( k s\right)^n } P\left( k \right)  
\ee
and the only non-vanishing coefficients are given by
\begin{align}
    \left.X^0_0 \right|_{\rm den \times den} =\; &b_1^X b_1^Y \, , \\
    \left.X^0_0 \right|_{\rm rsd \times rsd} =\; &f_X f_Y \frac{1 + 2 \cos^2 \theta}{15}\, , \\
    \left.X^0_2 \right|_{\rm rsd \times rsd} =\;
    &-\frac{f_X f_Y}{21} \left(1 + 11 \cos^2 \theta + \frac{18 \cos \theta \left( \cos^2 \theta -1 \right) r_X r_Y}{s^2} \right) \, , \\
    \begin{split}
        \left.X^0_4 \right|_{\rm rsd \times rsd} =\;
        &\frac{f_X f_Y}{35 s^4} \Big( 4 \left( 3 \cos^2 \theta - 1 \right) \left( r_X^4+ r_Y ^4 \right) \\
        &+ r_X r_Y \left(  \cos^2 \theta +3 \right) \left[3 \left( \cos^2 \theta +3 \right) r_X r_Y - 8 \left( r_X^2 + r_Y^2 \right) \cos \theta   \right] \Big)\, ,
    \end{split}\\
    \left.X^0_0 \right|_{\rm den \times rsd} =\; &\frac{b_1^X f_Y}{3} \, , \\
    \left.X^0_2 \right|_{\rm den \times rsd} =\; &-b_1^X f_Y \left( \frac{2}{3} - \left( 1 - \cos^2 \theta \right) \frac{r_X^2}{s^2}
    \right) \, .
\end{align}
In terms of the coordinates for the end-point line-of-sight convention we have to replace
\be \label{eq:LOS_coords}
r_X = \sqrt{r_Y^2 + s^2 + 2 r_Y s \mu_Y } \qquad \text{and} \qquad \cos \theta = \mu_Y  \frac{s}{r_X} + \frac{r_Y}{r_X}\, .
\ee
Now we expand the full-sky correlation function of \eq{eq:full_sky_2point} in terms of the small parameter $s/r_Y$. Since in the flat-sky approximation the density perturbation and redshift space distortions do not generate a dipole, we need to consider this expansion at least at linear order in $s/r_Y$. We therefore obtain 
\begin{align}
    \begin{split}
    \label{eq:xi_evo1}
        \xi_1^{\rm evo1} \left( s \right) &= \HH s\Bigg( \left\{ (1+z)\left[\frac{1}{3} {b_1^Y} {f'} + f \left(\frac{1}{3} {{b'}_1^X}  +\frac{1}{5} {f'}\right)\right]\right.\\
        &\hspace{1.6cm}-\left.\frac{1}{3} f^2  ({b_1^X}+{b_1^Y}) -\frac{1}{5} f^3  \right\} I_0^0 \left( s \right) D^2\\
        &\hspace{1.3cm}+\left\{\frac{4}{105} f  \Big(f (7 {b_1^X}+7 {b_1^Y}+6 f)-7 {{b'}_1^X} (1+z)\Big)\right.\\
        &\hspace{1.9cm}- \left.\frac{4}{105} {f'}  (1+z) (7 {b_1^Y}+6 f)\right\}  I_2^0 \left( s \right) D^2\Bigg)\, ,
    \end{split}\\
     \label{eq:xi_evo2}
    \xi_1^{\rm evo2} \left( s \right) &= \HH s \left(  {b_1^Y} {{b'}_1^X}  (1+z)-{b_1^X} {b_1^Y} f \right) I_0^0 \left( s \right) D^2\, , \\
    \xi_1^{\rm wa} \left( s \right) &= -\frac{4 f  (7 {b_1^Y}+3 f)}{35}  \frac{s}{r_Y} I_2^0 \left( s \right) D^2\, ,
    \label{eq:wa1}
\end{align}
where a prime denotes the derivative with respect to the redshift and all quantities are evaluated at the position $r_Y$.
We have separated the contributions in evolution terms, proportional to $\HH s$, and wide-angle corrections, proportional to $s/r_Y$. We have further split in two the evolution contributions since the term `$\rm evo2$' is also generated in real space and therefore will not have an impact on our comparison with the RayGalGroup simulation~\footnote{Later we will study the individual relativistic effects by subtracting out the real-space contributions.}. 

In the same way, density and redshift space distortions also leak into the octupole
\begin{align}
    \begin{split}
        \xi_3^{\rm evo}(s) =\; &\HH s  \left( \frac{2}{5} f^2 (b_1^X+b_1^Y)-\frac{2}{35} {f'} (1+z) (7 b_1^Y+6 f)\right.\\
        &\;\;\;\;\;\;\;\,-\left.\frac{2}{5} {b'}_1^X f (1+z)+\frac{12 f^3}{35} \right) I_2^0(s)D^2 \\
        &+ \HH s  \left( \frac{32}{315} {f'} f (1+z)-\frac{32 f^3}{315}\right)I_4^0(s)D^2\, , 
    \end{split}\\
    \xi_3^{\rm wa} \left( s \right) =\; &\frac{4 f s (7 b_1^Y+3 f)}{35 r_Y}  I_2^0 \left( s \right) D^2 +  \frac{16 f^2 s}{63 r_Y} I_4^0 \left( s \right) D^2\, .
\end{align}
We remark that the wide-angle term in \eq{eq:wa1} agrees with eq.~(2.14) and (2.15) of \citeref{Beutler2018:1810.05051v3} and eq.~(4.14) in \citeref{Reimberg2015:1506.06596v2}.

As shown in \eq{eq:hankel}, each multipole of the power spectrum is related to the same multipole of the correlation function through a Hankel transform. This will lead to two-dimensional integrals of the form
\be \label{I_def}
B_\ell = 4 \pi \left( - i\right) \int ds\, s^3 j_1 \left( k s \right) \int \frac{dq}{2 \pi^2} q^2 P\left( q \right) j_\ell \left( q s \right) \qquad \text{with} \quad \ell=0,2.
\ee
We remark that we have the factor $s^3$ (instead of $s^2$) because the evolution and the wide-angle effects arise from an expansion of the even multipole with respect to $s/r_Y$. The Bessel functions in \eq{I_def} carry parity and therefore oscillate with opposite phase, which makes them non-trivial to solve numerically~\cite{Castorina:2017inr,Castorina:2018nlb,Beutler2018:1810.05051v3}. To avoid this issue and to provide simpler expressions we show how the integrals in \eq{I_def} can be solved analytically. We start by performing the following integral
\bea \label{integral_s}
\int ds\,s^3 j_1 (k s) j_\ell(q s) &=&  - \left( \frac{\partial^2}{\partial k^2}  + \frac{2}{k} \frac{\partial}{\partial k } - \frac{2}{k^2} \right) \int ds\,s j_1( k s)  j_\ell( q s)\, ,
\nonumber \\
& =&- \frac{\pi}{2} \left( \frac{\partial^2}{\partial k^2}  + \frac{2}{k} \frac{\partial}{\partial k } - \frac{2}{k^2} \right)
\left\{
\begin{array}{cc}
\frac{{\Theta \left( k - q \right)} }{k^2} & \;\;\;\; \ell=0\, , \\
\frac{ k}{q^3}{ \Theta \left(q - k \right)} & \;\;\;\; \ell=2\, ,
\end{array}
\right.
\eea
where $\Theta$ denotes the Heaviside distribution and we have used eq.~(10.22.63) of \cite{NIST:DLMF} together with the identity (see e.g.~\cite{Assassi:2017lea,DiDio:2018unb})
\be
\left( \frac{\partial^2 }{\partial k^2} + \frac{2}{k} \frac{\partial}{\partial k} - \frac{\ell \left( \ell +1 \right) }{k^2} \right) j_\ell \left( k s \right) =- s^2 j_\ell \left( k s \right) \, .
\ee
Now, by performing the integral over $q$, we find 
\bea
B_0 &=& i \frac{d P\left( k  \right) }{dk} \, , \\
B_2 &=& -3 i \frac{P\left( k  \right) } {k} - i  \frac{d P\left( k \right) }{dk} \, .
\eea
With these solutions we can obtain directly the evolution and wide-angle corrections to the dipole of the power spectrum. For the evolution terms we have
\begin{align}
    P_1^{\rm evo1} \left( k , z \right) &= F^{\rm evo1,0} \left( z \right)  B_0\left( k \right)  +F^{\rm evo1,2} \left( z \right) B_2 \left( k \right)   \label{eq:evo1}\, , \\
    P_1^{\rm evo2} \left( k , z \right) &= F^{\rm evo2} \left( z \right) B_0 \left( k \right)\, , \label{eq:evo2} 
\end{align}
where
\begin{align}
    F^{\rm evo1,0} \left( z \right) =\; &\HH 
    \left( -\frac{1}{3} f^2  ({b_1^X}+{b_1^Y})
    +(1+z)\left[\frac{1}{3} {b_1^Y} {f'}
    +f \left(\frac{1}{3} {{b'}_1^X}  +\frac{1}{5} {f'}\right)\right]
    -\frac{1}{5} f^3  \right) D^2 \, ,\label{eq:evo} \\
    \begin{split}
        F^{\rm evo1,2} \left( z \right) =\; &\HH  \Bigg( \frac{4}{105} f  \left(f (7 {b_1^X}+7 {b_1^Y}+6 f)-7 {{b'}_1^X} (1+z)\right)\\
        &\;\;\;\;\;\;-\frac{4}{105} {f'}  (1+z) (7 {b_1^Y}+6 f)\Bigg)   D^2\, ,
    \end{split}\\
    F^{\rm evo2} \left( z \right) =\; &\HH  \left(  {b_1^Y} {{b'}_1^X}  (1+z)-{b_1^X} {b_1^Y} f \right) D^2\, .
\end{align}
Analogously for the wide-angle terms we have
\bea
P_1^{\rm wa} \left( k , z \right)& =& F^{\rm wa} \left( z \right) B_2 \left( k \right)\, ,
\label{eq:wadipole}
\eea
where we have introduced
\bea
F^{\rm wa} \left( z \right) &=&-\frac{4 f  (7 {b_1^Y}+3 f)}{35 r_Y}   D^2 \, .
\eea
These equations are consistent with eq.~(3.5) of \citeref{Beutler2018:1810.05051v3} if the window function is ignored. Note that our equations are not consistent with eq.~(29) of \citeref{Breton2018:1803.04294v2}, since they use the mid-point LOS definition in their estimator, while we use the end-point LOS in our FFT-based estimator discussed in \sect{sec:estimator}.

\subsubsection{Leading order}
\label{sec:leading_dipole}

Working in the weak field approximation shown in \eq{eq:firstorder},
the leading order contributions to the cross-power spectrum of two differently biased tracers $X$ and $Y$ are given by~\footnote{We remark that in our Fourier convention a radial derivative transforms as
$
\partial_r \rightarrow i \mu k \, .$ Therefore the Fourier transform of the linear galaxy number counts is given by
$$
\Delta \left( \bk \right) = \left[b_1 + \mu^2 f - i \mu \frac{\HH}{k} \left( f \mathcal{R} + \frac{3}{2} \Omega_m \right) \right] D \delta \left( k \right)\, .
$$}
\begin{equation}
    \begin{split}
        P^{(11)}(k,\mu,z) =\; &\Big[(b^X_1 + f\mu^2)(b^Y_1 + f\mu^2) + \Delta b_1  i\mu\frac{\mathcal{H}}{k} \frac{3}{2} \Omega_m \\
        &+ i \mu \frac{\HH}{k } f \left\{ \mathcal{R}^Y \left( b_1^X + f \mu^2 \right) -  \mathcal{R}^X \left( b_1^Y + f \mu^2 \right)\right\} \Big]D^2P(k) \\
        &+ \mathcal{O} \left( \HH^2/k^2 \right)\, ,
    \end{split}\label{eq:pk1firstorder}
\end{equation}
where $\Delta b_1 = b^X_1 - b^Y_1$, and we have introduced
\begin{equation}
    \mathcal{R} = 1 - b_e -f - \mathcal{H}^{-1} \partial_t\ln f - \left(2 - 5 s_m \right) \left( 1 - \frac{1}{\mathcal{H} r} \right).
\end{equation}
If the two halo populations have the same evolution and magnification biases (i.e.~$\mathcal{R}^X = \mathcal{R}^Y$) \eq{eq:pk1firstorder} simplifies to
\begin{equation}
    \begin{split}
        P^{(11)}(k,\mu,z) \stackrel{(\mathcal{R}^X=\mathcal{R}^Y)}{=} &\left[(b^X_1 + f\mu^2)(b^Y_1 + f\mu^2) + \Delta b_1  i\mu\frac{\mathcal{H}}{k} \left(f \mathcal{R} + \frac{3}{2} \Omega_m \right) \right]D^2 P(k)\\
        &+ \mathcal{O} \left( \HH^2/k^2 \right).
    \end{split}
\end{equation}
The relativistic terms contribute to the imaginary part of the cross-power spectrum.
The first term in the imaginary part on the right-hand side of \eq{eq:pk1firstorder} represents the Doppler contribution and the second term represents the leading potential contribution. Note that the potential term can be absorbed by the Doppler term at this order (assuming the Euler equation of \eqnb{eq:euler}), but we write it here explicitly, since we want to model the potential term in the RayGalGroup simulation without the Doppler term (as necessary for the redshift definition $z_1$ in \eqnb{eq:z1}), in which case these terms do not cancel. 

From \eq{eq:pk1firstorder} we can obtain directly the dipole
\begin{align}
    P_1(k,z) &= \frac{3}{2} \int d \mu\,\mu P^{(11)}(k,\mu) 
    \nonumber \\
    &= i \frac{\HH}{k} \left[  f \left(b_1^X \mathcal{R}^Y - b_1^Y \mathcal{R}^X \right)+ f^2 \frac{3}{5} \left(\mathcal{R}^Y - \mathcal{R}^X \right) + \frac{3}{2} \Delta b_1  \Omega_m \right] D^2 P \left( k  \right)\, . \label{eq:Doppler}
\end{align}
Similarly we can also compute the octupole
\begin{align}
    P_3 \left( k ,z\right) &= \frac{7}{2} \int d \mu P^{(11)} \left( k , \mu \right) \mathcal{L}_3 \left( \mu \right)\notag\\
    &= -\frac{2}{5} \left( \mathcal{R}^X - \mathcal{R}^Y \right) f^2P\left( k \right) \notag \\
    &=
    \left[ \frac{2}{5}  \left(b_e^X - b_e^Y \right)
    - 2  \left( 1 - \frac{1}{\HH r} \right) \left( s_m^X - s_m^Y \right) 
    \right] f^2D^2 P \left( k \right) \, .
\end{align}
If the two populations $X$ and $Y$ have the same evolution and magnification biases, the dipole reduces to
\be
P_1(k,z) \stackrel{(\mathcal{R}^X=\mathcal{R}^Y)}{=} i \Delta b_1  \frac{\mathcal{H}}{k} \left(f \mathcal{R} + \frac{3}{2} \Omega_m \right) D^2 P(k) \, ,
\label{eq:linear}
\ee
while the octupole vanishes.
As we can see, only the relativistic terms source the dipole within the flat-sky (or distant observer) approximation.

To perform the comparison with the RayGalGroup simulation we also need to compute the dipole induced by the so-called lightcone (LC) effect (see table 1 in \citeref{Breton2018:1803.04294v2}). This effect is present in the linear galaxy number counts (see \eqnb{eq:firstorder} and its interpretation \textbf{(5b)}).
The Fourier transform of the LC term is given by
\be
- \ndv \rightarrow i \mu \frac{\HH}{k} f D \delta\, ,
\ee
from which we can directly compute its contribution to the dipole of the power spectrum
\be
P_1^{\rm LC}( k ,z) = i\Delta b_1 \frac{\HH}{k} f D^2 P \left( k \right),
\label{eq:LC}
\ee
which is consistent with eq.~(32) of \citeref{Breton2018:1803.04294v2}.
As shown in \fig{fig:doppler_details} this effect is small in the RayGalGroup simulation and does not significantly impact our analysis.

Finally we also want to investigate the impact of the evolution bias, $b_e$. Using \eq{eq:Doppler} we can directly determine the impact of the evolution bias, simply by replacing 
\be
\mathcal{R} \rightarrow -b_e \, .
\ee
and neglecting the contribution sourced by the gravitational potential (proportional to $\Omega_M$), yielding 
\be
P_1^{b_e} \left(k,z \right) = i \frac{\HH}{k} \left[ 
f \left( b_1^Y b_e^X - b_1^X b_e^Y \right) + f^2 \frac{3}{5} \left( b_e^X - b_e^Y \right) 
\right] D^2 P \left( k \right) \, .
\label{eq:fevo}
\ee
All dipole contributions calculated in this section are compared in \fig{fig:doppler_details}. 

\subsubsection{Next-to-leading order}
\label{sec:Next-to-leading}

\begin{figure}[t]
\centering
\includegraphics[width=0.95\textwidth]{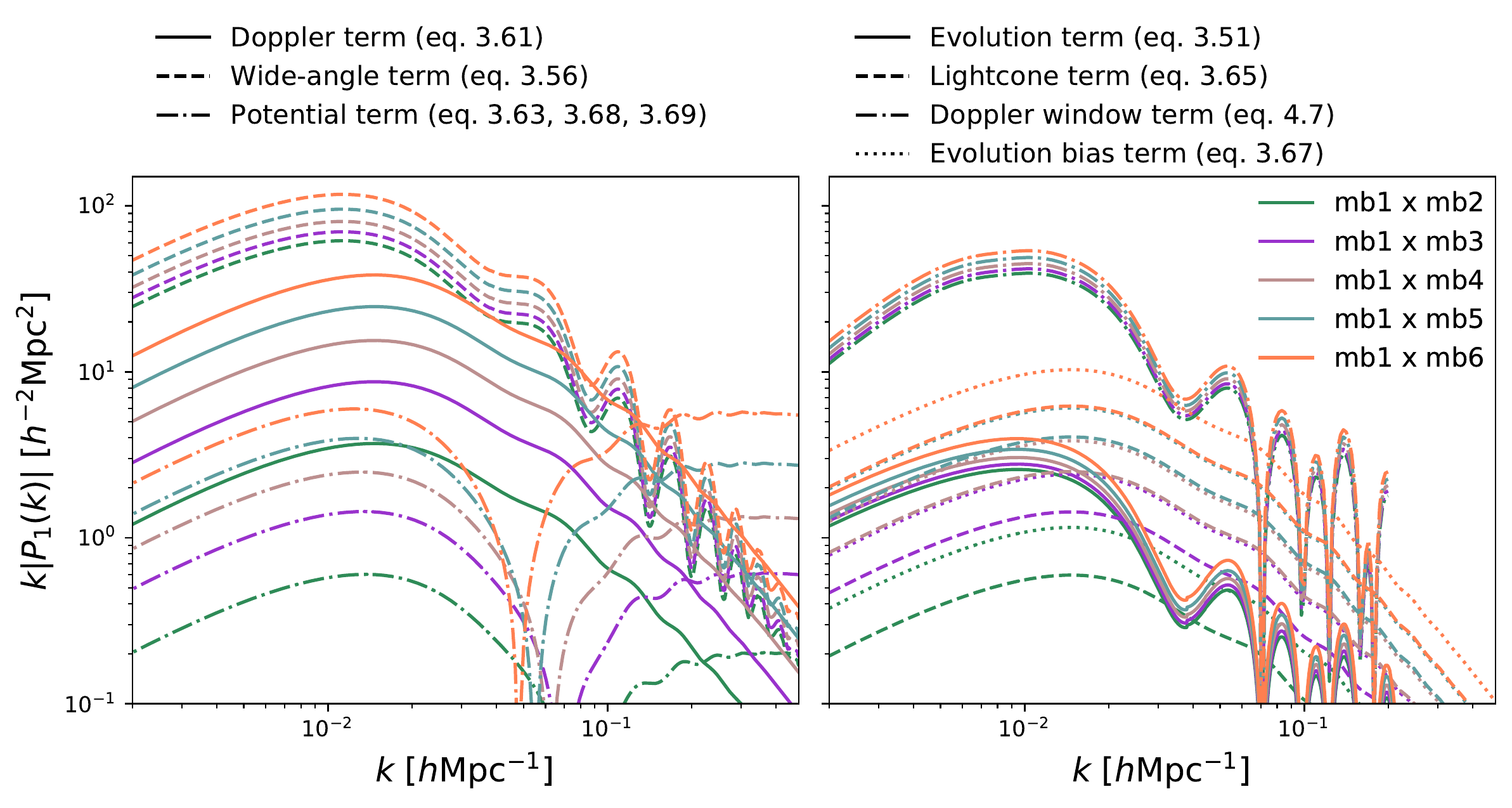}
\caption{Comparison of the dominant contributions to the power spectrum dipole. The left plot shows the relativistic Doppler term (first two terms in \eqnb{eq:Doppler}), the wide-angle term (\eqnb{eq:wadipole}) and the relativistic potential term (including the last term of \eqnb{eq:linear} and the $1$-loop contributions of eqs.~\ref{eq:second_order_pot1} and~\ref{eq:second_order_pot2}). Note that the Doppler term excludes lightcone and evolution bias contributions (which are included in \eqnb{eq:Doppler}), since those contributions are plotted separately in the plot on the right. The relativistic potential term is positive on large scales and changes sign around $k\sim0.06\kMpc$. The right plot shows the evolution bias (\eqnb{eq:evo1}), lightcone effects (\eqnb{eq:LC}), the Doppler window function contribution (\eqnb{eq:dipole_win}) and the evolution bias term (\eqnb{eq:fevo}). We see that wide-angle effects are of the same order as the relativistic Doppler term. The wide-angle contribution is not proportional to $\Delta b_1$ (but only evolves with $\beta = f/b_1$) and hence the wide-angle term dominates for halo mass cross-correlations with small $\Delta b_1$.
}
\label{fig:doppler_details}
\end{figure}

Beyond the linear dipole, we also consider 1-loop contributions. In section~7 of the accompanying paper~\cite{DiDio:2020}, we have derived the galaxy number counts sourced by the gravitational redshift as well as the linear and transverse Doppler effects. We named these three contributions $\Delta_1$, $\Delta_2$ and $\Delta_3$, respectively and include details in \appen{app:single_number_counts}.
Since the galaxy number counts do not depend linearly on the redshift perturbation we need to carefully define the different contributions. We note in particular that we later evaluate our power spectrum dipole model by comparing to the differences between dipole measurements using the different redshift definitions of eqs.~(\ref{eq:z1} - \ref{eq:z3}). This leads to non-trivial mixing terms that we need to account for.

We start with the gravitational redshift. Since the redshift definition $z_0$ in \eq{eq:z0} leads to a zero dipole, we can obtain the dipole sourced by the gravitational potential from $\Delta_1$ (i.e.~the galaxy density derived with the redshift prescription of \eqnb{eq:z1}). In this case, we can write the 1-loop contributions to the dipole as
\begin{align}
    P_1^{(22),\phi}(k) &= i\Omega_m\frac{\mathcal{H}}{k}\int\frac{d^3q}{(2\pi)^3}\left[\sum_{\Delta b =\{\Delta b_1,\Delta b_2, \Delta b_{12}\}}\Delta b J^{\Delta b}_{22}\left(\frac{q}{k},\hat{\vb{k}}\cdot\hat{\vb{q}}\right)\right]P(q)P(|\vb{k} - \vb{q}|) \, , \label{eq:second_order_pot1}\\
    P_1^{(13),\phi}(k) &= i\Omega_m P(k)\frac{\mathcal{H}}{k}  \left\{\!-\left( \Delta b_1 \frac{29}{210} + \Delta b_{12} \frac{3}{2} \right) k^2  \sigma_v^2 + \Delta b_1 \!\! \int\!\!\frac{d^3q}{(2\pi)^3} J^{\Delta b_1}_{13}\left(\frac{q}{k}\right)P(q) \!\right\} \, ,
    \label{eq:second_order_pot2}
\end{align}
where the kernels $J_{22}$ and $J_{13}$ are shown in appendix~\ref{app:Kernels} and $\Delta b_2 = b^X_2 - b^Y_2$, $\Delta b_{12} =b_2^X b_1^Y - b_1^X b_2^Y$. The 1-loop integrals run over all $k$, including scales beyond the validity of perturbation theory. For that reason we can not fully trust these terms. In particular, the contribution $P_1^{(13)}$ is more sensitive to the UV contribution of its integrand. Schematically we have
\be
P_1^{(13),\phi}(k) \sim  \int_0^k  f \left( k, q\ll k \right) d^3q   + \int_k^\infty  f \left( k, q\gg k \right)d^3q\, ,
\ee
where the UV contribution is determined by
\be
\int_k^\infty  f \left( k, q\gg k \right) d^3q  = \int_k^\infty d^3 q \frac{P \left( q \right)}{q^2}\, .
\ee
We have therefore isolated the contribution induced by the velocity dispersion 
\be
\sigma_v^2 \equiv \int \frac{d^3 q}{\left( 2 \pi \right)^3} \frac{P \left( q \right)}{q^2} = \frac{1}{2\pi^2}\int dk P(k)\, .
\ee
We can now introduce an EFT-inspired parameter $\sigma_0^2$ through
\be
\label{eq:EFT_par}
\sigma_v^2 \rightarrow \sigma_v^2 + \sigma_0^2 \, .
\ee
We used the RayGalGroup simulation to determine its amplitude, obtaining
\begin{equation}
    \frac{\sigma_v^2+ \sigma_0^2 } { \sigma_v^2} \sim 3\, .
    \label{eq:EFTpara}
\end{equation}
Now, for first time, we can derive the contribution of the transverse Doppler effect. Since this is already a second order effect, its leading contribution will be at 1-loop. Consistent with the notation of~\citeref{Breton2018:1803.04294v2} we consider the contribution of the transverse Doppler effect by  subtracting the dipole induced by the redshift perturbation of \eq{eq:z2} and the one defined in \eq{eq:z3}. We remark that this is not the same as computing the effect induced by the transverse Doppler effect on the redshift perturbation only since that would miss some mixing terms, in particular the terms proportional to $f^3$ in $P^{13}$. So we derive the dipole from the galaxy number counts $\Delta_3$ (see eqs.~\ref{eq:Delta_td_1} - \ref{eq:Delta_td_3}) and $\Delta_2$ (see eqs.~\ref{eq:Delta_dop_1} - \ref{eq:Delta_dop_3}), and then take the difference, resulting in 
\begin{align}
    P_1^{(22),\rm TD}(k) &= if^2\frac{\mathcal{H}}{k}\int\frac{d^3q}{(2\pi)^3}\left[\sum_{\Delta b =\{\Delta b_1,\Delta b_2\}}\Delta b T^{\Delta b}_{22}\left(\frac{q}{k},\hat{\vb{k}}\cdot\hat{\vb{q}}\right)\right]P(q)P(|\vb{k} - \vb{q}|) \, , \label{eq:second_order_TD1}\\
    P_1^{(13),\rm TD}(k) &= if^2P(k)\frac{\mathcal{H}}{k}   \Delta  b_1 \left[  \left( \frac{19}{21} - \frac{3}{10} f \right)  k^2 \sigma_v^2 - \frac{3k^2}{10f} \frac{v_{\rm obs}^2}{\HH^2} + \int\frac{d^3q}{(2\pi)^3}T^{\Delta b_1 }_{13}\left(\frac{q}{k}\right)P(q) \right] \, , \label{eq:second_order_TD2}
\end{align}
where we have again isolated the terms proportional to the velocity dispersion $\sigma_v^2$ and the kernels $T_{22}$ and $T_{13}$ are given in appendix~\ref{app:Kernels}. We will adopt the same EFT parameter introduced in \eq{eq:EFT_par}. Let us remark that there is a non-vanishing contribution due to the observer velocity, since the transverse Doppler effect is not linear in the velocity perturbation.
Indeed, the different redshift definitions in eqs.~(\ref{eq:z2} - \ref{eq:z4}) depend in principle on the velocity difference $\bv -\bv_{\rm obs}$. As shown in \appen{app:single_number_counts} this leads to an additional contribution in \eq{eq:second_order_TD2} proportional to  $v^2_{\rm obs}$.

In the same way, we could obtain the 1-loop contribution induced by the linear Doppler term, by taking the difference between the dipole of the galaxy number counts $\Delta_2$ (see eqs.~\ref{eq:Delta_dop_1} - \ref{eq:Delta_dop_3}) and $\Delta_1$ (see eqs.~\ref{eq:Delta_grav_1} - \ref{eq:Delta_grav_3}). However, even at linear order the dipole has large contribution from the wide-angle effect (see \fig{fig:doppler_details}), suggesting that such effects should be modelled consistently at 1-loop.

When implementing these equations we assumed a relation between higher-order bias terms and $b_1$ and $b_2$ (see \eqnb{eq:end:biases}) and we relate the second-order bias to the first-order bias through
\begin{equation}
    b_2(b_1) = 0.412 - 2.143b_1 + 0.929b_1^2 + 0.008b_1^3\, ,
    \label{eq:b2}
\end{equation}
which is calibrated to N-body simulations~\citep{Lazeyras2015:1511.01096v3}. Moreover, the 1-loop contributions derived in this section need to be integrated over the full redshift bin as indicated in \eq{eq:exp_val_estimator}.

\section{The RayGalGroup Simulation}
\label{sec:sims}

\begin{table}[t]
    \begin{center}
        \begin{tabular}{llllllll}
            Label & $N_{\rm ran}$ & $b_1$ & Halo mass & Particles & Haloes & $\overline{n}$ & $b_e$\\
            & & & [$h^{-1}M_{\odot}\times 10^{12}$] & & & [$h^3$Mpc$^{-3}$] &\\
            \hline
            mb1 & 5x & 1.08 & $1.88$ - $3.76$ & 100 - 200 & $5\,412\,928$ & $6.499\times 10^{-4}$ & 0.08\\
            mb2 & 6x & 1.22 & $3.76$ - $7.52$ & 200 - 400 & $3\,434\,248$ & $4.123\times 10^{-4}$ & 0.28 \\
            mb3 & 12x & 1.42 & $7.52$ - $15.04$ & 400 - 800 & $1\,901\,293$ & $2.283\times 10^{-4}$ & 0.49\\
            mb4 & 24x & 1.69 & $15.04$ - $30.08$ & 800 - 1600 & $957\,910$ & $1.150\times 10^{-4}$ & 0.74\\
            mb5 & 50x & 2.07 & $30.08$ - $60.16$ & 1600 - 3200 & $442\,193$ & $5.309\times 10^{-5}$ & 1.12\\
            mb6 & 100x & 2.59 & $60.16$ - $120.32$ & 3200 - 6400 & $182\,463$ & $2.191\times 10^{-5}$ & 1.82\\
            all & 5x & 1.28 & $1.88$ - $120.32$ & 100 - 6400 & $12\,331\,035$ & $1.480\times 10^{-3}$ & 0.31
        \end{tabular}
        \caption{Properties of the 6 different mass bins of the RayGalGroup simulation. The effective redshift is $z_{\rm eff} = 0.341$ (defined as the volume weighted mean) and the bias parameters are taken from~\citep{Breton2018:1803.04294v2}, except for the bias for the entire sample (last row), which we measured in this paper. The simulation uses a mass per particle of $1.88\times 10^{10}h^{-1}M_{\odot}$ and the volume between $z_{\rm min}= 0.05$ and $z_{\rm max} = 0.465$ is $V = 8.329h^{-3}\text{Gpc}^3$. From the volume we can estimate the fundamental mode to be $k_{\rm f} \approx \frac{2\pi}{V^{1/3}} = 0.0031\kMpc$. The evolution bias in the last column is evaluated at $z_{\rm eff}$. For the modeling we generally integrate over redshift and account for the redshift evolution of $b_1$ and $b_e$ as shown in appendix~\ref{app:zevo} and \sect{sec:evo_bias}, respectively.}\label{tab:simspecs}
    \end{center}
\end{table}

In this paper we make use of the publicly available RayGalGroup simulation~\citep{Breton2018:1803.04294v2} to test the scales on which the model discussed in the last section is valid. We use the cross-power spectrum dipole as our primary metric for this comparison. 
The RayGalGroup simulation is based on a dark matter only N-body simulation with a box size of $2\,625\Mpc$ and $4\,096^3$ particles. The haloes in this simulation have been detected using pFoF with a linking-length of $b=0.2$ and a minimal mass of $100$ particles. From this simulation a $4\pi$ light-cone catalog is produced including ray tracing and taking into account all relativistic effects (redshift and angular perturbations) at first order in the weak field approximation: Doppler, gravitational, transverse Doppler, ISW, and weak lensing. The cosmology of this simulation is $h=0.72$, $\Omega_m=0.25733$, $\Omega_b=0.043557099$, $\Omega_r=0.000080763524$, $n_s=0.963$, $\sigma_8=0.80100775$ and $w=-1$. We use these parameters as our fiducial cosmology when analyzing the simulation. The redshift range is limited to $0.05 < z < 0.465$ leading to an effective redshift of $z_{\rm eff} = 0.341$. We split this dataset into $6$ sub-samples divided by halo mass (see \tabl{tab:simspecs} for details). 

The simulation provides two sets of angular positions, $\beta$ and $\theta$, where the second accounts for lensing effects, while the first does not. We also have $5$ different redshifts available, which include different combinations of relativistic distortions~\footnote{While in eqs.~(38 - 43) of \citeref{Breton2018:1803.04294v2} the effects on the redshift perturbation are considered individually (and not summed up as we show in eqs.~\ref{eq:z0} - \ref{eq:z5}), our notation fully agrees with the documentation provided together with their publicly available halo catalogs.}:
\begin{align}
    \textbf{Real-space:} && z_0 &= \frac{a_0}{a} - 1\label{eq:z0}\\
    \textbf{+ Potential term:} && z_1 &= \frac{a_0}{a} \left( 1 - \frac{\phi_s - \phi_0}{c^2} \right) - 1\label{eq:z1}\\
    \textbf{+ Peculiar velocity:} && z_2 &= \frac{a_0}{a} \left( 1 - \frac{\phi_s - \phi_0}{c^2} - \frac{\vb{v}\cdot \vb{n}}{c} \right) -1\label{eq:z2}\\
    +\;\parbox{7.2em}{\textbf{Transverse\\ {Doppler term:}}} && z_3 &= \frac{a_0}{a} \left( 1 - \frac{\phi_s - \phi_0}{c^2} - \frac{\vb{v}\cdot \vb{n}}{c} + \frac{|\vb{v}\cdot\vb{v}|}{2c^2} \right) - 1\label{eq:z3}\\
    \textbf{+ ISW term:} && z_4 &= \frac{a_0}{a} \left( 1 - \frac{\phi_s - \phi_0}{c^2} - \frac{\vb{v}\cdot \vb{n}}{c} + \frac{|\vb{v}\cdot\vb{v}|}{2c^2} -  \frac{2}{c^2}\int^{\eta_0}_{\eta_s}\dot{\phi} d\eta \right) -1\label{eq:z4}\\
    && z_5 &= \frac{(g_{\mu\nu}k^{\mu}u^{\nu})_s}{(g_{\mu\nu}k^{\mu}u^{\nu})_0}-1.
    \label{eq:z5}
\end{align}
The quantities with a subscript `s' and with a subscript `0' are evaluated at the source and the observer position, respectively. The redshift defined in $z_5$ is based on the full covariant definition, including all cross-terms between the individual contributions in $z_1 - z_4$, which are ignored by the linear calculation. Therefore comparison of results obtained with $z_4$ and $z_5$ will allow us to propagate the assumptions which go into these redshift definitions to the final observables (mainly the power spectrum dipole). These two redshifts agree very well for all cases discussed in this paper as demonstrated in \fig{fig:dipole_diff} (bottom, right). 

\begin{figure}[t]
\centering
\includegraphics[width=0.45\textwidth]{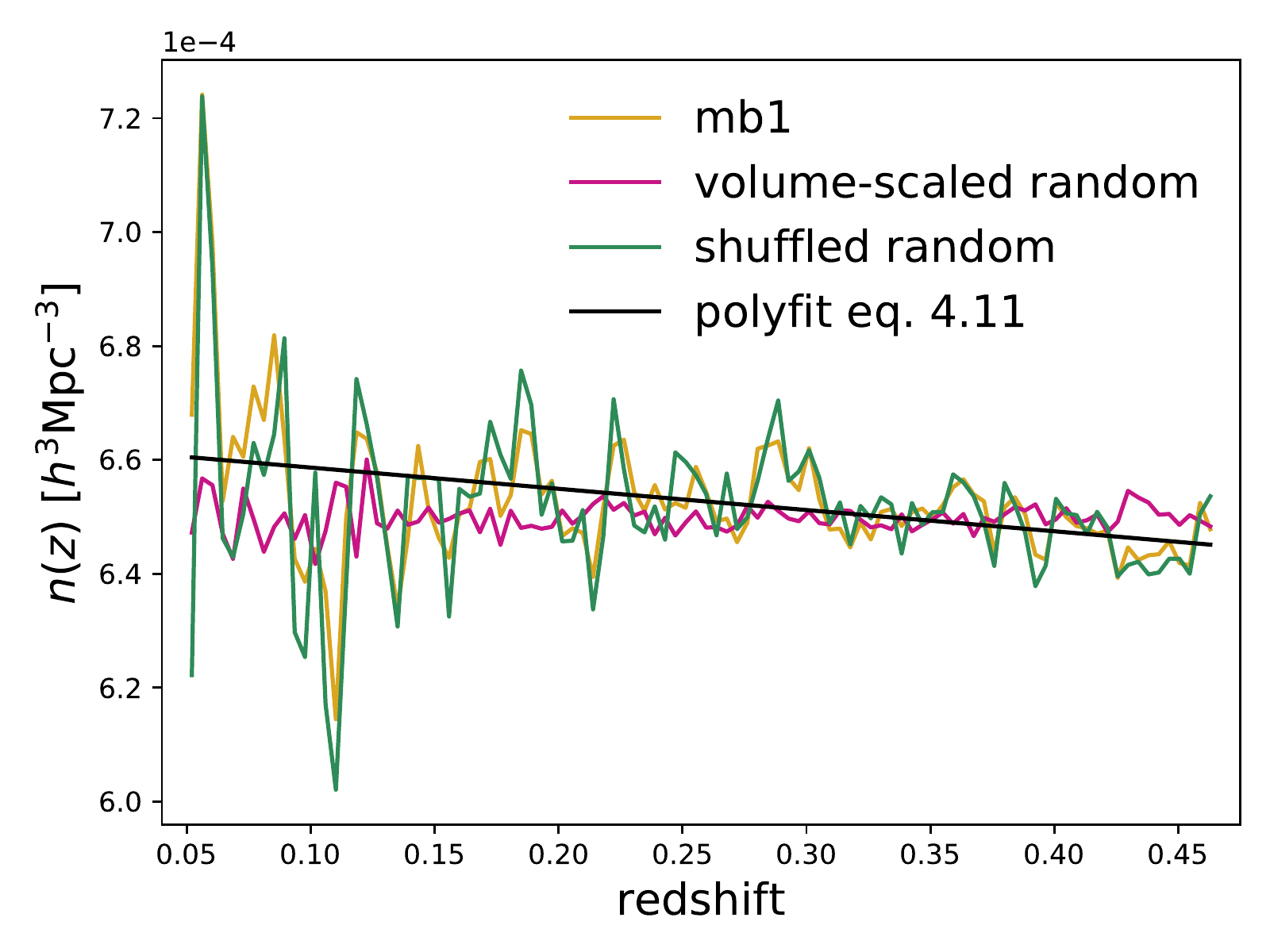}
\includegraphics[width=0.45\textwidth]{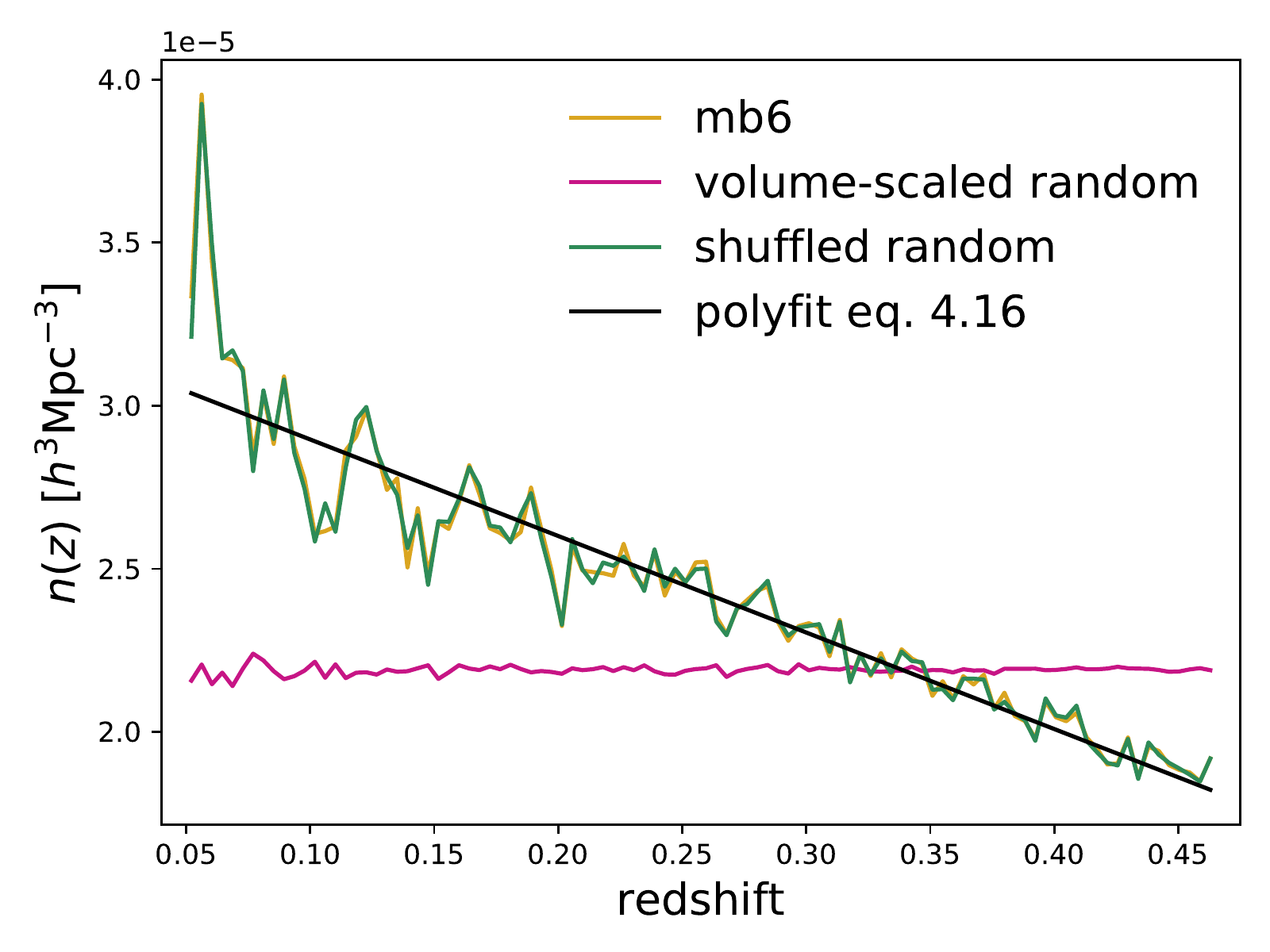}
\caption{Density distribution of haloes in the RayGalGroup simulation (yellow) and the corresponding random catalog using pure volume scaling (magenta) as well as a random catalog using shuffled redshifts of the simulation itself (green). The volume scaling works well for the low mass bin (left), but seems a mismatch in the highest mass bin (right). This mismatch is causing the excess power on large scales seen in \fig{fig:auto_ran_cmp}. When processing these densities for the evolution bias in \sect{sec:evo_bias} we use a polynomial fit shown as the solid black lines (see also eqs.~\ref{eq:poly1} - \ref{eq:poly6}).}
\label{fig:hist_z}
\end{figure}

\subsection{Window function}
\label{sec:window}

The measured halo power spectrum based on the estimator discussed in \sect{sec:estimator} is a convolution of the true underlying power spectrum with the survey window function. For that reason, we need to separately measure the survey window and convolve any power spectrum model, before comparing it with a measurement. 

The window function represents the power spectrum of the random catalog, which should follow the same distribution as the data catalog, but without any intrinsic clustering.
Given that the RayGalGroup simulation corresponds to a sphere with a pure volume scaling of the number of galaxies from redshift $z=0.05$ up to $z=0.465$, the random catalog can, in principle, be obtained analytically. However, as shown in \fig{fig:hist_z}, the highest mass bin of the RayGalGroup simulation does not exactly follow this distribution. The most likely reason is the evolution of the halo population within the redshift bin. We therefore construct random catalogs by selecting random positions on the sky and assigning redshifts by randomly sampling from the corresponding halo catalogs within the mass bins. This is the same method usually used to generate random catalogs in galaxy redshift surveys (see e.g.~\citeref{Reid2015:1509.06529v2} for BOSS). However, we note that this will erase large scale modes along the line-of-sight~\citep{Mattia2019:1904.08851v3}. For the rest of this paper, we will denote random catalogs generated using this method as `shuffled', while any result using the random catalogs based on volume scaling is labeled `volume'. The shuffled method will be our default.

\begin{figure}[t]
\centering
\includegraphics[width=0.65\textwidth]{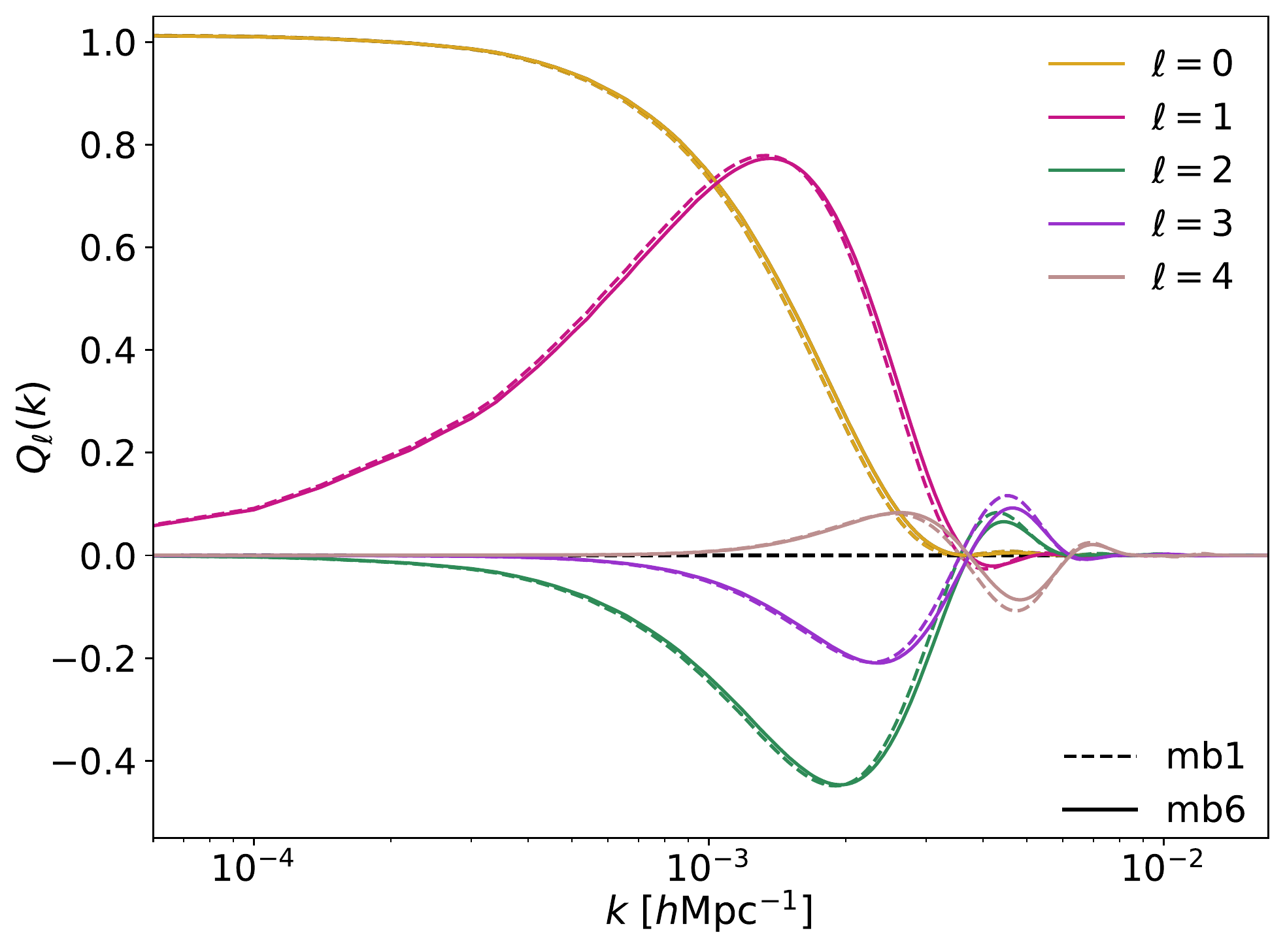}
\caption{Window function multipoles for the RayGalGroup simulation. The solid lines show the window function for the highest mass bin (mb6), while the dashed lines show the window functions for the lowest mass bin (mb1). The differences between the dashed lines and the solid lines reflect the different selection functions shown in \fig{fig:hist_z}, where mb6 includes the redshift evolution of the high mass halo densities. This variation in the window functions for the different halo mass bins is ignored in our analysis.}
\label{fig:cmp_window_z0_shuffle_mb6_mb1}
\end{figure}

Now we can proceed to measure the window function multipoles as described in appendix E of \citeref{Beutler2018:1810.05051v3}. Figure~\ref{fig:cmp_window_z0_shuffle_mb6_mb1} shows the first five window function multipoles for the highest (solid lines) and lowest (dashed lines) mass bin using the redshift definition $z_0$ (see \eqnb{eq:firstorder}). The differences between solid and dashed lines give an estimate of the impact of the slightly different survey geometries since the different mass bins have different random catalogs (when using the shuffle method).

One important point to note is that the main results of this paper will be derived from the differences between power spectra with different redshift definitions, including the relativistic (and Newtonian) contribution in turn (see eqs.~\ref{eq:z0} - \ref{eq:z5}). Since the survey geometry is the same for all redshift definitions, the window function will not impact most differential measurements. The only exception is the relativistic Doppler term, which does get additional contributions from the quadrupole and the Kaiser factor contributions to the monopole. These changes to the even multipoles are sourced by peculiar velocities, just like the relativistic Doppler term itself. The window function contribution to the power spectrum dipole is given by
\begin{align}
    \Delta P_1^{\rm win}(k) &= \hat{P}_1^{\rm red}(k) -  \hat{P}_1^{\rm real}(k)\, ,
    \label{eq:dipole_win}
\end{align}
where $\hat{P}_1^{\rm red}$ and $\hat{P}_1^{\rm real}$ are the convolved redshift-space and real-space power spectrum dipoles. Here we are only interested in the window function contributions of the even multipoles to the convolved dipole given by~\footnote{This corresponds to the case $n=0$ in eqs.~(3.5) and (3.6) of~\citep{Beutler2013:1312.4611v2}.}
\begin{equation}
    \hat{P}_1(k) = -3 i \int \mathrm{d}s\,s^{2} j_1 (ks) \,\hat{\xi}_1(s) - iQ_{1}(k) \int \mathrm{d}s\,s^{2} \,\hat{\xi}_0(s)
\end{equation}
with
\begin{align}
    \hat{\xi}_0(s) = \xi_0(s)Q_0(s)&+\frac{1}{5}\xi_2(s)Q_2(s)+\frac{1}{9}\xi_4(s)Q_4(s)\, ,
    \label{eq:xi00}\\
    \begin{split}
        \hat{\xi}_1(s) = \xi_0(s)Q_1(s)&+\xi_2(s)\left[\frac{2}{5}Q_1(s)+\frac{9}{35}Q_3(s)\right] + \frac{4}{21}\xi_4(s)Q_3(s)\, ,
    \end{split}\label{eq:xi01}
\end{align}
where $Q_{\ell}(s)$ are the window function multipoles in configuration-space~\citep{Wilson2015:1511.07799v2,Beutler2016:1607.03150v1}.

We first obtain the window function multipoles in Fourier-space, $Q_{\ell}(k)$, using the setup outlined in appendix E of \citep{Beutler2018:1810.05051v3} (see \fig{fig:cmp_window_z0_shuffle_mb6_mb1}), which we then Fourier-transform using the 1D Hankel transform defined in \eq{eq:hankel} to obtain $Q_{\ell}(s)$. For the even mulipoles in eqs.~(\ref{eq:xi00}) and (\ref{eq:xi01}) it is sufficient, for this analysis, to assume linear theory. The result is included in \fig{fig:doppler_details} as the dashed-dotted line (Doppler window term) on the right-hand plot and in the Doppler term of \fig{fig:dipole_diff} and \ref{fig:asym}.

The general effect of the window function is to correlate modes and smooth features at the scale of the fundamental mode. For the dipole, there are significant contributions from the even multipoles as shown in \fig{fig:doppler_details}. However, those contributions are a result of wide-angle effects. Without wide-angle effects, the window function would only correlate odd multipoles amongst each other and even multipoles amongst each other, which would significantly reduce the window function contributions to the dipole. Therefore, if these window function contributions significantly reduce the signal-to-noise of the dipole, one can always use an estimator which has less significant wide-angle effects (see e.g. eq.~4 of \citeref{Beutler2013:1312.4611v2})

\subsection{Galaxy density evolution bias}
\label{sec:evo_bias}

\begin{figure}[t]
\centering
\includegraphics[width=0.65\textwidth]{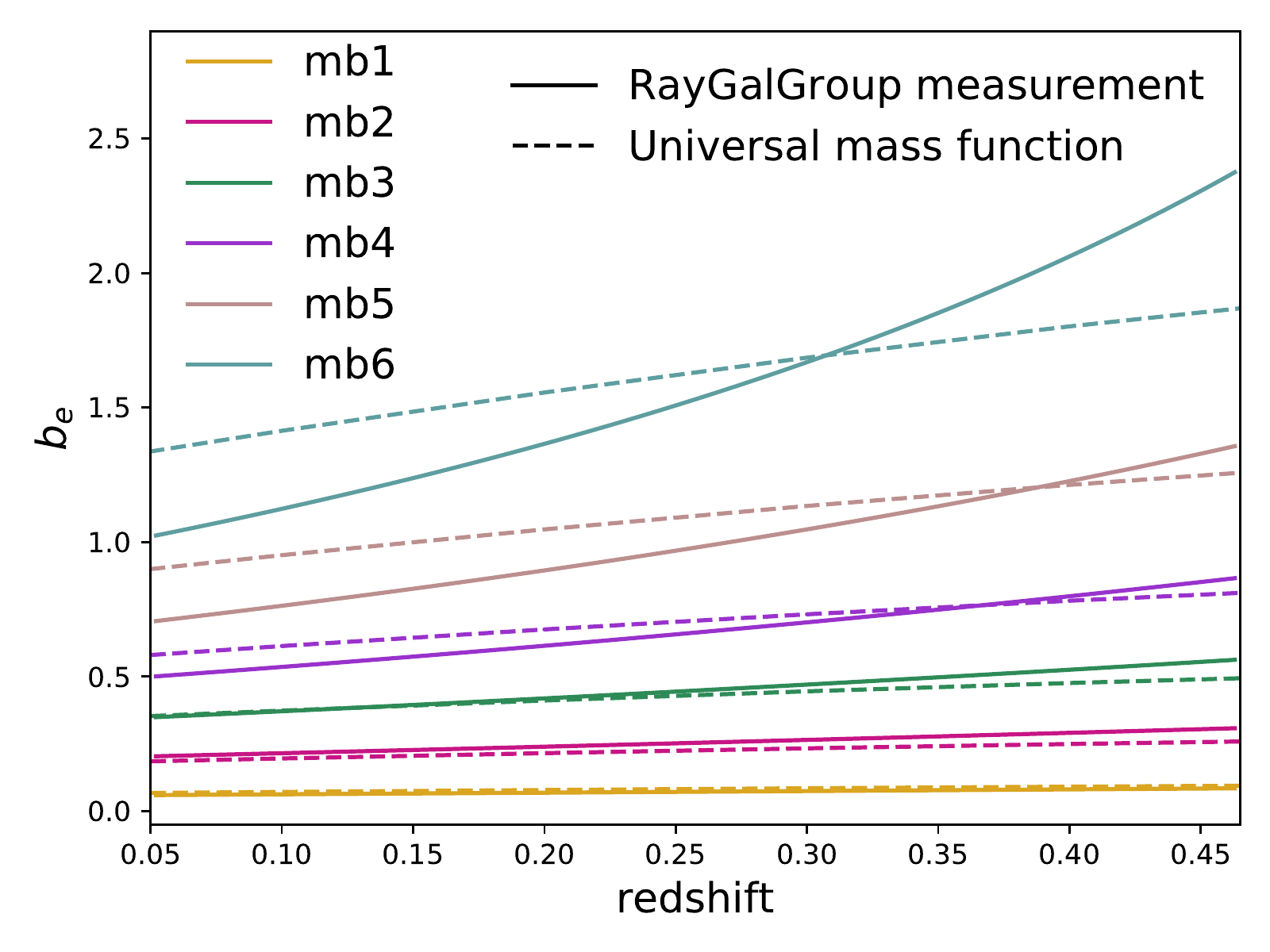}
\caption{Evolution bias $b_e$ calculated from the halo density (eqs.~\ref{eq:poly1} - \ref{eq:poly6}) following \eq{eq:bevolve} as a function of redshift for the six different halo mass bins of the RayGalGroup simulation (see \tabl{tab:simspecs}). The dashed lines correspond to the predictions assuming an universal mass function (see \eqnb{eq:beuniversal}).}
\label{fig:evolution_bias}
\end{figure}

Here we discuss the evolution bias, $b_e$, due to the changing galaxy tracer density within the redshift bin as given in \eq{eq:bevolve}. This is not related to the evolution effects discussed in \sect{sec:evo_wa}, which are sourced by the evolution of the linear galaxy bias $b_1$, linear growth rate $f$ and growth function $D$.

As we can see in \fig{fig:hist_z} the density of haloes is roughly constant in redshift for the lowest mass bin (mb1), while we notice some redshift dependence in the highest mass bin (mb6). Since the evolution bias quantifies the deviation from a conserved source number density in a comoving volume, we do expect non-vanishing values of $b_e$ for the different mass bins. Here we will calculate the evolution bias contributions for the different mass bins shown in \tabl{tab:simspecs}.

First we fit the (comoving) source densities $n(z)$ measured from each sample with a linear polynomial
\bea
n_{\rm mb1} \left( z \right) &=& \left( 6.62388 -0.372935 z \right) \times 10^{-4}\label{eq:poly1}
\, , \\
n_{\rm mb2} \left( z \right) &=& \left(4.4104 -0.845952 z \right) \times 10^{-4}
\, , \\
n_{\rm mb3} \left( z \right) &=& \left(2.5676 -0.837904 z  \right) \times 10^{-4}
\, , \\
n_{\rm mb4} \left( z \right) &=& \left(1.36626 -0.634672 z \right) \times 10^{-4}
\, , \\
n_{\rm mb5} \left( z \right) &=& \left( 0.68107 -0.441901 z\right) \times 10^{-4}
\, , \\
n_{\rm mb6} \left( z \right) &=& \left( 0.31923 -0.295892 z \right) \times 10^{-4}\label{eq:poly6}
\, ,
\eea
which are also included in \fig{fig:hist_z}.
Next we compute the evolution bias through~\eq{eq:bevolve}, obtaining
\begin{align}
    b_e^{\rm mb1} \left( z \right)  &= \frac{18.7615}{17.7615 - z}-1\, ,\\
    b_e^{\rm mb2}\left( z \right)  &=\frac{6.21353}{5.21353- z}-1\, , \\
    b_e^{\rm mb3}\left( z \right)  &= \frac{4.06431}{3.06431- z}-1\, , \\
    b_e^{\rm mb4}\left( z \right)  &=\frac{3.1527}{2.1527- z}-1\, , \\
    b_e^{\rm mb5}\left( z \right)  &=\frac{2.54123}{1.54123-z}-1\, , \\
    b_e^{\rm mb6} \left( z \right) &= \frac{2.07887}{1.07887- z}-1\, .
\end{align}
In \fig{fig:evolution_bias} we show the evolution bias $b_e$ as a function of redshift for the six different halo mass bins. We notice that the evolution bias in the lowest mass bin is almost constant. Indeed this reflects the redshift independence we see in \fig{fig:hist_z}. The dashed lines in \fig{fig:evolution_bias} correspond to the prediction of the evolution bias assuming a universal mass function~\citep{Jeong2011:1107.5427v2}
\begin{equation}
    b_e = (b_1 - 1)f\delta_c\, ,
    \label{eq:beuniversal}
\end{equation}
where we used $\delta_c\approx 1.686$. The assumption of a universal mass function has significant limitations, especially for massive haloes, but seems to work well for all mass bins in the RayGalGroup simulation, with excellent agreement at low halo mass.

With the measurements of $b_e$ presented in this section we can obtain the evolution bias contributions to the power spectrum dipole using \eq{eq:fevo}. The results are included in \fig{fig:doppler_details} (dotted lines in the right-hand plot).

While it is fairly straightforward to calculate the evolution bias for a halo catalog like the RayGalGroup simulation, it is naturally far more difficult to access such a quantity for a realistic and incomplete galaxy survey. The number density required for the derivative in \eq{eq:bevolve} is the true underlying density, which cannot easily be inferred from the density of observed galaxies in existing redshift surveys. However, from our analysis we can conclude that (1) the evolution bias contributions are sub-dominant for the halo masses investigated here and (2) using the universal mass function approach does seem promising for low mass haloes.

\section{Analysis}
\label{sec:analysis}

\begin{figure}[t]
\centering
\includegraphics[width=0.8\textwidth]{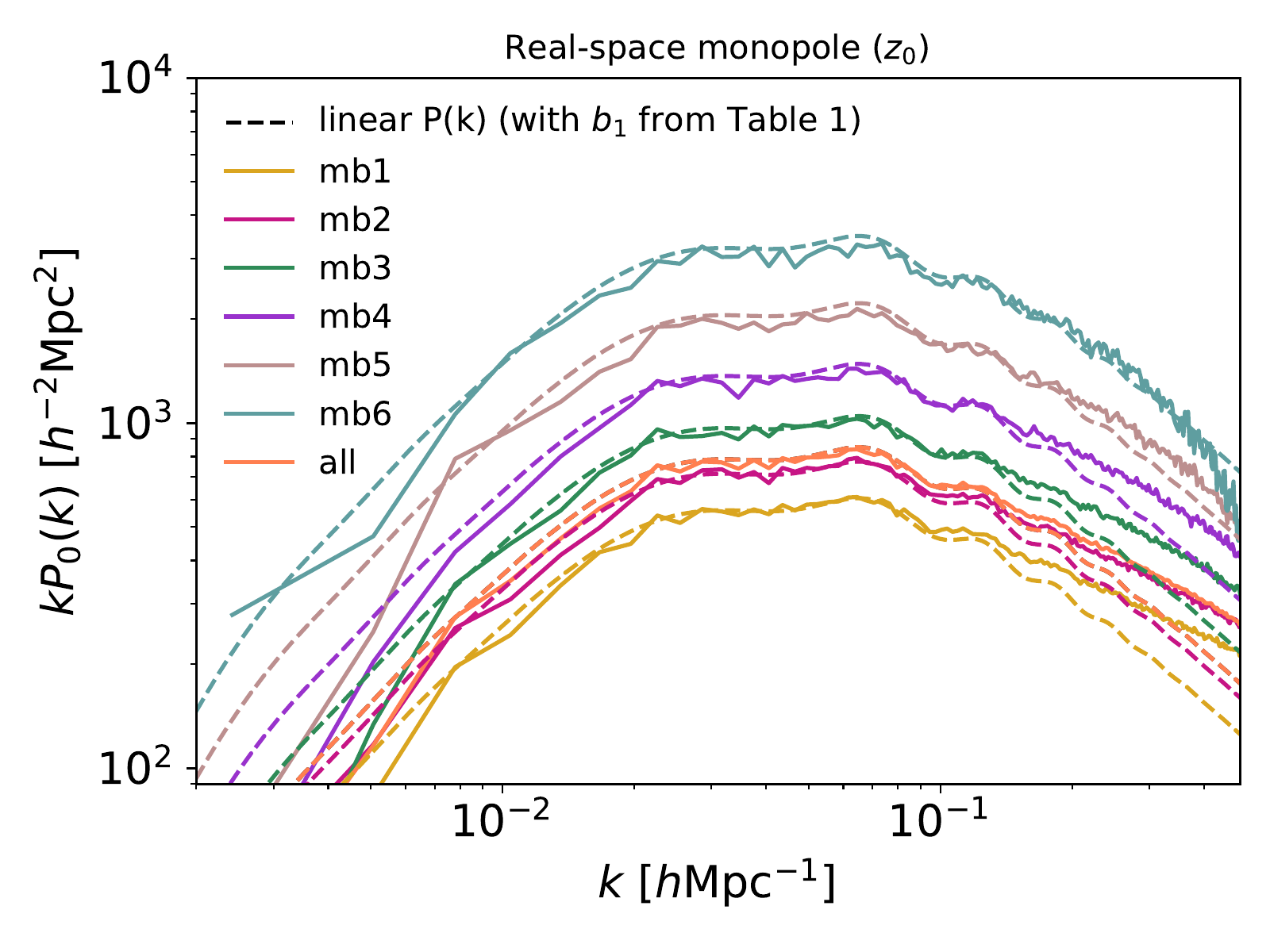}
\caption{Power spectrum monopole for the different mass bins of the RayGalGroup simulation (solid lines, see \tabl{tab:simspecs}). These measurements use the redshift definition $z_0$ (see \eqnb{eq:z0}),  the shuffled random catalogs and $\Delta k = 0.0024\kMpc$. The dashed lines represent linear power spectrum monopole models including the convolution with the survey window function and the linear halo bias parameters of \tabl{tab:simspecs}. }
\label{fig:bias}
\end{figure}

\begin{figure}[t]
\centering
\includegraphics[width=0.45\textwidth]{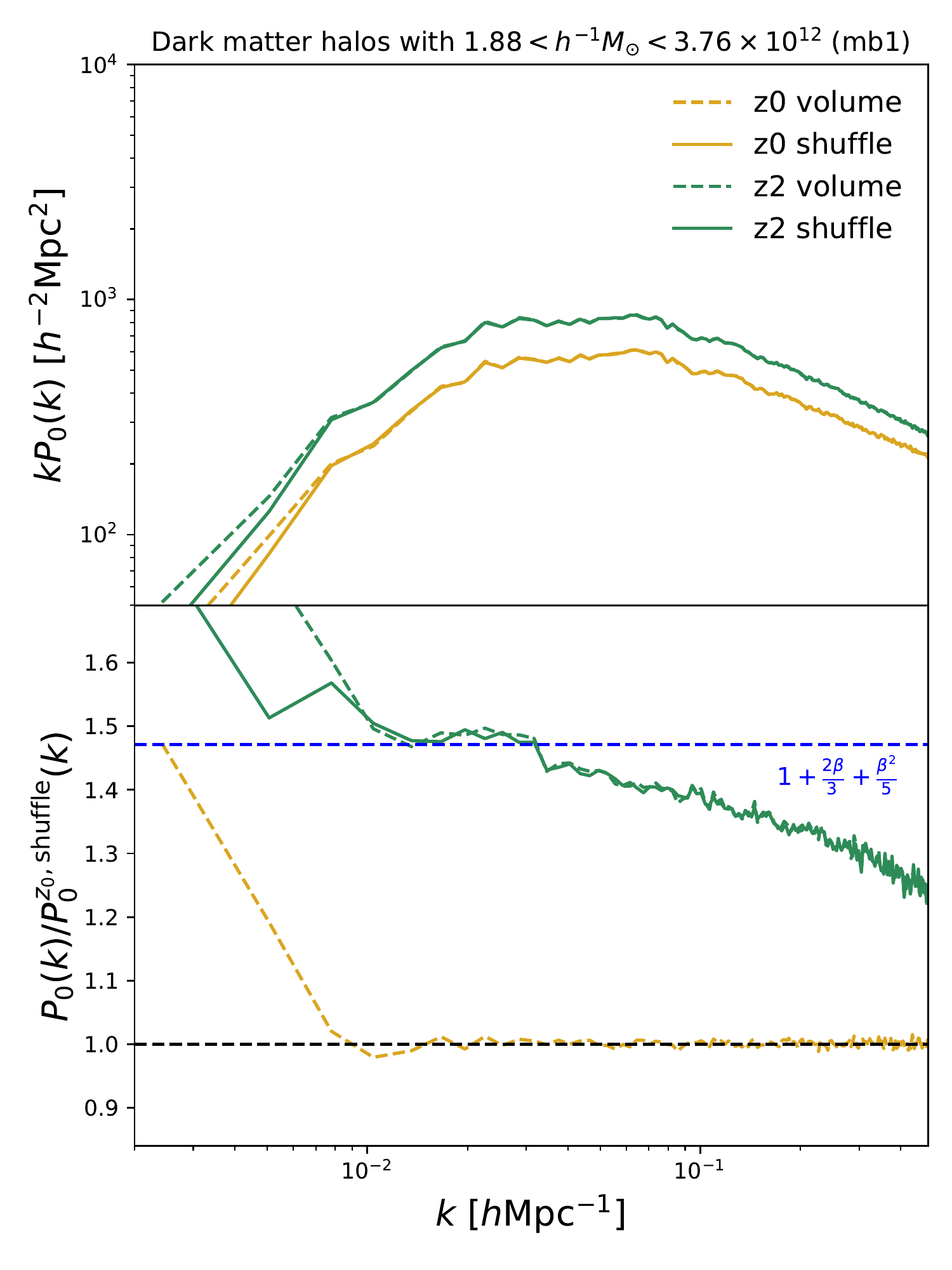}
\includegraphics[width=0.45\textwidth]{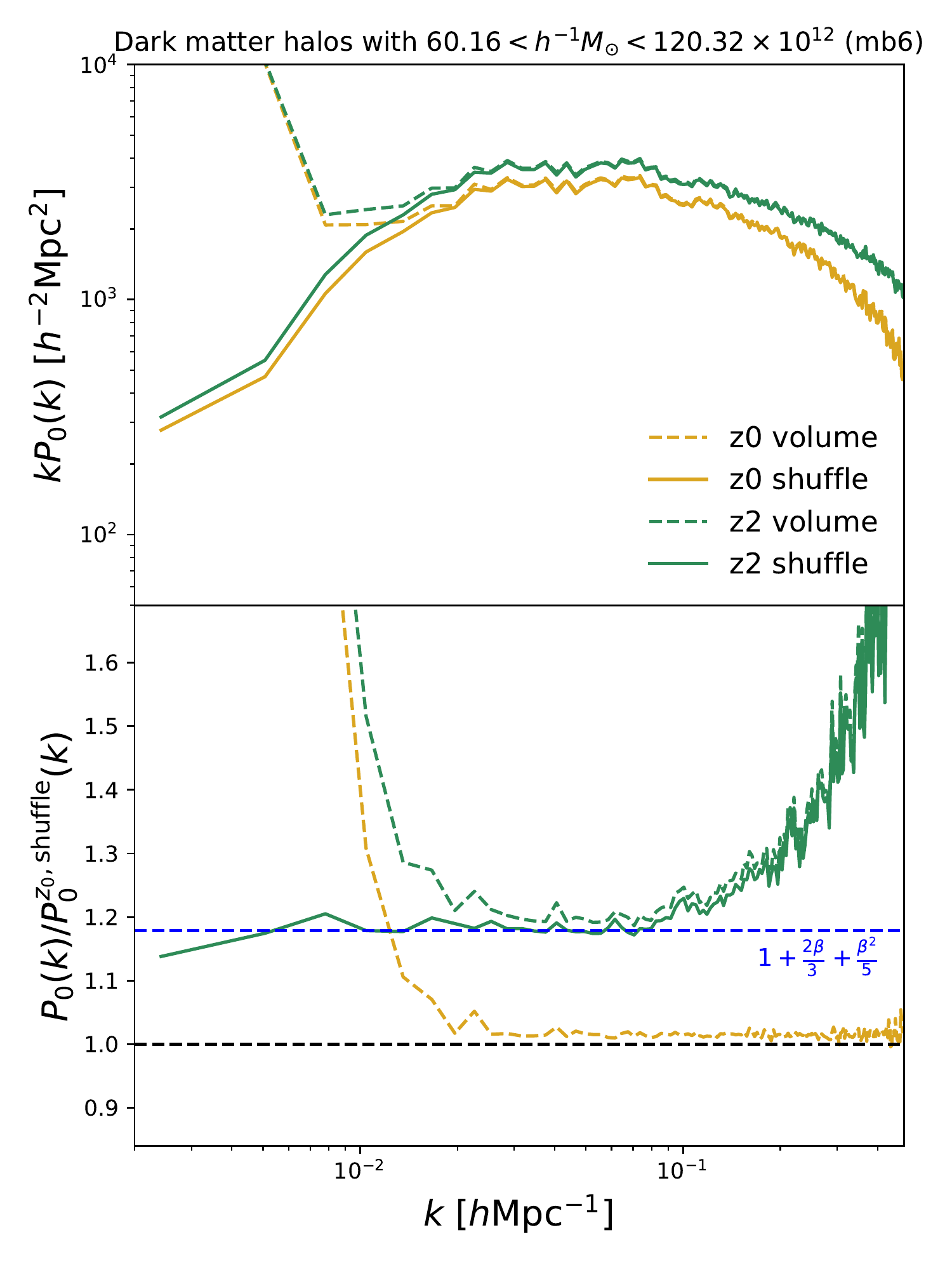}
\caption{Comparison of the power spectrum monopole for the real-space redshift definition $z_0$ in \eq{eq:z0} and $z_2$ in \eq{eq:z2}. Note that all other redshift definitions result in almost indistinguishable power spectra since the relativistic effects in the auto-power spectrum monopole are suppressed by $(\HH/k)^2$.  
The yellow dashed line shows the power spectrum using a random catalog with redshifts sampled according to the cosmic volume (volume), while the solid lines use random catalogs with redshifts sampled from the data catalog (shuffle). This comparison uses the lowest mass bin (mb1, left) and the highest mass bin (mb6, right). The lower panel shows the same measurements normalized to the power spectrum measured using the redshift definition $z_0$ and the shuffle method for the random catalog. The high mass bin with a volume scaled random catalog shows excess power on large scales, which is related to the mismatch shown in \fig{fig:hist_z}. The higher amplitude for the monopole measured using the redshift definition $z_2$ compared to $z_0$ is caused by the peculiar velocity term, of which the linear prediction (Kaiser factor) is included as the dashed blue line in the lower panel.}
\label{fig:auto_ran_cmp}
\end{figure}

\begin{figure}[t]
\centering
\includegraphics[width=0.45\textwidth]{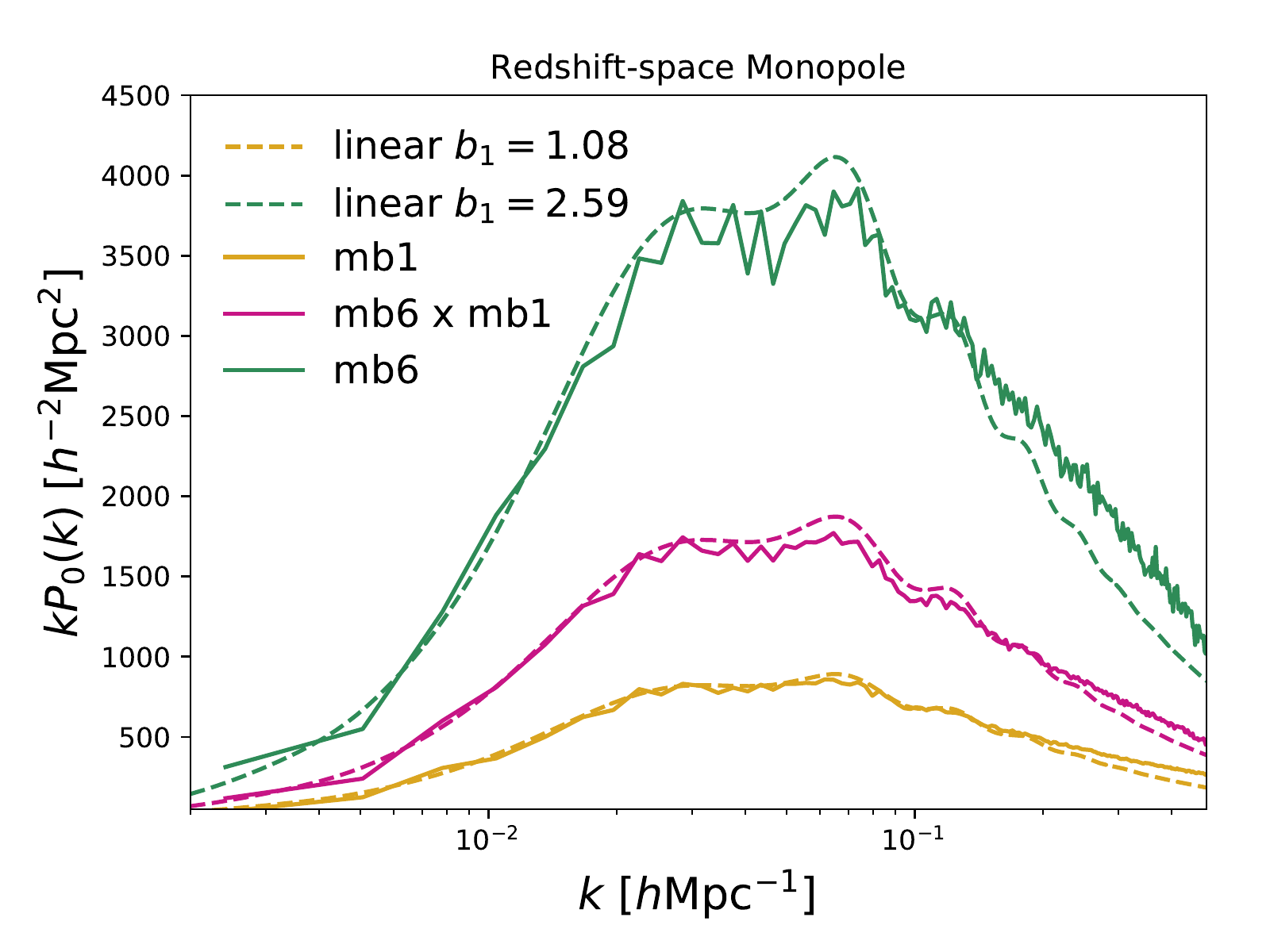}
\includegraphics[width=0.45\textwidth]{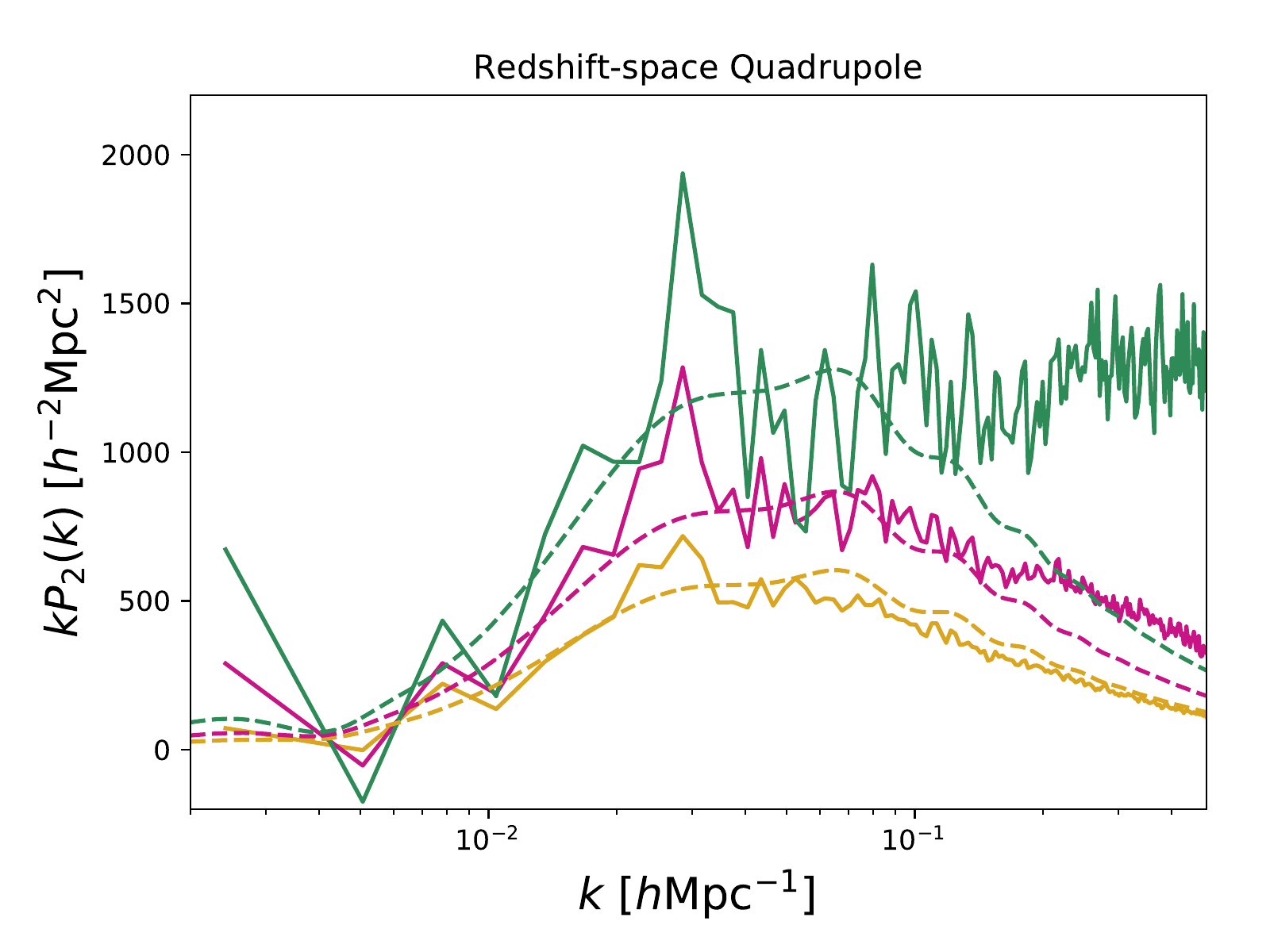}
\caption{The measured power spectrum monopole (left) and quadrupole (right) for the highest (mb6) and lowest (mb1) mass bins, including the cross-power spectrum (magenta). The dashed lines represent a linear model including the convolution with the survey window function and the bias parameters shown in \tabl{tab:simspecs}. These statistics are calculated using the redshift definition $z_5$ (see \eqnb{eq:z5}).}
\label{fig:linear}
\end{figure}

We will now analyze the different halo mass bins of the RayGalGroup simulation. We start with the auto-power spectrum to estimate the halo bias for the different samples. We then move on to the main analysis using the cross-power spectrum dipole.

\subsection{Auto-power spectrum measurements}

To measure the halo power spectrum we use the FFT-based estimator of~\citep{Bianchi2015:1505.05341v2} together with the shuffled random catalogs. Our measurements have a Nyquist frequency of $k_{\rm Ny} = 0.96\kMpc$ and hence we will limit all model comparisons to half that frequency. Given that here we only intend to measure the linear halo bias, we only focus on very large scales where linear theory roughly holds.

Figure~\ref{fig:bias} shows the power spectrum monopole for the different mass bins using the real-space redshift definition $z_0$ (see \eqnb{eq:z0}). The difference in amplitude is well described by the bias parameters given in Table~\ref{tab:simspecs}, which are taken from the original analysis of~\citep{Breton2018:1803.04294v2} (derived from the correlation function). The model power spectra plotted in this figure are based on a linear power spectrum extracted from \textit{CLASS}~\citep{Lesgourgues2011:1104.2932v2} using the fiducial cosmological parameters of the simulation as well as a convolution with the survey window function~\citep{Beutler2013:1312.4611v2, Wilson2015:1511.07799v2, DAmico2019:1909.05271v1}.

Figure~\ref{fig:auto_ran_cmp} shows a comparison between the power spectrum monopole using the $z_0$ and $z_2$ redshift definitions given in eqs.~(\ref{eq:z0}) and (\ref{eq:z2}), for the low mass bin (left) and high mass bin (right). All other redshift definitions would not show any significant differences since the relativistic effects are strongly suppressed in the auto-correlation. We also included the power spectrum measured with a random catalog based on a volume scaling of the number of galaxies, which has a significant impact on the largest scales (dashed lines). The blue dashed lines in the lower panels show the linear (Kaiser) prediction, which describes the low-$k$ measurement of the high mass bin (\fig{fig:auto_ran_cmp}, right) rather well. In the low mass bin (\fig{fig:auto_ran_cmp}, left), where redshift-space distortions have a much larger impact (since $\beta \propto 1/b_1$), the linear model seems to fail at all scales, suggesting that non-linear contributions matter even on the largest scales.

Figure~\ref{fig:linear} shows the power spectrum monopole and quadrupole of the high and low mass bins together with a linear Kaiser model. The model describes the multipoles well up to $k\sim0.1\kMpc$. We refer to \citeref{Breton2018:1803.04294v2} for further tests of these simulations including a comparison of the auto-correlation functions for different mass bins with the CosmicEmu emulator~\citep{Heitmann2015:1508.02654v1}.

\subsection{Cross-power spectrum dipole measurements}
\label{sec:dipole}

\begin{figure}[t]
\centering
\includegraphics[width=0.45\textwidth]{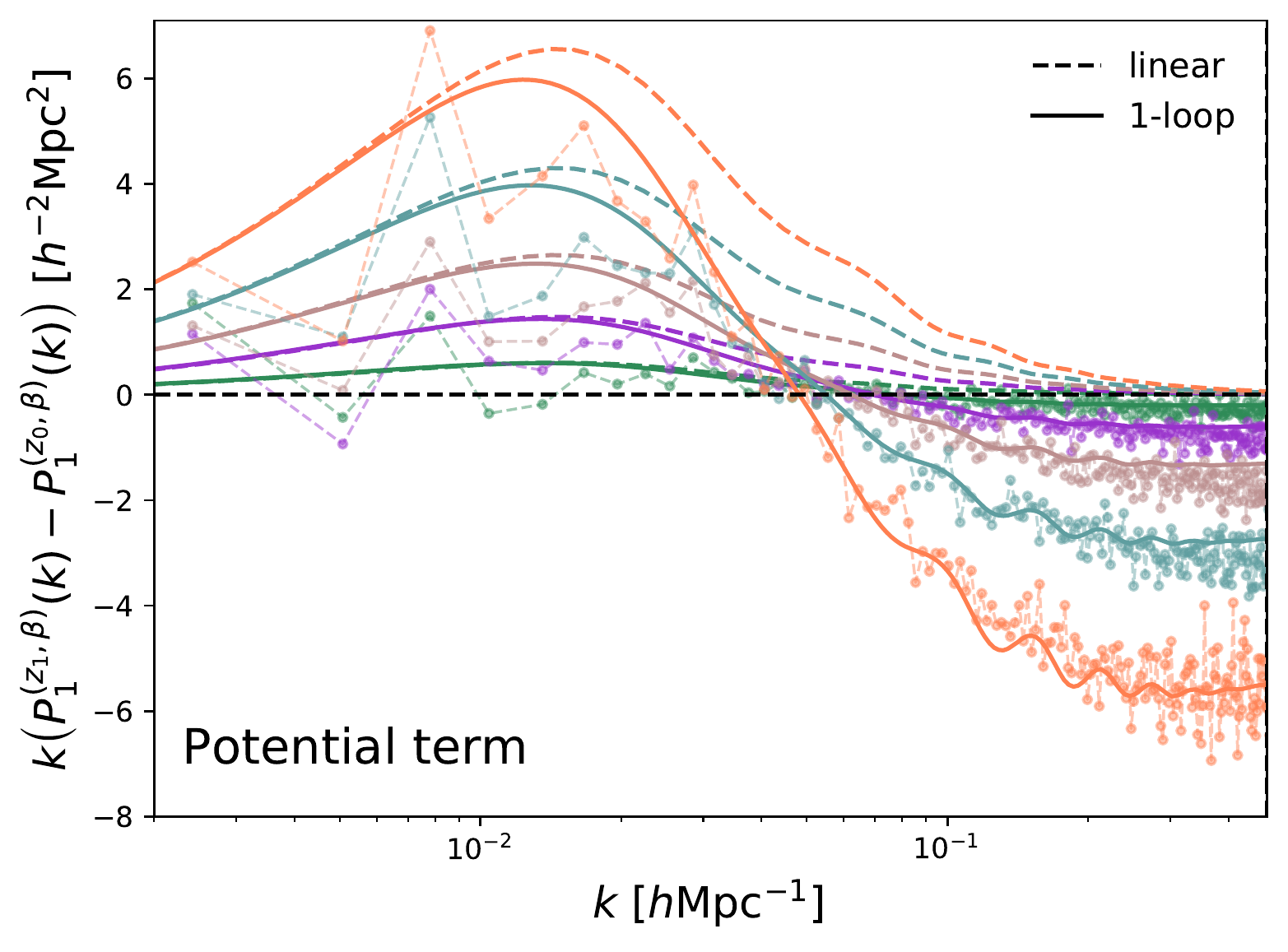}
\includegraphics[width=0.45\textwidth]{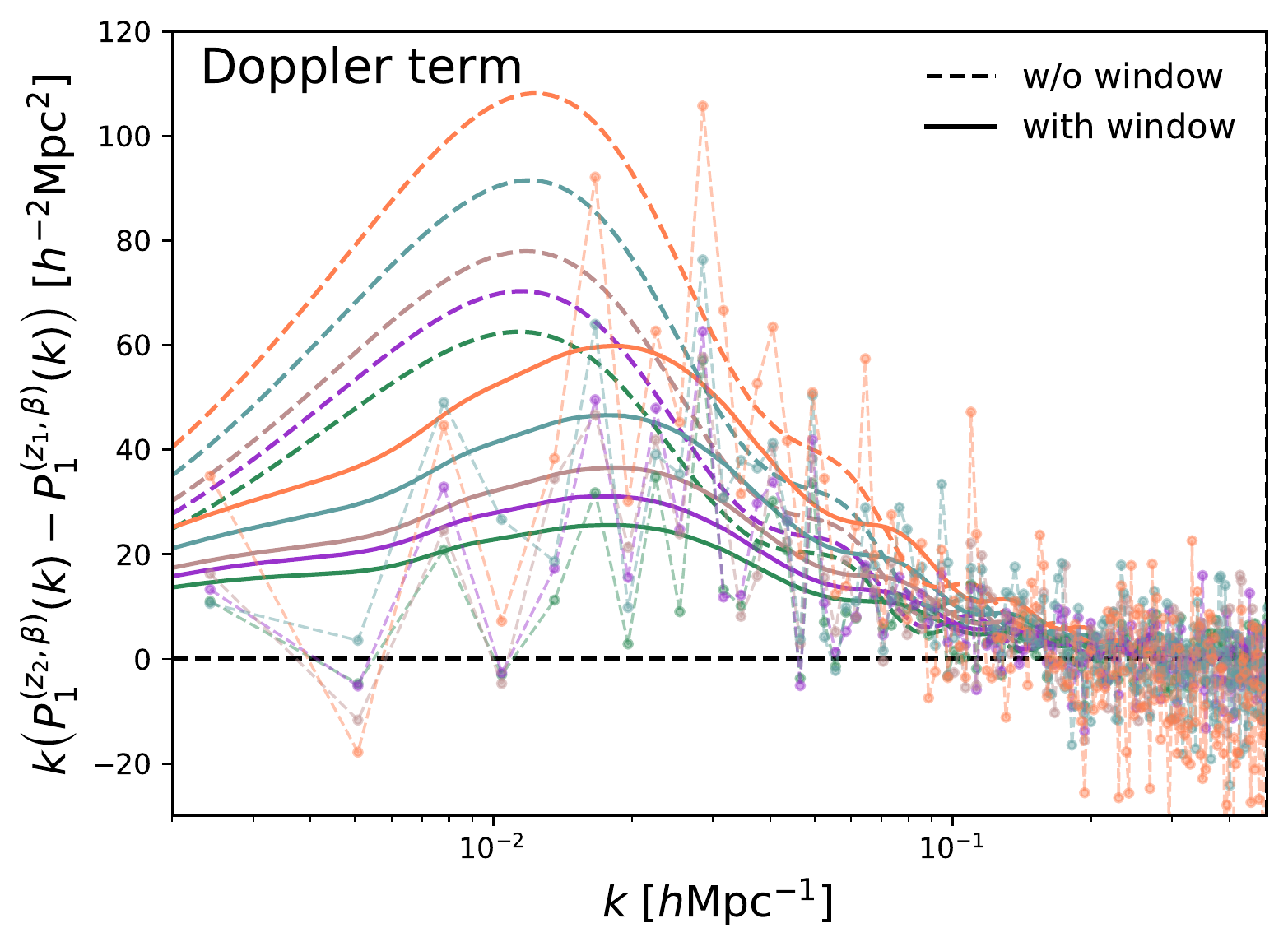}\\
\includegraphics[width=0.45\textwidth]{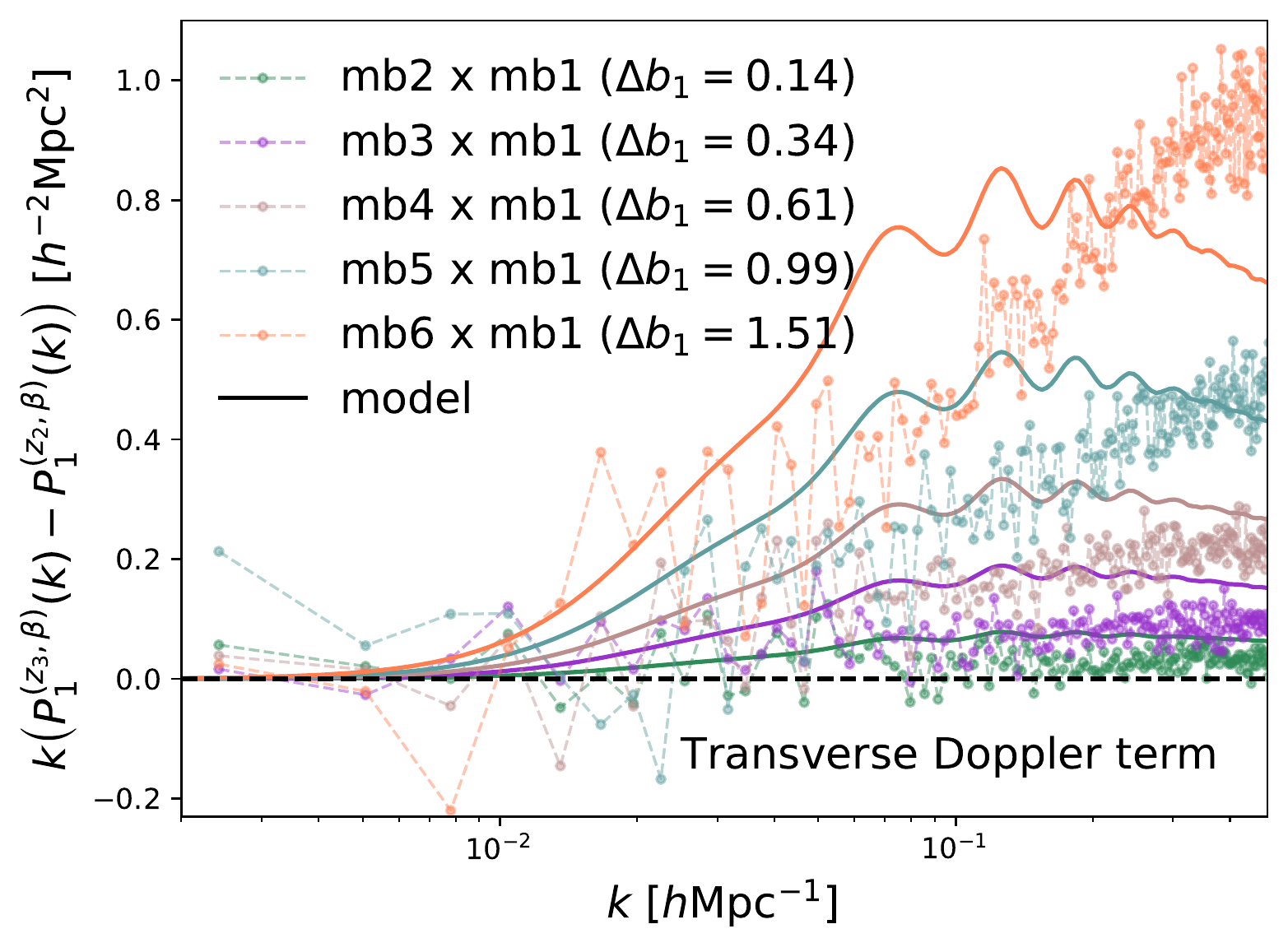}
\includegraphics[width=0.45\textwidth]{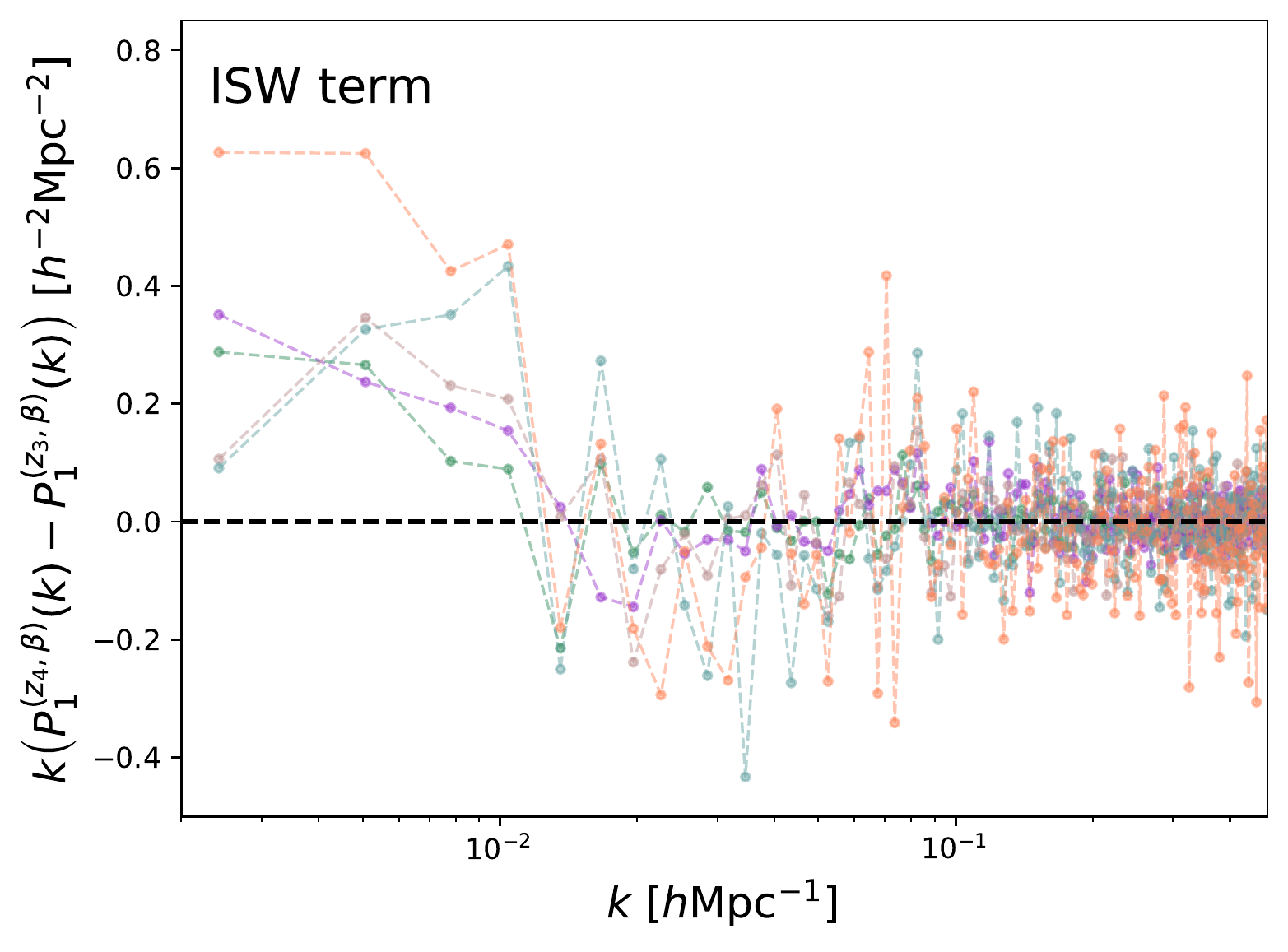}\\
\includegraphics[width=0.45\textwidth]{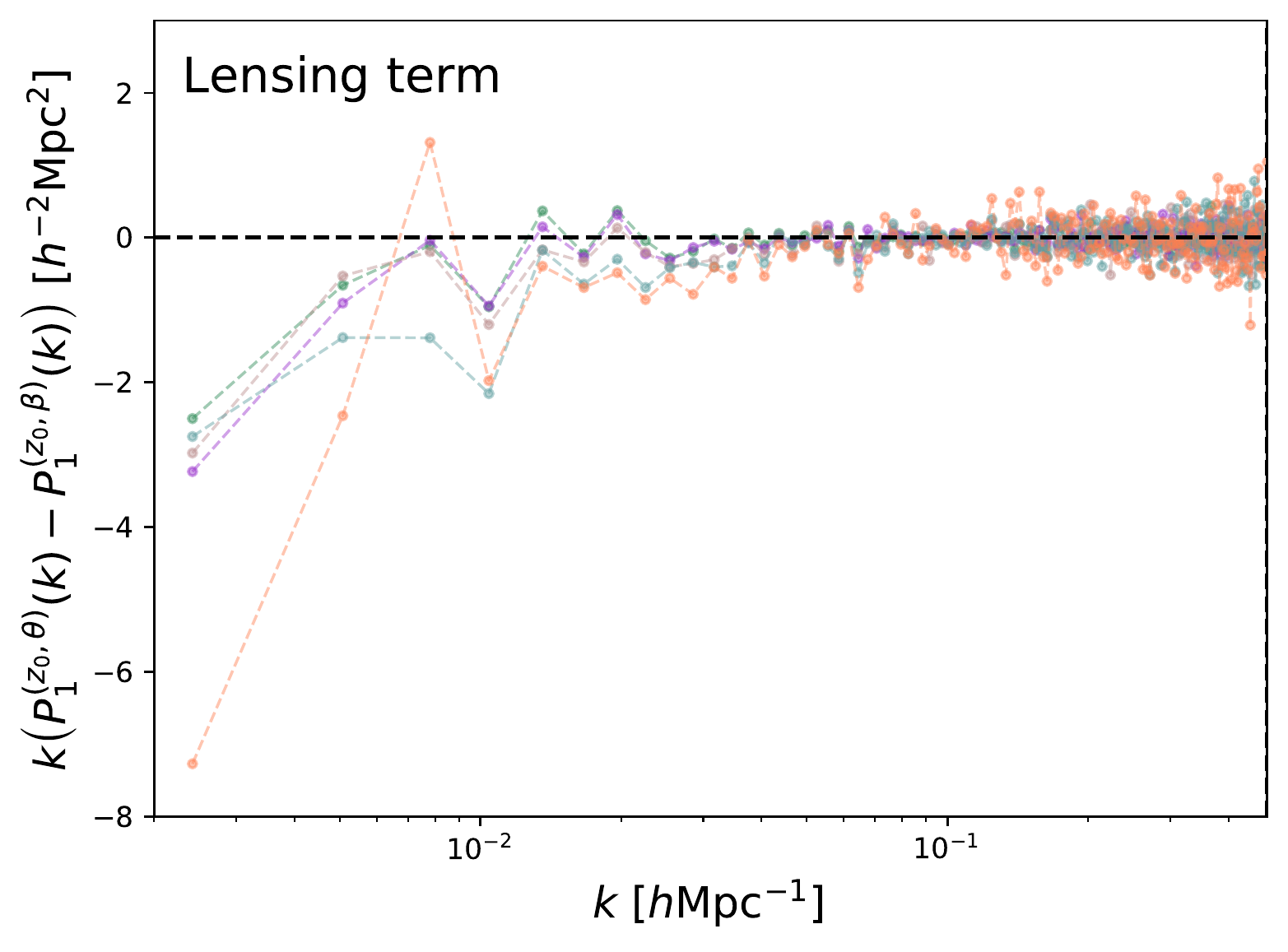}
\includegraphics[width=0.45\textwidth]{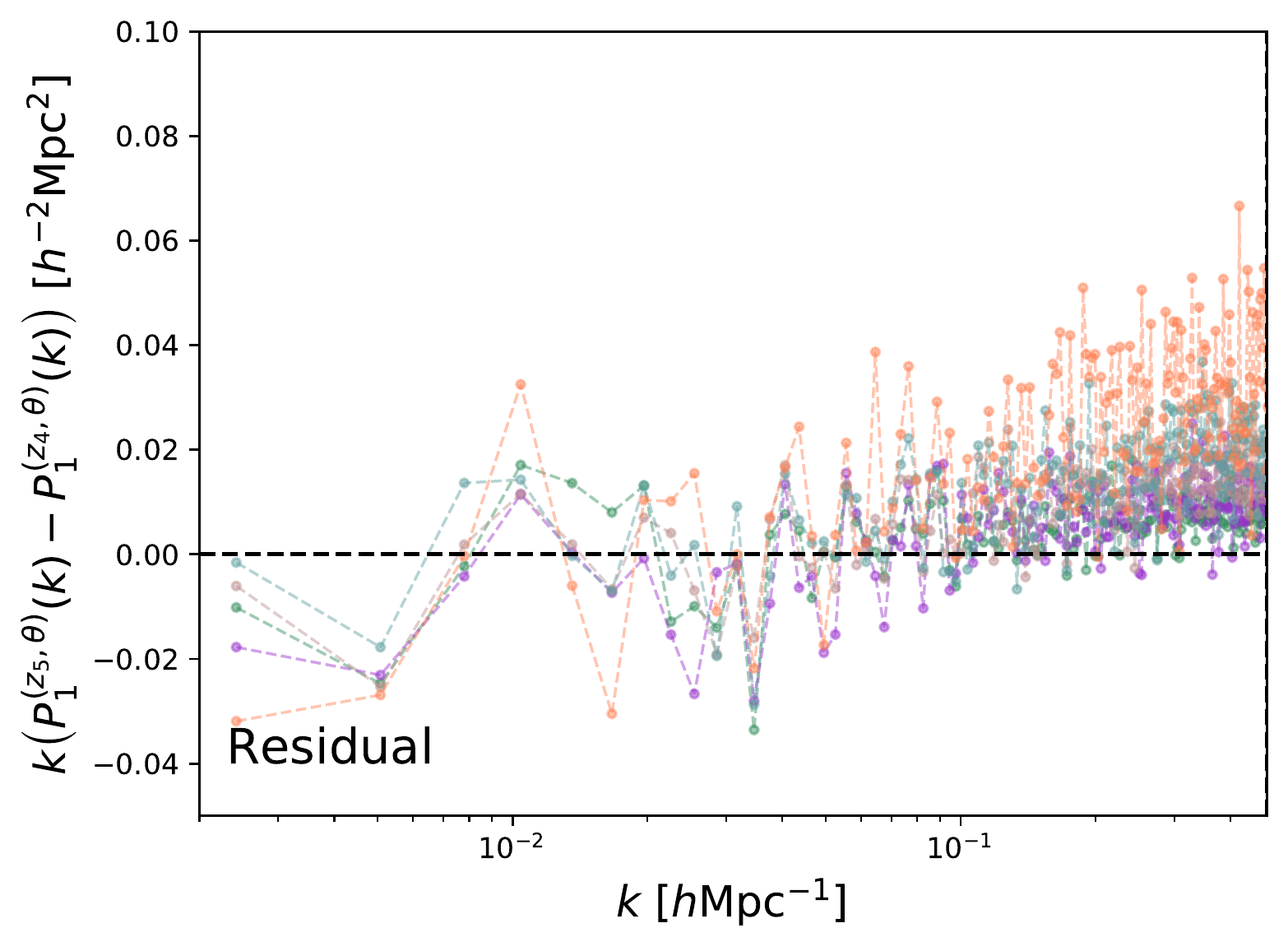}
\caption{Here we are plotting the differences between cross-power spectrum dipole measurements in the RayGalGroup simulation between different redshift definitions (see eqs.~\ref{eq:z0} - \ref{eq:z5}, dashed lines with markers). The measurement in the simulations is compared to the PT-based model we discussed in \sect{sec:theory} (solid lines). The colors correspond to different combinations of the halo mass-selected sub-samples. The plot on the top left shows the potential term, obtained by taking the differences between the cross-power spectrum dipoles measured using the redshift and angle definitions $(z_1,\beta)$ and $(z_0,\beta)$, respectively. The plot on the top right shows the Doppler term, which carries by far the largest signal in these simulations. The middle left plot shows the transverse Doppler term and the plot on the middle right shows the ISW contribution. The bottom row shows the lensing term using $(z_0,\theta) - (z_0,\beta)$ on the left and the residual obtained through $(z_5,\theta) - (z_4,\theta)$ on the right.
}
\label{fig:dipole_diff}
\end{figure}

Using the estimator discussed in \sect{sec:estimator} we now measure the cross-power spectrum dipole and plot the differences between these measurements in \fig{fig:dipole_diff}. These measurements, like the auto-power spectrum measurements in the previous section, have a Nyquist frequency of $k_{\rm Ny} = 0.96\kMpc$ and we will make use of these measurements up to half that frequency. We plot the differences between dipole measurements because (1) this isolates the individual (relativistic) contributions to the power spectrum dipole and (2) this efficiently removes sample variance and allows us much more precise measurements.

The top left plot of \fig{fig:dipole_diff} shows the difference of the power spectrum dipole using the redshift definitions $z_0$ and $z_1$ (see \eqnb{eq:z0} and \ref{eq:z1}) as well as the unlensed angular position defined by $\beta$, which isolates the relativistic contributions from the difference in the gravitational potential. The measurements are compared with the dipole model (solid lines) we developed in \sect{sec:theory}, specifically the second term in~\eq{eq:Doppler} as well as eqs.~(\ref{eq:second_order_pot1}) and (\ref{eq:second_order_pot2}).

The plot on the top right shows the Doppler term together with the model of \eq{eq:Doppler} (first term on the right-hand side). The significantly increased noise level for this measurement is caused by the inclusion of velocity fluctuations that do not cancel out in the difference of dipole measurements with redshift $z_2$ and $z_1$. The plot on the middle left shows the transverse Doppler term where the solid lines correspond to the 1-loop model of eqs.~(\ref{eq:second_order_TD1}) and (\ref{eq:second_order_TD2}). The ISW term on the middle right seems not to be detectable and we also cannot find a significant signal for the lensing term on the bottom left. The difference between $(z_5,\beta) - (z_4,\beta)$ is shown on the bottom right, which represents a measure for the accuracy of the redshift definitions in $z_1$ to $z_4$ and suggests that we can trust these measurements up to $\Delta P\gtrsim 0.05\,h^{-3}$Mpc$^3$, significantly below the noise level of all terms of interest (note that this does not test the validity of the weak field approximation).

\subsection{Asymmetry of the dipole estimator}
\label{sec:asym}

\begin{figure}[t]
\centering
\includegraphics[width=0.45\textwidth]{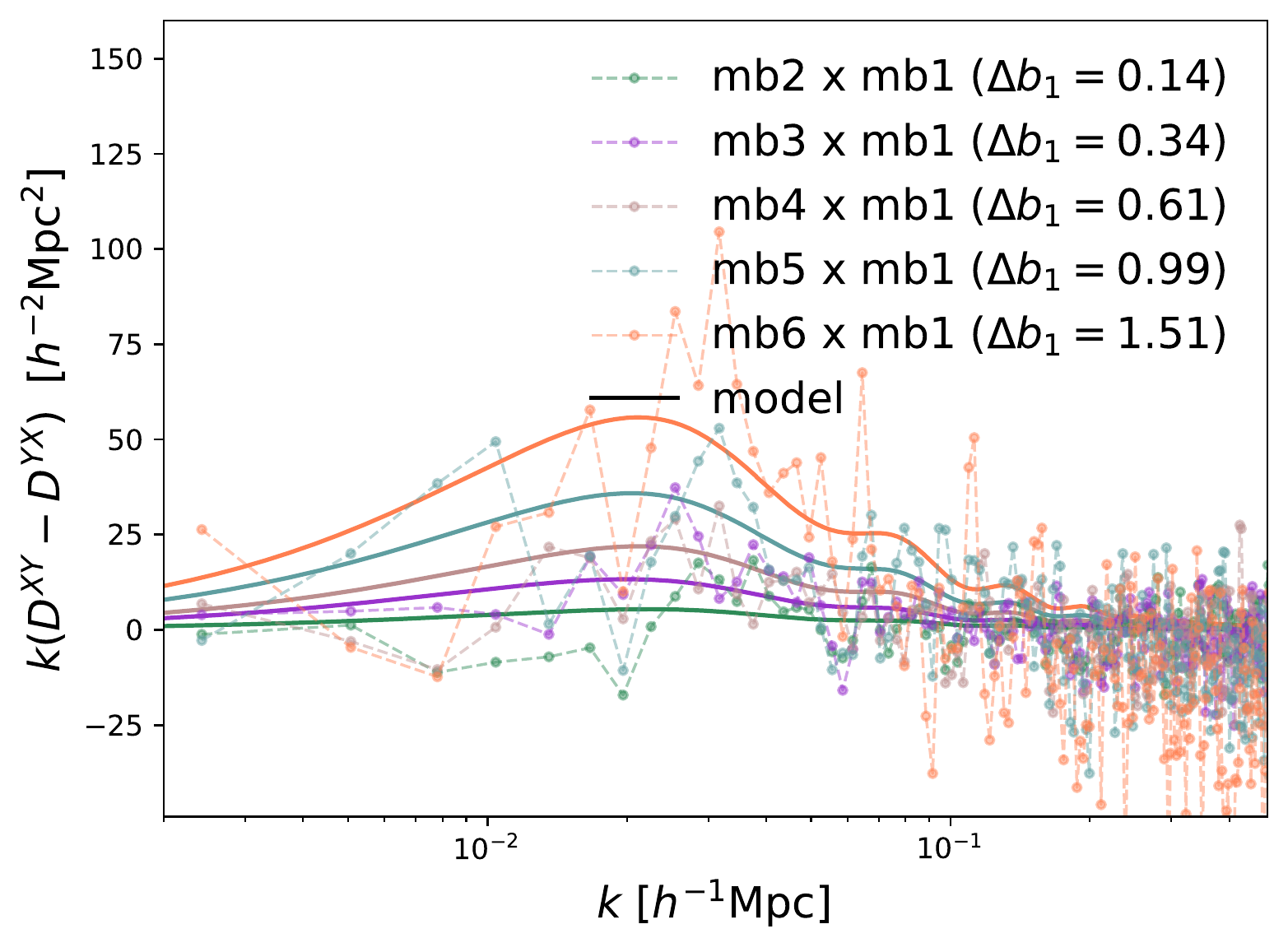}
\includegraphics[width=0.45\textwidth]{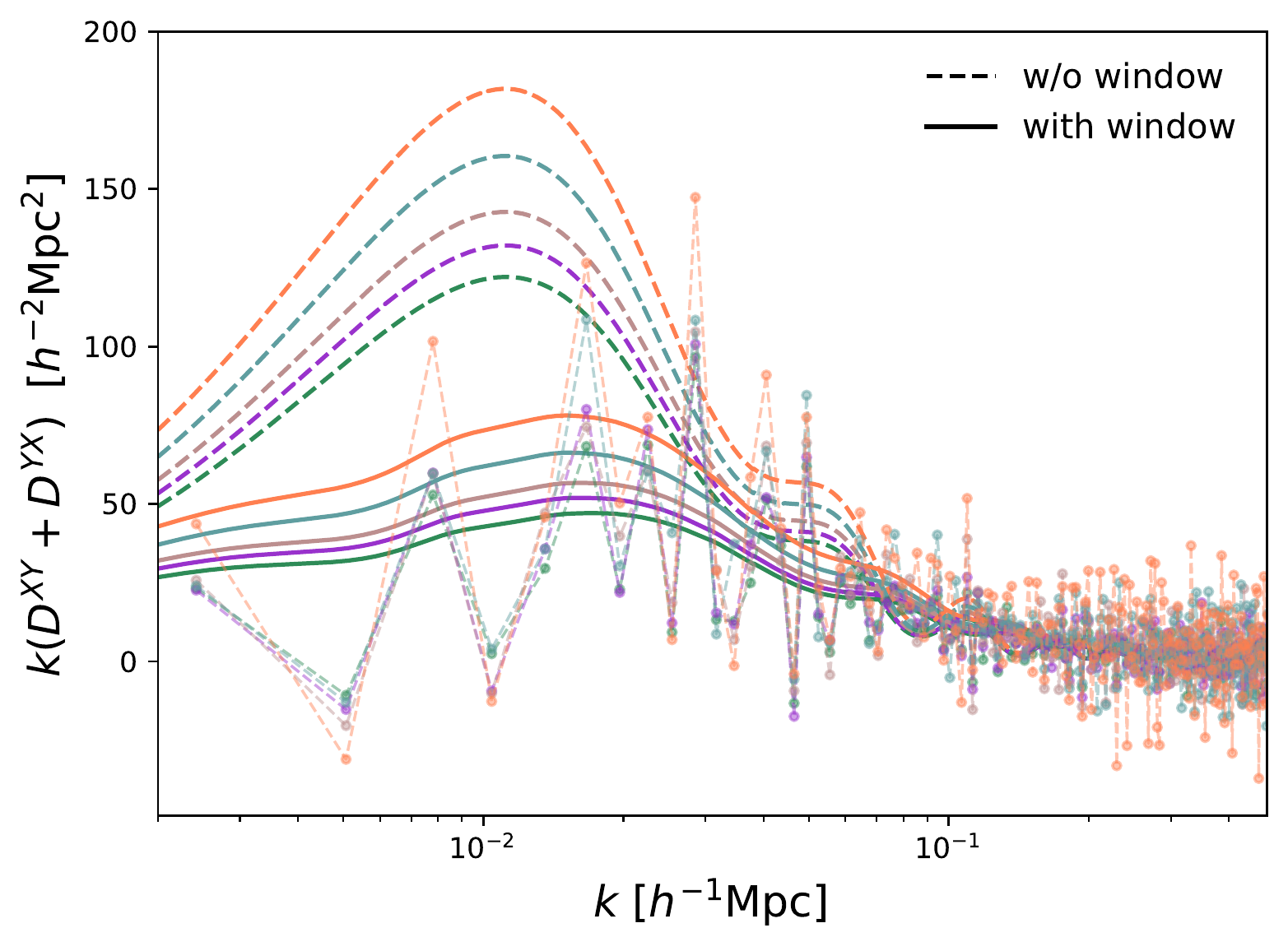}
\caption{These plots highlight the asymmetries in the estimator used in our analysis and defined in \eq{eq:pkestimator} meaning $P_1^{XY}$ vs. $P_1^{YX}$. The details are discussed in \sect{sec:asym}. The asymmetric terms only show up in $z_2$ and hence the y-axis quantities are combinations of dipole power spectrum differences e.g. $D^{XY} = P^{XY,(z2,\beta)}_1 - P^{XY,(z1,\beta)}_1$. The left plot shows the sum of these differences using the two possible cross-correlations (see \eqnb{eq:asym2}). This quantity has twice the relativistic terms, no window function contribution, wide-angle contributions proportional to the bias differences ($b_1^Y - b_1^X$) and contributions from the evolution term. The plot on the right shows the sum defined in \eq{eq:asym1}, which contains only the wide-angle, evolution and window function contributions, with all relativistic terms canceling out. The impact of the window function is shown by the  difference between the dashed and solid lines.
}
\label{fig:asym}
\end{figure}

Here we note that the estimator as defined in \sect{sec:estimator} is not symmetric in all dipole contributions, meaning $P^{XY}_1 \neq P^{YX}_1$. Eq.~(\ref{eq:wadipole}) demonstrates this asymmetry for the wide-angle term, which has $P^{XY}_1(k) \propto b_1^Y/r_Y$ and $P^{YX}_1(k) \propto b_1^X/r_X$ for the two possible cross-correlations. A similar asymmetry exists in the evolution terms given in eqs.~(\ref{eq:evo1}) and (\ref{eq:evo2}). The relativistic terms on the other hand are anti-symmetric, meaning they change signs in $P^{XY}_1$ and $P^{YX}_1$, while the window function contributions and lightcone effects are symmetric. 

This opens the interesting option to isolate the individual dipole contributions. The quantity
\begin{equation}
    D^{XY} + D^{YX} = \left(P^{XY,(z2,\beta)}_1 - P^{XY,(z1,\beta)}_1\right) + \left(P^{YX,(z2,\beta)}_1 - P^{YX,(z1,\beta)}_1\right)
    \label{eq:asym1}
\end{equation}
only contains the wide-angle, evolution and window function contributions, while all relativistic terms cancel out. On the other hand 
\begin{equation}
    D^{XY} - D^{YX} = \left(P^{XY,(z2,\beta)}_1 - P^{XY,(z1,\beta)}_1\right) - \left(P^{YX,(z2,\beta)}_1 - P^{YX,(z1,\beta)}_1\right)
    \label{eq:asym2}
\end{equation}
has twice the relativistic Doppler and no window function contribution, while the wide-angle term is proportional to the bias difference ($b^Y_1 - b_1^X$). Like the wide-angle term, the bias dependence of the evolution term is non-symmetric with the details shown in \eq{eq:evo}. Both of these quantities are plotted in \fig{fig:asym}.

\section{Discussion}
\label{sec:discussion}

The Doppler term in \fig{fig:dipole_diff} (top right) is measured as the difference between the cross-power spectrum dipole of ($z_1$,$\beta$) and ($z_2$,$\beta$), where the redshift definitions are given in eqs.~(\ref{eq:z1}) and (\ref{eq:z2}) and $\beta$ describes the angular distribution excluding any lensing effects. The Doppler term has contributions from the evolution bias, wide-angle effects, lightcone effects, window function effects, and the relativistic Doppler term. As shown in \fig{fig:doppler_details} the window function is the dominant contribution in our case, but depends on the survey geometry and might look very different for realistic galaxy surveys with a changing number density. The wide-angle term dominates over the relativistic Doppler term for samples with smaller $\Delta b_1$, highlighting the importance of this correction. The evolution bias and lightcone effects are usually sub-dominant. Our model for the Doppler term is limited to linear theory for all $4$ contributions but is consistent with the measurements in the RayGalGroup simulation.

The potential term in \fig{fig:dipole_diff} (top left) is measured as the difference between the cross-power spectrum dipole of ($z_0$,$\beta$) and ($z_1$,$\beta$). On the largest scales, this signal is almost one order of magnitude smaller than the Doppler term, while it becomes comparable on small scales ($k\gtrsim0.1\kMpc$). For the potential terms, we include 1-loop corrections and our model is consistent with the measurement up to the largest scales included in this comparison ($k_{\rm max}=0.48\kMpc$) without any notable deviations for any of the mass bins. Linear theory (dashed lines), is only valid on the largest scales, clearly demonstrating the necessity for 1-loop corrections when modeling this term.

The transverse Doppler term in \fig{fig:dipole_diff} (middle left) is measured as the difference between the cross-power spectrum dipole of ($z_2$,$\beta$) and ($z_3$,$\beta$). This signal is almost another magnitude smaller than the potential term. The leading contribution for this term comes from 1-loop, but our model does show some differences compared to the measurements in the RayGalGroup simulation. The exact source of this deviation is unclear but it leads us to conclude that our modeling of the power spectrum dipole is limited to signals with $\Delta P_1 \gtrsim 1h^{-3}$Mpc$^3$.

The ISW and lensing terms in \fig{fig:dipole_diff} (middle right and bottom left), are more complicated to model since they have contributions to the even multipoles even within the weak field approximation. This means that in Fourier-space we would expect additional window function contributions from the even multipoles (just as we saw for the Doppler term). A self-consistent model for these signals would, therefore, require detailed modeling of all multipoles, not just the dipole. We do not implement such an analysis framework since our measurements do not indicate any detection of the ISW or lensing effect. Galaxy surveys at higher redshift might have significantly enhanced ISW and lensing signals, which might make it necessary to include models for these signals to be able to fully exploit a dipole measurement.

Finally we note that because of the small value of $\Omega_m$ in the RayGalGroup simulation, the relativistic Doppler and potential terms are generally expected to be larger in the real Universe. Planck measured $\Omega_m=0.3158$~\citep{Collaboration]2018:1807.06209v1}, which is $\sim20\%$ larger compared to the RayGalGroup simulation with $\Omega_m=0.25733$. 

\subsection{Comparison to Breton et al.~\citep{Breton2018:1803.04294v2}}

Here we will compare our findings with \citeref{Breton2018:1803.04294v2}, which studied the same simulation in configuration-space using linear perturbation theory to model the dipole.

For the potential term, \citeref{Breton2018:1803.04294v2} found that linear theory works on large scales, but fails below $30 - 60\,h^{-1}$Mpc. We find a very similar result, showing that the linear model is only consistent with the simulation on the largest scales ($k > 0.01\kMpc$). However, when including 1-loop corrections, we find excellent agreement with the power spectrum dipole measured in the simulation up to the maximum scale of our analysis ($k_{\rm max}=0.48\kMpc$). This success relies on one additional (EFT motivated) nuisance parameter, $\sigma^2_0$, which we constrain with the simulation itself as shown in eqs.~(\ref{eq:EFT_par}) and (\ref{eq:EFTpara}).

Another finding of \citeref{Breton2018:1803.04294v2} was that the residual term reaches a similar size as the potential term on small scales, suggesting that cross-terms and non-linearities of the mapping are important. In our case, the residual term is an order of magnitude below the potential term even on the largest scales. The most likely explanation is that at $k_{\rm max}=0.48\kMpc$ we are not probing the small scales $s \lesssim 5\Mpc$ at which \citeref{Breton2018:1803.04294v2} made this observation.

\section{Forecasts}
\label{sec:forecast}

From \fig{fig:dipole_diff} we can see that for a survey with the characteristics of the RayGalGroup simulation we should expect the Doppler term to dominate on large scales, while on small scales ($k\gtrsim0.1\kMpc$) the potential term can become comparable. The details, however, do depend on the redshift and densities of the galaxy samples and their $\Delta b_1$. In particular, the relative signal strength shown in \fig{fig:dipole_diff} does not apply to high redshift samples like the extended Baryon Oscillation Spectroscopic Sample (eBOSS,~\citep{Ahumada2019:1912.02905v1}) since for such samples the peculiar velocities will be smaller, reducing the signal in the Doppler term, while the larger distances will increase the integrated signals like lensing.

Here we want to focus on the future Dark Energy Spectroscopic Instrument (DESI, ~\citep{DESI2016:1611.00036v2}) dataset and in particular, the DESI-BGS sample, which will cover the low redshift ($z < 0.5$) Universe over $14\,000\deg^2$. We will first discuss the analytic dipole covariance matrix, before discussing the DESI forecasts in detail.

\subsection{Analytic covariance of the dipole}
\label{sec:dipole_cov}

To simplify the analysis we derive the covariance matrix in the flat-sky approximation, which neglects wide-angle effects. Here we are interested in the detectability of relativistic effects in upcoming surveys, while additional signals such as wide-angle effects can be included in the model or avoided by using more advanced power spectrum estimators. In the flat-sky approximation we can define the dipole estimator as
\be
P_1^{XY} \left( k \right) = \frac{3}{2 V} \int \frac{d\Omega_{\hat \bk}}{2 \pi} \mathcal{L}_1\left( \mu \right)\Delta^X \left( \bk \right) \Delta^Y \left(- \bk \right)\, ,
\ee
where $V$ is the survey volume.
We can, therefore, compute the covariance of the dipole estimator through
\begin{align}
    \left\langle P_1^{XY} \left( k \right) \left( P_1^{XY} \left( q \right) \right)^* \right\rangle_c =\; &\frac{9}{4V^2} \int \frac{d\Omega_{\hat \bk}}{2\pi} \frac{d\Omega_{\hat \bq}}{2\pi} \mu_k \mu_q \langle \Delta^X \left( \bk \right) \Delta^Y \left( - \bk \right) \Delta^X \left( - \bq \right) \Delta^Y \left( \bq \right) \rangle \label{eq:cov} \notag \\
    \begin{split}
        =\; &\frac{9}{4} \frac{\left( 2 \pi \right)^6}{V^2}\delta_D^{(3)} \left( 0 \right) \frac{\delta_D \left( k - q \right)}{k^2}\\
        &\times \int \frac{d \Omega_{\hat \bk}}{\left( 2 \pi \right)^2} \mu^2 \left[P^{XX}\left( \bk \right)  P^{YY} \left( - \bk \right) - P^{XY}\left( \bk \right)  P^{YX} \left( - \bk \right)  \right] \notag
    \end{split}\\
    \begin{split}
        =\; &\frac{9}{4}  \frac{\left( 2 \pi \right)^6}{V^2} \frac{\delta_D^{(3)} \left( 0 \right)}{2\pi} \frac{\delta_D \left( k - q \right)}{k^2} \sum_{\ell_1 \ell_2}
        \left[ P_{\ell_1}^{XX} \left( k \right) P_{\ell_2}^{YY} \left( k \right) \right. \\ 
        &-\left. P_{\ell_1}^{XY} \left( k \right) P_{\ell_2}^{YX}(k) \right]
        \int d\mu\, \mu^2 \mathcal{L}_{\ell_1} \left( \mu \right)  \mathcal{L}_{\ell_2} \left( -\mu \right) \, .
    \end{split}
\end{align}
The integral over the Legendre polynomials is non-zero only for $\ell_2 = \left\{\ell_1-2, \ell_1, \ell_1+2\right\}$. Hence, odd and even multipoles of the power spectrum are not combined in the variance. Since relativistic effects do not source even multipoles of the power spectrum (within the weak-field approximation), we do expect that the signal-to-noise of the dipole is not suppressed by the large sample variance of the standard Newtonian terms of the even multipoles. At tree-level, we can show the contributions of the even multipoles in \eq{eq:cov}. Beyond linear order we have a power spectrum of the form
\begin{equation}
    \begin{split}
        P^{XY}\left( k \right) =\; &\left( b_X + \left[ T^{(3)}_{21} f + T^{(2)}_{21} f b_X + \left( T_{21} f b_X^2+ T_{2 2} f^2 b_X \right)\right] \mu^2 \right. \\
        &+
        \left. 
        \left[T^{(2)}_{42} f^2+ T_{42} f^2 b_X \right] \mu^4 + T_{63} f^3 \mu^6 \right) \\
        & \times 
        \left( b_Y + \left[ T^{(3)}_{21} f + T^{(2)}_{21} f b_Y + \left( T_{21} f b_Y^2+ T_{2 2} f^2 b_Y \right)\right] \mu^2
        \right. \\
        &+\left.
        \left[T^{(2)}_{42} f^2+ T_{42} f^2 b_Y \right] \mu^4
        + T_{63} f^3 \mu^6 \right)
        P\left( k \right)\, ,
    \end{split}
\end{equation}
where $T^{(i)}_{jk}$ are arbitrary~\footnote{They can be computed from \eq{eq:3or_newt}. However their amplitude is irrelevant in order to show that the even multipoles of the power spectrum do not contribute to the covariance of the dipole.} coefficients.
Given this general form of the power spectrum we can show 
\be
\sum_{\ell_1 \ell_2 \ {\rm even}}
\left[ P_{\ell_1}^{XX} \left( k \right) P_{\ell_2}^{YY} \left( k \right) 
-
 P_{\ell_1}^{XY} \left( k \right) P_{\ell_2}^{YX} \left( k \right) \right]
 \int d\mu \mu^2 \mathcal{L}_{\ell_1} \left( \mu \right)  \mathcal{L}_{\ell_2} \left( -\mu \right)
 =0 \, .
\ee
This leads to
\begin{align}
    \begin{split}
        \left\langle P_1^{XY} \left( k \right) \left( P_1^{XY} \left( q \right) \right)^* \right\rangle_c =\; & -
        \frac{9}{4} \frac{\left( 2 \pi \right)^6}{V^2} \frac{\delta_D^{(3)} \left( 0 \right)}{2\pi} \frac{\delta_D \left( k - q \right)}{k^2}\\
        &\times \sum_{\ell_1 \ell_2 \;{\rm odd}} \left[ P_{\ell_1}^{XX} \left( k \right) P_{\ell_2}^{YY} \left( k \right) - P_{\ell_1}^{XY} \left( k \right) P_{\ell_2}^{YX} \left( k \right) 
        \right] \\
        & \times \int d\mu\,\mu^2 \mathcal{L}_{\ell_1} \left( \mu \right)  \mathcal{L}_{\ell_2} \left( \mu \right) \notag
    \end{split}\\
    \begin{split}
        =\; &-\frac{\delta_D^{(3)} \left( 0 \right)}{2\pi}  \frac{\left( 2 \pi \right)^6}{V^2}\frac{\delta_D \left( k - q \right)}{k^2}\left( \frac{9}{10} \left[ P_1^{XY} \left( k \right)\right]^2 + \frac{18}{35} P_1^{XY}\left(k \right)P_3^{XY}\left(k \right)\right.\\
        &\left.+ \frac{23}{70} \left[ P_3^{XY} \left( k \right) \right]^2 +\frac{20}{77} P^{XY}_{3} \left( k \right) P^{XY}_5  \left( k \right)
        + \frac{59}{286} \left[ P_5^{XY} \left( k \right) \right]^2\right)\, ,
    \end{split}
\end{align}
where we have considered that the relativistic effects at one loop will source only $P_1 \left( k \right)$, $P_3 \left( k \right)$ and $P_5 \left( k \right)$. The expression for all odd  multipoles induced by relativistic effects are derived in appendix~\ref{app:odd_multipoles}.

So far we did not include the shot-noise contribution to the covariance. Assuming a standard Poisson noise, without any correlation between the different galaxy populations, we simply need to replace the monopole as follows
\be
P_0^{XX} \left( k \right) \rightarrow P_0^{XX} \left( k \right)+ \frac{1}{\bar{n}_X}
\qquad 
\text{and}
\qquad 
P_0^{YY} \left( k \right) \rightarrow P_0^{YY} \left( k \right)+ \frac{1}{\bar{n}_Y}\, ,
\ee
where $\bar{n}_X$ and $\bar{n}_Y$ are the densities of population $X$ and $Y$, respectively.
Therefore, shot-noise leads to the following contributions to the covariance
\begin{equation}
    \begin{split}
        \frac{3}{2} \frac{(2\pi)^6}{V^2} \frac{\delta_D^{(3)}(0)}{2\pi} \frac{\delta_D(k - q)}{k^2} 
        &\Bigg[ \frac{1}{\bar{n}_X} P_0^{YY}(k) + \frac{1}{\bar{n}_Y} P_0^{XX}(k) + \frac{1}{\bar{n}_X\bar{n}_Y} \\
        &+ \frac{2}{5} \left(  \frac{1}{\bar{n}_X} P_2^{YY}(k) + \frac{1}{\bar{n}_Y} P_2^{XX}(k) \right)  \Bigg] \, .
    \end{split}
\end{equation}
For a finite volume we can approximate the 3-dimensional Delta-Dirac distribution as $\delta_D^{(3)} \left( 0 \right) \simeq V/\left( 2 \pi \right)^3$. Hence, the covariance can be written as
\begin{equation}
    \begin{split}
        \left\langle P_1^{XY} \left( k \right) \left( P_1^{XY} \left( q \right) \right)^* \right\rangle_c =\; &\frac{\left( 2 \pi \right)^2}{V} \frac{\delta_D \left( k - q \right)}{k^2}\left( 
        -\frac{9}{10} \left[ P_1^{XY} \left( k \right)\right]^2 - \frac{18}{35} P_1^{XY}\left(k \right)P_3^{XY}\left(k \right)\right. \\
        &\left. - \frac{23}{70} \left[ P_3^{XY} \left( k \right) \right]^2 -\frac{20}{77} P^{XY}_{3} \left( k \right) P^{XY}_5  \left( k \right) - \frac{59}{286} \left[ P_5^{XY} \left( k \right) \right]^2 \right. \\
        &\left. + \frac{3}{2\bar{n}_X} P_0^{YY} \left( k \right) + \frac{3}{2\bar{n}_Y} P_0^{XX} \left( k \right) + \frac{3}{2\bar{n}_X\bar{n}_Y} \right.\\
        &\left. + \frac{3}{5\bar{n}_X} P_2^{YY}\left( k \right) +   \frac{3}{5\bar{n}_Y} P_2^{XX}\left( k \right)\right)     
        \\
        \equiv \;& 
        \frac{\left( 2 \pi \right)^2}{V} \frac{\delta_D \left( k - q \right)}{k^2} \sigma_{P_1}^2 \left( k \right) 
        \, .
    \end{split}\label{eq:variance_final}
\end{equation}
The covariance matrix derived here does not include window function effects, which as we discussed above can contribute to the dipole power spectrum and increase the variance. However, deconvolution of the power spectrum as well as estimators with reduced wide-angle effects can reduce such contributions~\citep{BeutlerMcDonald2020}. 

\subsection{Forecast for DESI}
\label{sec:desiforecast}

\begin{table}[t]
    \begin{center}
        \begin{tabular}{lll}
            z & $\bar{n}$ & $b_1$\\
            & [1/sq.deg./dz] & \\
            \hline
            0.05 & 1114.3 & 1.0 \\
            0.15 & 3694.1 & 1.1 \\
            0.25 & 4166.4 & 1.2 \\
            0.35 & 2865.6 & 1.5 \\
            0.45 & 1031.3 & 2.0 \\
            0.55 & 136.1 & 2.5
        \end{tabular}
        \caption{The density and linear bias distribution of the DESI-BGS sample taken from~\citeref{DESI2016:1611.00036v2} and used for our Fisher matrix forecasts.}
        \label{tab:DESI}
    \end{center}
\end{table}

Using the analytic covariance matrix developed in the last section we now employ it to investigate the possibility for a detection of relativistic effects with the future DESI experiment. Here we focus on the DESI-BGS sample, which provides the most promising avenue for such a detection. The density and bias distributions for the DESI-BGS sample are given in \tabl{tab:DESI}.

\begin{figure}[t]
\centering
\includegraphics[width=0.45\textwidth]{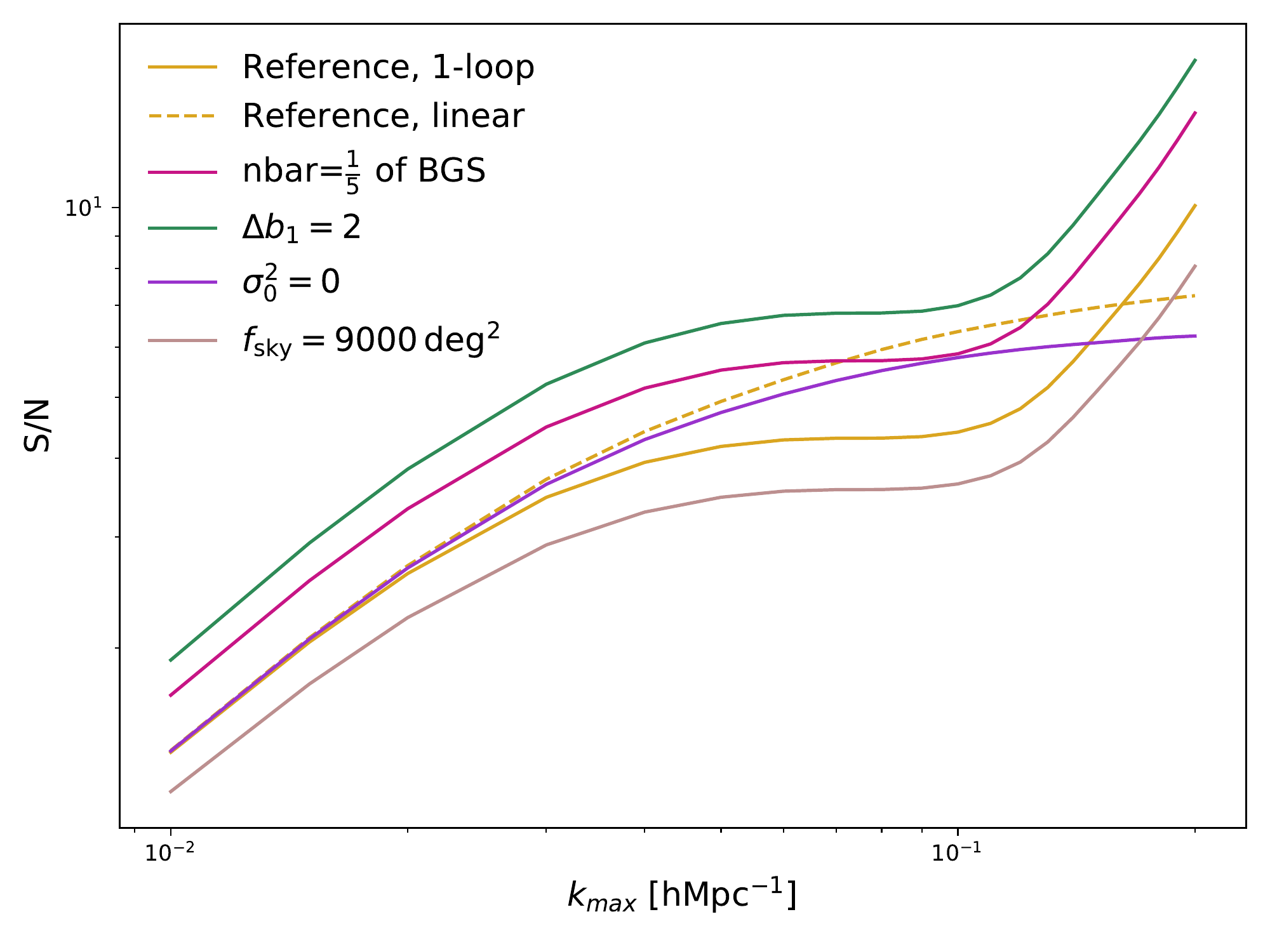}
\includegraphics[width=0.45\textwidth]{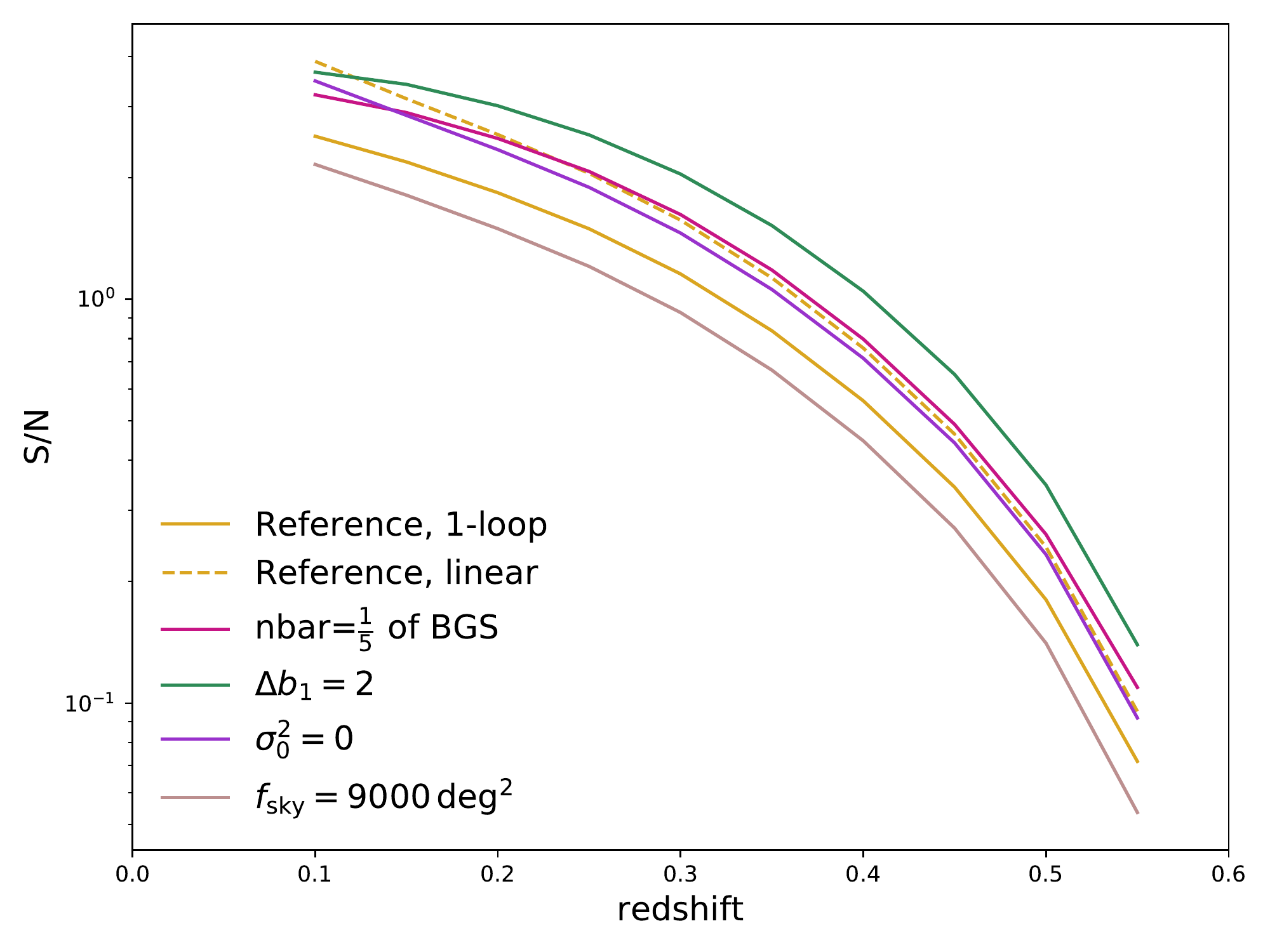}
\caption{The cumulative signal-to-noise for the cross-correlation of two DESI-BGS sub-samples. The solid yellow line shows the reference model with a density $1/10$ of the nominal BGS density and $\Delta b_1 = 1$ (see other assumptions listed in \sect{sec:desiforecast}). The plots also include a case with only linear contributions to the signal and variance (dashed yellow line), a case with $2$ times higher density for both sub-samples (solid red line), a case with $\Delta b_1 = 2$ (solid green line), a case where the EFT parameter is set to $\sigma_0^2 = 0$ (solid magenta line) and a case where the BGS sample only covers $9000\deg^2$ (solid brown line). Left: The cumulative signal-to-noise as a function of $k_{\rm max}$. Right: The same signal-to-noise calculation as a function of redshift, assuming $k_{\rm max} = 0.1\kMpc$.}
\label{fig:BGS_forecast}
\end{figure}

Using these density and bias distributions we can calculate the cumulative signal-to-noise ratio for the relativistic dipole as
\be
\left( \frac{S}{N} \right)^2 = \frac{1}{4 \pi^2} \sum^{z_{\rm bins}}_{i}V(z_i)\int_{k_{\rm min}}^{k_{\rm max}} dk\, k^2 \frac{ \left| P_1^{XY} \left( k,z_i \right) \right|^2 }{\sigma_{P_1}^2 \left( k ,z_i\right) }\, .
\label{eq:SN}
\ee
We start with a conservative reference model using the following assumptions:
\begin{enumerate}
    \item We can select two sub-samples from the BGS sample with $\Delta b_1 = 1$ over the entire BGS redshift range.
    \item The two sub-samples have $1/10$ of the density of the nominal BGS sample.
    \item The BGS sample will cover $14\,000\deg^2$ of the sky.
    \item The k-integration of \eq{eq:SN} uses $k_{\rm min} = 0.001\kMpc$.
    \item Zero evolution and magnification bias ($b_e=0$, $s_m=0$).
    \item The second order bias parameter is given  by the linear bias through \eq{eq:b2}.
    \item The EFT parameter $\sigma_0^2$ is set according to \eq{eq:EFTpara}, which provided the best description of the measurements in the RayGalGroup simulation.
    \item We use the covariance matrix developed in \sect{sec:dipole_cov} which does not account for window function contributions and neglects 1-loop corrections, while our dipole model does include such corrections
    \item We assume the redshift bins of $\Delta z = 0.05$ in \eq{eq:SN} to be independent.
\end{enumerate}
These assumptions lead to the signal-to-noise ratio as a function of $k_{\rm max}$ shown in \fig{fig:BGS_forecast} (left) where the dashed yellow line shows the signal-to-noise using a pure linear power spectrum dipole model, while the yellow solid line uses the 1-loop model discussed in \sect{sec:theory}. This model also contributes to the noise through the covariance calculation of \eq{eq:variance_final}. Comparing the dashed yellow and solid yellow lines shows that ignoring 1-loop corrections can lead to incorrect forecasts even at $k=0.1\kMpc$. This is consistent with our findings in \sect{sec:analysis} and \fig{fig:dipole_diff} where linear theory fails to describe the gravitational redshift on scales $k \gtrsim 0.1\kMpc$. Figure~\ref{fig:BGS_forecast} shows that the signal-to-noise ratio is flattening when approaching $k\simeq 0.1\kMpc$ before increasing again on smaller scales. This is caused by the zero crossing of the gravitational redshift signal driven by the 1-loop corrections.

At $k_{\rm max}=0.1\kMpc$ our reference setup based on the assumptions listed above leads to a signal-to-noise of $4.4\sigma$, which increases to $6.3\sigma$ when pushing to $k_{\rm max} = 0.2\kMpc$. The distribution of the signal-to-noise ratio as a function of redshift is shown on the right hand side of \fig{fig:BGS_forecast} using $\Delta z = 0.05$ and $k_{\rm max} = 0.1\kMpc$. We can see that most of the signal is located in the first few redshift bins.

Now we will discuss several variations of the reference model discussed above. First we vary the density of the sample. Assuming we can divide the BGS sample into two sub-samples with $1/5$ of the nominal BGS density, the signal-to-noise increases to $5.9\sigma$ at $k_{\rm max}=0.1\kMpc$ (solid red line in \fig{fig:BGS_forecast}). If rather than increasing the density we could get a larger dipole signal by increasing the bias difference to $\Delta b_1=2$, the signal-to-noise increases to $7\sigma$ at $k_{\rm max}=0.1\kMpc$ (solid green line in \fig{fig:BGS_forecast}) and $10.1\sigma$ at $k_{\rm max}=0.2\kMpc$. The EFT parameter $\sigma_0^2$ contributes to the signal-to-noise at high $k$. Setting it to $\sigma^2_0=0$ rather than the default $\sigma_0^2$ given by \eq{eq:EFTpara}, increases the signal-to-noise from $4.4\sigma$ to $5.8\sigma$ at $k_{\rm max} = 0.1\kMpc$. For higher $k$, a larger EFT parameter leads to a larger signal-to-noise as shown by the solid magenta line in \fig{fig:BGS_forecast}. We note, however, that our forecasts include the even multipole contributions to the covariance only at linear order, therefore we might underestimate the covariance at small scales when linear even multipoles do not agree with simulations (see figure~\ref{fig:linear}). 

While we consider our reference forecasts outlined in the $9$ points above as conservative, we also want to investigate two cases, which we consider our worst-case scenarios. First, we consider a smaller sky coverage of the BGS sample. The sky coverage of the BGS depends on the redshift efficiency of the DESI instrument during gray time~\footnote{The BGS sample will take the twilight observing timeslots.}, which is still uncertain.  Reducing the nominal BGS sky coverage from $14\,000\deg^2$ to $9000\deg^2$ leads to a reduction of the signal-to-noise from $4.4\sigma$ to $3.5\sigma$ (solid brown line in \fig{fig:BGS_forecast}). A second case considers the largest scales included in the analysis. The Doppler signal is located on very small $k$ with a peak around $k=0.01\kMpc$. The default $k_{\rm min} = 0.001\kMpc$ might be difficult to achieve due to the small volume of the sample. However, using $k_{\rm min} = 0.01\kMpc$ instead only reduces the signal-to-noise from $4.4\sigma$ to $4.2\sigma$, since sample variance limits the large scale contributions to the total signal-to-noise. The fact that our analysis does not rely on extremely large scales, does make it less sensitive to observational systematics. It also implies that assumption 9 listed above (independent redshift bins) should have a small impact. Indeed, repeating the analysis with $\Delta z = 0.1$ only reduces the signal-to-noise from $4.4\sigma$ to $4.2\sigma$~\footnote{Here we expect the redshift correlation as well as the signal evolution within the redshift bin to contribute to the reduction of the signal-to-noise.}.

We also note that our assumption of two sub-samples with constant $\Delta b_1$ over the entire redshift range is unrealistic since BGS is magnitude limited, which means the high redshift end will only contain bright, highly biased galaxies. However, as shown in the left plot of \fig{fig:BGS_forecast}, most of the signal-to-noise is located at low redshift, where a separation into two samples with significantly different bias should be possible. 

We also investigated the cross-correlations between the other DESI samples (LRGs, ELGs, and QSOs), but found that the lower expected density of these samples significantly reduces the signal-to-noise. The cross-correlation of the ELG and LRG samples only yields a $1.7\sigma$ detection at $k_{\rm max} = 0.2\kMpc$. Similarly, cross-correlation of the BGS and ELG sample can yield a signal-to-noise of $1.6\sigma$. We therefore conclude that sub-samples of BGS provide the most promising case for a first detection of the relativistic signal in galaxy clustering.

\section{Conclusion}
\label{sec:conclusion}

In this paper, we measure the auto and cross-power spectrum multipoles for different halo mass bins in the RayGalGroup simulation. We probe haloes from $1.88\times10^{12}M_{\odot}$ (lower limit of mb1) to $1.20\times10^{14}M_{\odot}$ (upper limit of mb6) at an effective redshift of $z_{\rm eff} = 0.341$. We model relativistic corrections as well as wide-angle, evolution, window and lightcone effects. Our model includes all relativistic corrections up to third-order including third-order bias expansion. We consider all terms which depend linearly on $\mathcal{H}/k$ (weak field approximation). While we restrict all terms proportional to $v_{\parallel}$ (Doppler term) to linear theory, 
we include 1-loop corrections to the matter power spectrum for the potential term and transverse Doppler effect~\citep{DiDio:2018zmk,DiDio:2020}. Our main results are:
\begin{enumerate}
    \item[(1)] We, for the first time, compare perturbation theory-based models with ray-tracing based simulations in Fourier-space using the power spectrum multipoles. The Fourier-basis is the natural choice for such a comparison, since the dominant signal peaks on large scales, where modes in Fourier-space are uncorrelated. 
    \item[(2)] When using the end-point line-of-sight definition, as required by many FFT-based estimators, the wide-angle effects in the power spectrum dipole are of the same order as the relativistic Doppler term and hence need to be included in the modeling.
    \item[(3)] We show that the standard cross-power spectrum estimator is asymmetric in the wide-angle and evolution terms, which allows us to isolate their contributions to the power spectrum dipole.
    \item[(4)] We, for the first time, compare the measurements of the transverse Doppler effect with a 1-loop model. While our model does predict the right order of magnitude of the effect, it fails to capture its scale dependence accurately.
    \item[(5)] We demonstrate that PT-based models including 1-loop corrections can model the potential term (gravitational redshift) up to the largest scales included in our comparison ($k = 0.48\kMpc$) without indication of a breakdown of the model for any of the halo mass combinations. Our model includes one additional EFT-inspired free fitting parameter given in \eq{eq:EFTpara}.
    \item[(6)] We estimate the evolution bias contributions for the different halo mass bins in the RayGalGroup simulation and their contributions to the power spectrum dipole. Similar estimates with real galaxy survey datasets are difficult since usually, these datasets are not complete in halo mass. However, we found these contributions to be sub-dominant for all halo mass bins investigated in our analysis.
    \item[(7)] We forecast that the DESI-BGS sample will be able to detect relativistic effects in the galaxy power spectrum dipole with $4.4\sigma$ significance if we can get two sub-samples with $\Delta b_1=1$ and $1/10$ of the density of the nominal BGS sample, which we consider a conservative choice. This detection significance can reduce to $3.5\sigma$ if DESI-BGS will only cover $9000\deg^2$ rather than the nominal $14\,000\deg^2$. Conversely it can increase to $10\sigma$ for a more optimistic case where we can achieve twice as high a signal with $\Delta b_1 = 2$ and can leverage the dipole measurement up to $k_{\rm max} = 0.2\kMpc$.
\end{enumerate}
In this paper, we demonstrated that we can model the relativistic and non-relativistic contributions to the power spectrum dipole up to very small scales ($k = 0.48\kMpc$). In general, we do not need to know how to model the even multipoles when extracting the relativistic signal, except for the couplings of the multipoles due to the window function. This, however, does not impose significant limitations, since (1) enough of the signal is located on large scales $k < 0.1\kMpc$, where we have good models for the even multipoles and (2) one can in principle apply deconvolution to remove the window function contributions to the dipole~\citep{BeutlerMcDonald2020}.

Most relativistic effects vanish in the auto-correlation analysis. However, future galaxy survey experiments will rely much more on cross-correlations, in which case relativistic effects can matter. Most relativistic effects are still limited to the odd multipoles, but due to the coupling with the window function, it might become essential to model relativistic effects accurately even when extracting signals from the even multipoles. This is of particular relevance for signals located on large scales like primordial non-Gaussianity.

Our forecasts predict that relativistic effects should be detectable with the next generation of galaxy redshift surveys. Detecting these effects in the galaxy distribution allows new tests of gravity~\citep{Bonvin2018:1803.02771v2} on the largest scales, providing an interesting additional science case for galaxy survey experiments.

\acknowledgments

We would like to thank the authors of~\citeref{Breton2018:1803.04294v2} for making their simulations publicly available and in particular to Michel-Andres Breton for helpful discussions. We thank Obinna Umeh, Michel-Andres Breton, Morag Scrimgeour and Emanuele Castorina for comments on a draft of this manuscript and Patrick McDonald for helpful discussions. FB is a Royal Society University Research Fellow.
ED (No.~171494 and~171506) acknowledges financial support from the Swiss National Science Foundation.

\bibliographystyle{JHEP}
\bibliography{GRpk}{}

\newpage

\appendix

\section{Number counts}
\label{app:single_number_counts}
 
Here we summarize the galaxy number counts induced by the different contributions to the redshift perturbation up to third order in perturbation theory. These are derived in detail in \citeref{DiDio:2020}.
We consider the following redshift perturbations
 \bea
 \delta z_1 &=& - \left( 1 +z \right) \Psi \, , \\
\delta z_2 &=&- \left( 1 +z \right)\left(  \Psi + \ndv \right) \, , \label{eq:Doppler_pert}\\
\delta z_3 &=&- \left( 1 +z \right)\left(  \Psi + \ndv - \frac{v^2}{2} \right) \, .
 \label{eq:td_pert} \eea
 The related number counts read as follow.
 Considering only the gravitational redshift
 \bea
  \label{eq:Delta_grav_1}
\Delta_1^{(1)}\left( \bn , z\right) &=& \delta_g + \HH^{-1} \partial_r \Psi
\, ,\\
\Delta_1^{(2)}\left( \bn , z\right) &=& \Delta_1^{(1\rightarrow 2)} +  \HH^{-1} \partial_r \left( \delta_g \Psi \right)
 \label{eq:Delta_grav_2}
\, ,\\
\Delta_1^{(3)}\left( \bn , z\right) &=& \Delta_1^{(2\rightarrow 3)}
\, .
 \label{eq:Delta_grav_3}
\eea
Including the linear Doppler term in \eq{eq:Doppler_pert} we get
\begin{align}
    \Delta_2^{(1)} \left( \bn , z \right) =& \; \delta_g + \HH^{-1} \partial_r \ndv 
    + \HH^{-1} \partial_r \Psi - \HH^{-1}\dot \ndv + \ndv \tilde{\mathcal{R}}\, , \label{eq:Delta_dop_1} \\
    \begin{split}
        \Delta_2^{(2)} \left( \bn , z \right) =&\; \Delta_2^{( 1 \rightarrow 2)} + \HH^{-1} \partial_r \left( \delta_g \left( \Psi + \ndv \right) \right) - \partial_t \left( \delta_g \ndv \right) + \delta_g  \ndv  \tilde{\mathcal{R}} \\
        &+ \HH^{-1} \ndv \partial_r \ndv \left(\tilde{\mathcal{R}} - 1\right)\, , \label{eq:Delta_dop_2}
    \end{split}\\
    \begin{split}
        \Delta_2^{(3)} \left( \bn , z \right) =&\; \Delta_2^{( 2 \rightarrow 3)} + \HH^{-1} \delta_g \ndv \partial_r \ndv \left( \tilde{\mathcal{R}} -1 \right) 
        + \HH^{-2} \partial_r \left(\ndv \Psi \partial_r \delta_g \right)
        - \HH^{-2} \partial_r \left( \ndv \dot \delta_g \right) \ndv \\
        &- \HH^{-2} \partial_r \delta_g \dot \ndv \ndv
        +\frac{ \HH^{-1}}{2} \ndv^2 \partial_r \delta_g  \left(b_{e } + 3 \tilde{\mathcal{R}} - \frac{2}{\HH r}  \right) \, ,\label{eq:Delta_dop_3}
    \end{split}
\end{align}
and finally including also the transverse Doppler effect of \eq{eq:td_pert} we get
\bea
  \label{eq:Delta_td_1}
\Delta_3^{(1)} \left( \bn , z \right) &=& \Delta_2^{(1)} 
\, , \\
  \label{eq:Delta_td_2}
\Delta_3^{(2)} \left( \bn , z \right) &=& \Delta_2^{(2)}  - \HH^{-1} \bv \cdot \partial_r \bv
\, , \\
  \label{eq:Delta_td_3}
\Delta_3^{(3)} \left( \bn , z \right) &=& \Delta_2^{(3)}  - \HH^{-1} \left[  \bv \cdot \partial_r \bv\right]^{(3)} - \frac{\HH^{-1}}{2} \partial_r \left( \delta_g v^2 \right) 
\, ,
\eea
where $v^2 = \left| \bv \cdot \bv \right|$ and
 \be
\tilde {\mathcal{R} }= -1 -b_{e} +\frac{\dot \HH}{\HH^2} + \frac{2}{\HH r} \, .
 \ee
To compare our theoretical model for the transverse Doppler effect at 1-loop with the  RayGalGroup simulations, we also need to consider the impact of the peculiar velocity of the observer, which for this catalog is set to $v_{\rm obs} = 0.00224794$ (with the $c=1$).
From eqs.~(\ref{eq:Delta_td_1} - \ref{eq:Delta_td_3}) we have
\be
 \Delta_3  \simeq \Delta_3^{(1)}+ \Delta_3^{(2)}+ \Delta_3^{(3)} = \Delta_2 - \frac{\HH^{-1}}{2} \partial_r v^2 - \frac{\HH^{-1}}{2} \partial_r \left( \delta_g v^2 \right) \, .
\ee
We can account for the observer peculiar velocity $v_{\rm obs}$ by replacing
 \be
 v^2 \rightarrow v^2 + v_{\rm obs}^2 - 2 \bv \cdot \bv_{\rm obs} \, ,
 \ee
obtaining
\begin{equation}
    \begin{split}
        \Delta_3 =\; &\Delta_2 - \frac{\HH^{-1}}{2} \partial_r v^2 + \HH^{-1} \bv_{\rm obs} \cdot \partial_r \bv - \frac{\HH^{-1}}{2} \partial_r \left( \delta_g v^2 \right)\\
        &- \frac{\HH^{-1}}{2} v_{\rm obs}^2 \partial_r  \delta_g + \HH^{-1} \bv_{\rm obs} \cdot \partial_r \left( \delta_g \bv \right) \, .
    \end{split}
\end{equation}
To compute the dipole of the power spectrum, we need to consider the following additional 2-point correlations induced by a non-vanishing observer velocity
 \bea
\HH^{-1} \bv_{\rm obs} \cdot \langle  \Delta_2^{\rm Newt} \partial_r \bv \rangle \, , \\
\HH^{-1} \bv_{\rm obs} \cdot \langle  \Delta_2^{\rm Newt} \partial_r \left( \delta_g \bv \right)\rangle 
 \, , \\
 \HH^{-1} v_{\rm obs}^2  \langle  \Delta_2^{\rm Newt} \partial_r  \delta_g \rangle 
 \, .
 \eea
The first two correlations have to vanish by isotropy since the 2-point function cannot point in a given direction. The contribution to the dipole induced by the third term is
\be
P_1^{\rm TD} \left( k \right)  \supset -  \frac{3i }{10 } \Delta b_1 \frac{k}{\HH} f v^2_{\rm obs} \, .
\ee

\section{Kernels}
\label{app:Kernels}
 
In this section, we summarize the kernels needed to express the 1-loop dipole shown in \sect{sec:Next-to-leading}.
To keep the notation simple we introduce the following variables:  $r=q/k$ and $x=\hat \bk \cdot \hat \bq$.
 
\subsection{Gravitational potential}
\label{app:Kernels_gravpot}
  
\begin{align}
    \begin{split}
        J_{22}^{\Delta b_1} \left( r, x \right) =\; &\frac{1}{{196 r^2 \left(r^2-2 r x + 1\right)^2}}\Big\{ 8 r^4 \left(10 x^2-17\right)+8 r^3 x \left(41-20 x^2\right) \\
        &+ r^2 \left(420 x^4-536 x^2+39\right)+42 r x \left(5-8 x^2\right)+63 x^2+28 \Big\}\, , 
    \end{split}\\
    J_{22}^{\Delta b_2} \left( r, x \right) =\;  &\frac{r \left(9-30 x^2\right)+21 x}{28 r \left(r^2-2 r x+1\right)}\, , \\
    J_{22}^{\Delta b_{12}} \left( r, x \right) =\; &\frac{3}{4} \left(\frac{1}{r^2-2 r x+1}+\frac{1}{r^2}\right)\, , \\
    J_{13}^{\Delta b_1} \left( r  \right) =\; &\frac{-15 (r^2-1)^3 (8 r^2+1) \log \left(\frac{\left| r-1\right| }{r+1}\right)-240 r^7+610 r^5-448 r^3+30 r}{1680 r^5} \, .
\end{align}

\subsection{Transverse Doppler}
\label{app:Kernels_TD}
  
\begin{align}
    \begin{split}
        T_{22}^{\Delta b_1} \left( r, x \right) =\; &-\frac{(r-x) \left(r \left(8 r^2-16 r x+42 x^2-13\right)-21 x\right)}{42 r^2 \left(r^2-2 r x+1\right)^2} \\
        &-f \frac{3 (r-x) \left(r \left(2 x^2-1\right)-x\right)}{10 r^2 \left(r^2-2 r x+1\right)^2}\, , 
    \end{split}\\
    T_{22}^{\Delta b_2} \left( r, x \right) =\; &\frac{r-x}{2 r^3-4 r^2 x+2 r}\, ,\\
    T_{13}^{\Delta b_1} \left( r \right) =\; &\frac{2 r \left(r^2-3\right) \left(3 r^2+1\right)-3 \left(r^2-1\right)^3 \log \left(\frac{r+1}{\left| 1-r\right| }\right)}{56 r^5} \, .
\end{align}

\section{Odd power spectrum multipoles}
\label{app:odd_multipoles}

To compute the covariance of the dipole induced by relativistic effects we need to derive all the non-vanishing odd multipoles at 1-loop in perturbation theory. 
Differently from the kernels derived in appendix~\ref{app:Kernels}, here we do not separate the terms induced by different physical origins. We only split the contributions proportional to the gravitational potentials from the ones induced by peculiar velocities, similarly to the approach adopted in \citeref{DiDio:2018zmk}. We remark that we have also assumed the Euler equation. 

\subsection{Dipole ($\ell=1$)}

\begin{align}
    \begin{split}
        P_1^{(13),\phi} \left( k \right) 
        =\; &
        i\Omega_m P(k)\frac{\mathcal{H}}{k}  \left\{\!-\left( \frac{3 \Delta b_{12}}{2}+\frac{1}{350} \Delta b_{1} f (15 f-133)+\frac{9 \Delta b_{12} f}{10} \right) k^2  \sigma_v^2\right. 
        \\
        &+ \left. \Delta b_1 \!\! \int\!\!\frac{d^3q}{(2\pi)^3} J^{1,\Delta b_1}_{13}\left(\frac{q}{k}\right)P(q) \!\right\} \, ,
    \end{split} \\
    P_1^{(22),\phi} \left( k \right) =\; &i\Omega_m \frac{\mathcal{H}}{k}  \int \frac{d^3 q}{\left( 2 \pi \right)^3} \left[ \sum_{\Delta b = \left\{\Delta b_1 ,\Delta b_2 ,\Delta b_{12}    \right\}}\!\!\!\!\!\!\! \Delta b J^{1,\Delta b}_{22} \left(\frac{q}{k}, \hat \bk \cdot \hat \bq \right) \right] P \left( q \right) P \left( \left| \bk - \bq  \right|\right) \, , \\
    \begin{split}
         P_1^{(13),v} \left( k \right) =\; &i \frac{\HH}{k} f \Delta b_1 P\left( k \right)  \left\{\left( \frac{2}{15} f^2 \left(1-\frac{3 (\HH+\dot\HH R)}{\HH^2 R}\right)+\frac{9 f \left(2 H^2 R-4 \HH-3 \dot\HH R\right)}{25 \HH^2 R}\right. \right. \\
        &- \left. \left. \frac{31 (2 \HH+\dot\HH R)}{105 \HH^2 R}\right) k^2 \sigma^2_v + \!\! \int\!\!\frac{d^3q}{(2\pi)^3} I^{1,\Delta b_1}_{13}\left(\frac{q}{k}\right)P(q)\right\} \, ,
    \end{split}\\
    P_1^{(22),v} \left( k \right) =\; &i \frac{\mathcal{H}}{k} f  \int \frac{d^3 q}{\left( 2 \pi \right)^3} \left[ \sum_{\Delta b = \left\{\Delta b_1 ,\Delta b_2 ,\Delta b_{12}\right\}}\!\!\!\!\!\!\! \Delta b I^{1,\Delta b}_{22} \left(\frac{q}{k}, \hat \bk \cdot \hat \bq \right) \right] P \left( q \right) P \left( \left| \bk - \bq  \right|\right) \, ,
 \end{align}
with
\begin{align}
    J_{13}^{1,\Delta b_1} \left( r \right) = &\;-\frac{f \left(15 \left(r^2-1\right)^3 \left(r^2+8\right) \log \left(\frac{r+1}{\left| r-1\right| }\right)-30 r^7-160 r^5+574 r^3+240 r\right)}{2800 r^5}\, ,\\
    \begin{split}
        I_{13}^{1,\Delta b_1} \left( r \right) = & \;\frac{1}{8400 \HH^2 r^5 R}\left\{5 (2 \HH+\dot \HH R) \left(30 \left(2 r^2+1\right) \left(r^2-1\right)^3 \log \left(\frac{(r+1)}{\left|r-1 \right|}\right)\right. \right. \\
        &+\left. \left.4 r \left(-30 r^6+65 r^4-26 r^2+15\right)\right)\right.\\
        &- \left. 6 f \left(15 \left(r^2-1\right)^3 \left(r^2+2\right) \log \left(\frac{r+1}{\left| r-1\right| }\right)-30 r^7+20 r^5+94 r^3+60 r\right)\right. \\
        &\times\left. \left(2 \HH^2 R-4 \HH-3 \dot\HH R\right)\right\} \, ,
    \end{split}\\
    \begin{split}
        J^{1,\Delta b_1}_{22} \left( r, x \right) = &\;-\frac{9 f^2 \left(r^2 \left(12 x^4-2 x^2-5\right)+2 r x \left(6-11 x^2\right)+8 x^2-3\right)}{140 r^2 \left(r^2-2 r x+1\right)^2}\\
        &+\frac{f}{140 r^2 \left(r^2-2 r x+1\right)^2} \left(-32 r^4 \left(2 x^2+1\right)+32 r^3 x \left(4 x^2+5\right)\right.\\
        &+ \left. r^2 \left(37-4 x^2 \left(57 x^2+46\right)\right)+6 r x \left(64 x^2-19\right)-135 x^2+48\right) \\
        &-\frac{\left(2 r^2-4 r x+3 x^2-1\right) \left(r \left(2 r x^2+r-3 x\right)+1\right)}{7 r^2 \left(r^2-2 r x+1\right)^2} \, ,
    \end{split}\\
    J^{1,\Delta b_2}_{22} \left( r, x \right) = &\;\frac{3  f \left(4 r \left(2 r x^2+r-3 x\right)+3\right)}{20 r^2 \left(r^2-2 r x+1\right)}\, , \\
    J^{1,\Delta b_{12}}_{22} \left( r, x \right) =\; &\frac{ \left(3 r \left(2 r x^2+r-3 x\right)+3\right)}{4 r^2 \left(r^2-2 r x+1\right)}\, , \\
    \begin{split}
        I^{1,\Delta b_{1}}_{22} \left( r, x \right) = &\;\frac{f^2}{35 \HH^2 r^2 R \left(r^2-2 r x+1\right)^2} \left(\HH^2 R \left(r^2 \left(6-5 \left(2 x^4+x^2\right)\right)+r x \left(25 x^2-7\right)-10 x^2+1\right)\right. \\
        &+ \left.3 \HH \left(r^2 \left(4 x^4+2 x^2-1\right)-10 r x^3+4 x^2+1\right)\right.\\
        &+\left. 3 \dot\HH R \left(r^2 \left(4 x^4+2 x^2-1\right)-10 r x^3+4 x^2+1\right)\right)\\
        &-\frac{f \left(2 r x^2+r-3 x\right) \left(r \left(8 r^2-16 r x+42 x^2-13\right)-21 x\right) \left(2 \HH^2 R-4 \HH-3 \dot\HH R\right)}{210 \HH^2 r^2 R \left(r^2-2 r x+1\right)^2} \\
        &+\frac{\left(r \left(3-10 x^2\right)+7 x\right) \left(r \left(-16 r^2+32 r x-42 x^2+5\right)+21 x\right) (2 \HH+\dot\HH R)}{294 \HH^2 r^2 R \left(r^2-2 r x+1\right)^2}\, ,
    \end{split}\\
    I^{1,\Delta b_{2}}_{22} \left( r, x \right) = &\; \frac{f \left(2 r x^2+r-3 x\right) \left(2 \HH^2 R-4 \HH-3 \dot\HH R\right)}{10 \HH^2 r R \left(r^2-2 r x+1\right)}-\frac{\left(6 r x^2+r-7 x\right) (2 \HH+\dot\HH R)}{14 \HH^2 r R \left(r^2-2 r x+1\right)}\, , \\
    I^{1,\Delta b_{12}}_{22} \left( r, x \right) =\; &-\frac{\left(r \left(2 x^2-1\right)-x\right) (2 \HH+\dot\HH R)}{2 \HH^2 r R \left(r^2-2 r x+1\right)}\, .
\end{align}

\subsection{Octupole ($\ell=3$)}

\begin{align}
    \begin{split}
        P_3^{(13),\phi} \left( k \right) =\; &i\Omega_m P(k)\frac{\mathcal{H}}{k}  \left\{\!-\left( \frac{1}{525} \Delta b_1 f (140 f+507)+\frac{3 \Delta b_2 f}{5}\right) k^2  \sigma_v^2 \right.\\
        &+\left.\Delta b_1 \!\! \int\!\!\frac{d^3q}{(2\pi)^3} J^{3,\Delta b_1}_{13}\left(\frac{q}{k}\right)P(q) \!\right\} \, ,
    \end{split}\\
    P_3^{(22),\phi} \left( k \right) =\; &i\Omega_m \frac{\mathcal{H}}{k}  \int \frac{d^3 q}{\left( 2 \pi \right)^3} \left[ \sum_{\Delta b = \left\{\Delta b_1 ,\Delta b_2   \right\}}\!\!\!\!\!\!\! \Delta b J^{3,\Delta b}_{22} \left(\frac{q}{k}, \hat \bk \cdot \hat \bq \right) \right] P \left( q \right) P \left( \left| \bk - \bq  \right|\right)\, , \\
    \begin{split}
        P_3^{(13),v} \left( k \right) =\; &i \frac{\HH}{k} f \Delta b_1 \left\{-\left(\frac{2 f^2 \left(\HH^2 R+2 \HH+2 \dot\HH R\right)}{15 \HH^2 R}+\frac{46 f \left(3 R \left(\HH^2+\dot\HH\right)+4 \HH\right)}{525 \HH^2 R}\right) k^2 \sigma^2_v \right.\\
        &+\left. \!\! \int\!\!\frac{d^3q}{(2\pi)^3} I^{3,\Delta b_1}_{13}\left(\frac{q}{k}\right)P(q)\right\}\, ,
    \end{split}\\
    P_3^{(22),v} \left( k \right) =\; &i \frac{\mathcal{H}}{k} f  \int \frac{d^3 q}{\left( 2 \pi \right)^3} \left[ \sum_{\Delta b = \left\{\Delta b_1 ,\Delta b_2     \right\}}\!\!\!\!\!\!\! \Delta b I^{3,\Delta b}_{22} \left(\frac{q}{k}, \hat \bk \cdot \hat \bq \right) \right] P \left( q \right) P \left( \left| \bk - \bq  \right|\right)\, ,
\end{align}
with
\bea
J^{3,\Delta b_1}_{13} \left( r \right) &=& -\frac{f \left(15 \left(r^2-1\right)^3 \left(7 r^2-4\right) \log \left(\frac{r+1}{\left| r-1\right| }\right)-2 r \left(105 r^6-340 r^4+391 r^2+60\right)\right)}{1400 r^5} \, ,
\\
I^{3,\Delta b_1}_{13} \left( r \right) &=&
\frac{f \left(15 \left(r^2-1\right)^3 \left(3 r^2+1\right) \log \left(\frac{r+1}{\left| r-1\right| }\right)-90 r^7+210 r^5-118 r^3+30 r\right) }{2800 \HH^2 r^5 R}
\nonumber \\
&& \times \left(3 R \left(\HH^2+\dot\HH\right)+4 \HH\right)\, ,
\\
J^{3,\Delta b_{1}}_{22} \left( r, x \right) &=& \frac{f^2 \left(r^2 \left(-138 x^4+123 x^2-5\right)+r x \left(153 x^2-113\right)-42 x^2+22\right)}{60 r^2 \left(r^2-2 r x+1\right)^2}
\nonumber \\
&&
+\frac{ \left(r^2-2 r x+1\right)^{-2}}{70 r^2}f \left(r^4 \left(16-48 x^2\right)+96 r^3 x^3
\right.
\nonumber \\
&&
\left.
\qquad \qquad
+r^2 \left(-216 x^4+122 x^2-31\right)+18 r x \left(11 x^2-6\right)-45 x^2+16\right)\, ,
\\
J^{3,\Delta b_{2}}_{22} \left( r, x \right) &=& 
\frac{3 f \left(r^2 \left(6 x^2-2\right)-4 r x+1\right)}{10 r^2 \left(r^2-2 r x+1\right)}\, ,
\\
I^{3,\Delta b_{1}}_{22} \left( r, x \right) &=&
\frac{f \left(r \left(3 x^2-1\right)-2 x\right) \left(r \left(8 r^2-16 r x+42 x^2-13\right)-21 x\right) \left(3 R \left(\HH^2+\dot\HH\right)+4 \HH\right)}{210 \HH^2 r^2 R \left(r^2-2 r x+1\right)^2}
\nonumber \\
&&+
\frac{f^2  \left(r^2-2 r x+1\right)^{-2}}{90 \HH^2 r^2 R} \left(3 \HH^2 R \left(r \left(r \left(10 x^4-15 x^2+7\right)-5 x^3+x\right)+2\right)
\right.
\nonumber \\
&&
\left.
\qquad
+2 \HH \left(r^2 \left(46 x^4-27 x^2+1\right)+5 r x \left(5-13 x^2\right)+21 x^2-1\right)
\right.
\nonumber \\
&&
\left.
\qquad
+2 \dot\HH R \left(r^2 \left(46 x^4-27 x^2+1\right)+5 r x \left(5-13 x^2\right)+21 x^2-1\right)\right)\, ,
\\
I^{3,\Delta b_{2}}_{22} \left( r, x \right) &=&
-\frac{f \left(r \left(3 x^2-1\right)-2 x\right) \left(3 R \left(\HH^2+\dot\HH\right)+4 \HH^2\right)}{10 \HH^2 r R \left(r^2-2 r x+1\right)}\, .
\eea

\subsection{Triakontadipole ($\ell=5$)}

\bea
P_5^{(13),\phi} \left( k \right) &=&
- i\Omega_m P(k)\frac{\mathcal{H}}{k} \frac{4}{21} f^2 \Delta b_1 k^2  \sigma_v^2\, ,
\\
P_5^{(22),\phi} \left( k \right) &=&
i\Omega_m \frac{\mathcal{H}}{k}   \Delta b_1 \int \frac{d^3 q}{\left( 2 \pi \right)^3} J^{5,\Delta b_1}_{22} \left(\frac{q}{k}, \hat \bk \cdot \hat \bq \right)  P \left( q \right) P \left( \left| \bk - \bq  \right|\right)\, , \quad
\\
P_5^{(13),v} \left( k \right) &=&0\, ,
\\
P_5^{(22),v} \left( k \right) &=&
i \frac{\mathcal{H}}{k} f \Delta b_1  \int \frac{d^3 q}{\left( 2 \pi \right)^3}  I^{5,\Delta b_1}_{22} \left(\frac{q}{k}, \hat \bk \cdot \hat \bq \right) P \left( q \right) P \left( \left| \bk - \bq  \right|\right)\, , \quad
\eea
with
\begin{align}
    J^{5,\Delta b_1}_{22} \left( r, x\right) &= 
    \frac{ f^2 \left(r^2 \left(-30 x^4+33 x^2-5\right)+r x \left(27 x^2-23\right)-6 x^2+4\right)}{21 r^2 \left(r^2-2 r x+1\right)^2}\, , \\
    I^{5,\Delta b_1}_{22} \left( r, x\right) &= \frac{2  f^2 \left(r^2 \left(10 x^4-9 x^2+1\right)+r x \left(7-11 x^2\right)+3 x^2-1\right) \left(3 \HH^2 R+2 \HH+2 \dot\HH R\right)}{63 H^2 r^2 R \left(r^2-2 r x+1\right)^2}\, .
\end{align}

\section{Redshift integrals}
\label{app:zevo}

From fig.~A.1 of \citeref{Breton2018:1803.04294v2} we can extrapolate the following redshift dependence of the halo bias for the different mass bins in the RayGalGroup simulation:
\begin{align}
    b_1^{\rm mb1} \left( z \right)  &= 0.910495 + 0.305421 z + 0.538537 z^2 \, ,\\
    b_1^{\rm mb2}\left( z \right)  &= 0.982831 + 0.736614 z - 0.0573006 z^2\, ,\\
    b_1^{\rm mb3}\left( z \right)  &= 1.1708 + 0.344533 z + 1.14409 z^2\, ,\\
    b_1^{\rm mb4}\left( z \right)  &= 1.36054 + 0.727176 z + 0.666657 z^2\, ,\\
    b_1^{\rm mb5}\left( z \right)  &= 1.61733 + 0.736704 z + 1.61102 z^2\, ,\\
    b_1^{\rm mb6} \left( z \right) &= 1.99799 + 1.58423 z + 0.463353 z^2\, .
\end{align}
When we calculate the model we use this redshift evolution within the redshift bin to redshift average the model following \eq{eq:zaverage}. 

\end{document}